%% file: Dufour_He_2025_Compare_Inequality_Dependent_Samples_arXiv.tex
\documentclass[titlepage,11pt,amstex]{article}
\usepackage[utf8]{inputenc}
\usepackage{amssymb}
\usepackage{amsfonts}
\usepackage{amsmath}
\usepackage{mathptmx}
\usepackage{eurosym}
\usepackage{graphicx}
\usepackage{rotating}
\usepackage[justification=centering]{caption}
\usepackage{comment}
\usepackage{booktabs}
\usepackage{placeins}

\input{Dufour_arXivPreamble.tex}
\input{tcilatex}

\begin{document}

\title{%
\vartitle%
\thanksvar }
\author{%
\varAuthors%
}
\date{%
\vardate%
}
\maketitle

\begin{quote}
\forthcoming%
\end{quote}

\slideFontSize%

\newpage

\begin{Abstract}
\end{Abstract}

\newpage

\tableofcontents%

\newpage

\listoftables%

\listoffigures%

\listoftheorems
\addcontentsline{toc}{section}{\listttheoremnameb}%

\begin{TOCLists}
%
\end{TOCLists}

\newpage

\pagenumbering{arabic} \setcounter{section}{0} \setcounter{page}{1} 
\renewcommand{\thefootnote}{\arabic{footnote}}%

\pagestyle{headings} 

\section{\sectitlesize Introduction \label{Sec: Introduction}}

\resetcountersSection


Researchers often seek to compare inequality levels across different areas
or over time. In practice, inequality measures are estimated from samples,
making it essential to investigate the robustness of these comparisons
through statistical inference. However, most inference procedures assume
either independence between samples or complete dependence (matched pairs of
observations), as detailed in \cite{Mills1997}, \cite{Cowell2015}, \cite%
{Dufour-Flachaire-Khalaf(2019)}, \cite{Dufour2023b}, \cite{Ibragimov2025},
and references therein.

However, much of the income data is dependent on the fact that there is an
overlap of data between the two samples. The overlap refers to paired
observations, such as the incomes of the same household interviewed in two
consecutive periods. For instance, labor force surveys in many countries
are, by design, rotated panel data. Other examples include the Current
Population Survey (CPS), the Panel Study of Income Dynamics (PSID), and the
Consumer Expenditure Survey (CEX) for U.S. income data. In this paper, we
utilize the Survey of Household and Income Wealth (SHIW) from Italy,
featuring an overlap between consecutive waves. Two issues emerge when using
such data: first, some individual income observations are correlated across
periods; second, income data for some units in one of the periods is missing.

Several studies have provided inference procedures with overlapping samples. 
\cite{Zheng2001} proposed an asymptotic method for comparing interpolated
Gini indices and generalized entropy measures+. \cite{Zheng2004} extends
results to the Foster\emph{-}Greer\emph{-}Thorbecke poverty measure. \cite%
{Biewen2002} presented a percentile bootstrap for comparing indices in a
family which are smooth functions of population moments. The above studies
focus on specific measurements and do not cover the popular Gini index and
quantile-based measurements, such as the Lorenz curve. Moreover, some
indices are covered by only one of the two methods (\emph{i.e.}, the
asymptotic and bootstrap methods).

In this paper, we provide inference procedures for a broad class of
inequality, poverty, and welfare measures within a unified framework. We
focus on estimators in the form of asymptotically linear Gaussian
functionals (defined below) previously considered in \cite[henceforth DFK]%
{Dufour-Flachaire-Khalaf(2019)}. Nevertheless, we differ from DFK in two
central ways. First, DFK focus on hypothesis tests based on permutation
methods, while this is not the case here. Second, DFK assume independent
samples, while we allow dependent samples. 
Nevertheless, we do not have such a restriction. So, we can
study a broader family of indices, including quantile-based measurements
such as the Lorenz curve.

Our methods are closely related to the literature on influence functions
(IF), dating back to \cite{Hampel1974}. As one of the earliest applications
in econometrics, \cite{Cowell1996} investigated the robustness of various
inequality measurements. In contrast, we employ the IF to derive the
asymptotic distribution and estimate the asymptotic variance with great
computational advantage; see, for example, \cite{Cowell2015}. Furthermore,
this paper extends the estimation of asymptotic variances to dependent
samples. We make the following contributions.

\emph{First}, we focus on a wide range of indices covering most inequality,
poverty, and risk analysis measurements, including the well-known Gini index
and Lorenz curve.

\emph{Second}, we develop methods applicable without any assumption on the
dependence between the compared samples. This is done by using the
intersection approach previously considered by \cite{Dufour1998} in a
finite-sample framework. We introduce asymptotic and bootstrap intersection
methods (IMs) for assessing changes in indices. These methods utilize
information within each sample and handle arbitrary dependence between two
samples. Furthermore, our method provides solutions to a generalized
Behrens-Fisher problem, which involves comparing the means of two
distributions with arbitrary dependence across the two samples.

\emph{Third}, we examine overlapping samples, a specific form of dependent
samples where sample dependence arises solely from overlap (\emph{i.e.},
matched pairs). Such a framework is applicable to split or rotating panels
in survey data. We propose asymptotic and bootstrap inference for index
differences, and we present a consistent and numerically positive (definite)
estimator for the asymptotic variance.

\emph{Fourth}, we conduct a series of simulation experiments to examine the
performance of the proposed methods. Our confidence intervals based on
overlapping samples have coverage rates close to the nominal level with
reasonable widths. By contrast, conventional inference -- which ignores
sample dependence -- can fail to adequately cover the actual value
sufficiently frequently or become overly conservative with large widths. We
then analyze the impacts of heavy tails and find that the asymptotic
inference performs well for realistic distributions but poorly for
distributions with heavy tails. The bootstrap method can alleviate this
issue, except when the variance is substantial or nonexistent. Wed show that
intersection methods can yield reliable results in all scenarios,
particularly when overlapping samples are not fulfilled, and can be
reasonably efficient in some instances, such as extremely heavy-tailed
distributions.

\emph{Fifth}, we apply the proposed methods to investigate the dynamic
change of financial inequality from 2012 to 2014. The results reveal
distinct patterns of internal inequality within the two regions,
highlighting the practical importance of negative values in inequality
measurement.

The paper is organized as follows. Section \ref{sec:index} reviews some
commonly used inequality measures which are asymptotic linear. Section \ref%
{sec: Intersection methods} describes inference based on intersection
methods. Section \ref{Sec: Overlapping samples} discusses the special case
of overlapping samples. We assess the performance of the proposed procedures
through Monte Carlo experiments in Section \ref{sec:MC}. In Section \ref%
{sec:App-SHIW}, we apply the methods to analyze the dynamic change in
household financial inequality in Italy. We briefly discuss the extensions
to clustered data to broaden the scope of our analysis in Section \ref%
{sec:clusters}. The paper concludes in Section \ref{sec:conclusion}. All the
proofs are given in the appendix.

\newpageSlides

\section{\sectitlesize Asymptotically linear Gaussian functionals \label%
{sec:index}}

\resetcountersSection

To extend existing results to a more general form of welfare indices, we
focus on the class of asymptotically linear functionals as defined below.
Such a collection of parameters encompasses many welfare measurements,
allowing us to present the asymptotic distribution conveniently.

To facilitate our discussion, we introduce some notation. Let $F\in \mathcal{%
F}$ be a cumulative distribution function (CDF), and $\theta :\mathcal{F}%
\rightarrow \mathbb{R}$ a functional of distribution $F$. For a sample $%
\{X_{i}\}_{i=1}^{n}$, we denote by $\hat{F}$ the empirical distribution
function (EDF) for $F$, and estimate $\theta (F)$ by $\theta (\hat{F})$. We
denote by $o_{p}(1)$ a variable which converges (in probability) to zero as $%
n\rightarrow \infty $. Let $X_{n}\overset{d}{\longrightarrow }X$ be a
sequence of variables $X_{n}$ convergence in distribution to $X$. We
consider a random sample of independent and identically distributed (iid)
observations $\{X_{i}\}_{i=1}^{n}$ drawn from the distribution $F$, written
as $\{X_{i}\}_{i=1}^{n}\overset{iid}{\sim }F$. We now define asymptotically
linear Gaussian (ALG) functionals.

\begin{definition}
\label{def:ALG}%
\captiondefinition{\definitionname}{Asymptotically linear Gaussian functional}
Given a sample $\{X_{i}\}_{i=1}^{n}$ drawn from a distribution $F$, an
estimator $\theta (\hat{F})$ for the functional $\theta (F)$ is \emph{%
asymptotically linear Gaussian}, if it satisfies 
\begin{equation}
\sqrt{n}[\theta (\hat{F})-\theta (F)]=\frac{1}{\sqrt{n}}\sum_{i=1}^{n}\psi
(X_{i};\theta ,F)+o_{p}(1)\underset{n\rightarrow \infty }{\overset{d}{%
\longrightarrow }}\mathrm{N}[0,\sigma _{\theta }^{2}(F)]  \label{eq:ALG}
\end{equation}%
where $\sigma _{\theta }^{2}(F):=\mathrm{Var}[\psi (X_{i};\theta ,F)]$.
\end{definition}

In many cases, the function $\psi $ in \eqref{eq:ALG} turns out to be the
(asymptotic) influence function (IF) of the estimator $\theta (\hat{F})$,
which is defined as follows: 
\begin{equation}
\psi (x;\theta ,F)=\lim_{t\rightarrow 0}\dfrac{\theta ((1-t)F+t\delta _{x})}{%
t}
\end{equation}%
where $\delta _{x}$ denotes a point mass at $x$. The IF has been widely used
to analyze the robustness of an estimator, as it quantifies the effect on an
estimator when a single data point in the sample is contaminated; see \cite%
{Wasserman2006} for further details. On the other hand, we utilize the IF as
a convenient device to derive asymptotic distributions, especially the
asymptotic covariance matrices.

Definition \ref{def:ALG} relies on a high-level condition, but in practice
we often impose more primitive assumptions. Table \ref{tab:ALGIndex} lists
various ALG estimators for several popular indices. For illustrative
purposes, we provide examples of the Gini index and the Lorenz curve (LC)
below, whose sufficient conditions are provided in the referenced papers.

\begin{table}[tb]%

\caption{A brief list of welfare indices allow asymptotic linear estimators.}
\label{tab:ALGIndex}

\begin{center}
Table \thetable

List of estimators for welfare indices as asymptotic linear Gaussian
functionals
\end{center}

\hspace{-0.5\totalhormargin} \begin{minipage}{\paperwidth}
\centering%

\small%

\begin{center}
\begin{tabular}{|c|m{6cm}|c|}
\hline
\rule[-1ex]{0cm}{4ex} & Welfare indices & References \\ \hline
\rule[-1ex]{0cm}{4ex} Inequality & Gini index, Lorenz curve, generalized
Lorenz curve, generalized entropy class, Atkinson index, coefficient of
variation, Pietra ratio & \cite{Cowell2015} \\ \cline{2-3}
\rule[-1ex]{0cm}{4ex} & S-Gini index, E-Gini index & \cite{Barrett2009} \\ 
\cline{2-3}
\rule[-1ex]{0cm}{4ex} & extended Lorenz curves, extended Gini indices & \cite%
{Dufour2024} \\ \hline
\rule[-1ex]{0cm}{4ex} Poverty & Foster-Greer-Thorbecke poverty measure, Sen
poverty index, Sen-Shorrocks-Thon Poverty Index & \cite{Cowell2015} \\ 
\cline{2-3}
\rule[-1ex]{0cm}{4ex} & Gini poverty index & \cite{Barrett2009} \\ \hline
\rule[-1ex]{0cm}{4ex} Risk measurement & Value-at-Risk, expected shortfall & 
\cite{Zhang2021a} \\ \hline
\end{tabular}
\end{center}

\end{minipage}%

\medskip 
\noindent
\footnotesize%
Note -- The table displays a collection of estimators in the form of
asymptotically linear Gaussian functionals in Definition \ref{def:ALG}.
Relevant conditions and the expressions of the influence functions are given
in the referenced papers.

\normalsize%

\end{table}%

\begin{example}
\label{Eg:Gini} 
\captionresult{\examplename}{Gini index}
The Gini index, also known as the Gini coefficient, measures the degree of
inequality using a scalar value ranging from $0$ to $1$, where $0$
represents perfect equality and $1$ indicates complete inequality. For an
income variable $X\geq 0$ with distribution $F$, there are over a dozen
formulas for the Gini index; see, for instance, \cite{Yitzhaki2013}. We will
use the following definition: 
\begin{equation}
I(F)=\dfrac{2\int tF(t)dF(t)}{\mu (F)}-1.
\end{equation}%
One can estimate $I(F)$ by plugging in the EDF $\hat{F}$, which yields 
\begin{equation}
I\big(\hat{F}\big)=\frac{\frac{2}{n^{2}}\sum_{i=1}^{n}(i-0.5)X_{(i)}}{\mu %
\big(\hat{F}\big)}-1
\end{equation}%
where $X_{(1)}\leq \cdots \leq X_{(n)}$ are the ordered statistics from the
sample $\{X_{i}\}_{i=1}^{n}$. \cite{Davidson2009} has proposed a
bias-corrected estimate $\widetilde{I}=\left( \frac{n}{n-1}\right) I\big(%
\hat{F}\big)$. The author then applies the delta method to demonstrate that
both estimators satisfy the Definition \ref{def:ALG} with $\psi
(x;L,F)=h(x;L,F)-\mathbb{E}[h(X;L,F)]$, where 
\begin{equation}
h(x;I,F)=\frac{2xF(x)-2\int t\mathbf{1}(t<x)\,dF(t)-[I(F)+1]x}{\mu (F)}.
\label{eq:Gini-IF-aux}
\end{equation}%
We can estimate this function by substituting the EDF $\hat{F}$, which can
be efficiently computed based on order statistics as follows. 
\begin{equation}
h\big(X_{(i)};I,\hat{F}\big)=\frac{\left( \frac{2i}{n}\right) X_{(i)}-\frac{2%
}{n}\sum_{j=1}^{i}X_{(j)}-(\hat{I}+1)X_{(i)}}{\frac{1}{n}\sum_{i=1}^{n}X_{i}}%
.
\end{equation}
\end{example}

\begin{example}
\label{Eg:LC} 
\captionresult{\examplename}{Lorenz curve ordinates}
Consider an income variable $X\geq 0$ with distribution $F$. The Lorenz
curve (LC) is defined as follows [\cite{Gastwirth1971}]: 
\begin{equation}
L(p;F)=\dfrac{\int_{0}^{p}Q(u;F)\,du}{\mu (F)}  \label{eq:LC-def}
\end{equation}%
where $\mu (F)=\int x\,dF>0$ is the mean and $Q(p;F)=\inf \{x:F(x)\geq p\}$
is the $p$-th quantile. A point on the curve, such as $L(p)=m$, indicates
that the $100p\%$ poorest households receive $100m\%$ of the total income.
We obtain the following estimator by substituting the EDF $\hat{F}$: 
\begin{equation}
L\big(p;\hat{F}\big)=\dfrac{\frac{1}{n}\sum_{i=1}^{n}X_{i}\mathbf{1}\big[%
X_{i}\leq Q\big(p;\hat{F}\big)\big]}{\mu \big(\hat{F}\big)}\,.
\end{equation}%
\cite{Beach1983} and \cite{Cowell2015} showed by the delta method that $L%
\big(p;\hat{F}\big)$ satisfies Definition \ref{def:ALG} with $\psi
(x;L,F,p)=h(x;L,F,p)-\mathbb{E}[h(X;L,F,p)]$, where 
\begin{equation}
h(x;L,F,p)=\frac{\big[x-Q(p;F)\big]\mathbf{1}\big[x\leq Q(p;F)\big]-x\,L(p;F)%
}{\mu (F)}\,.  \label{eq:LC-IF-aux}
\end{equation}%
\end{example}

There are often multiple ways to establish the ALG form in \eqref{eq:ALG}.
For instance, using the delta method, \cite{Davidson2009} showed that the
Gini index is an ALG functional. However, one can also achieve results by
using more advanced techniques, such as empirical process theory [\cite%
{Davidson2010a}]. Likewise, the S-Gini [\cite{Donaldson1980}] and E-Gini [%
\cite{Chakravarty1988}], which are two popular extensions of the Gini index,
satisfy Definition \ref{def:ALG} using either the delta method [\cite%
{Davidson2010a}] or empirical process theory [\cite{Barrett2009}]. In this
paper, we only assume the existence of an ALG form as defined in Definition %
\ref{def:ALG}, thereby avoiding various complex conditions.

One merit of the ALG functional in Definition \ref{def:ALG} is its
computational efficiency in calculating estimators for the asymptotic
variances of the indices. Specifically, we can estimate the asymptotic
variance by its sample analog: 
\begin{equation}
\hat{\sigma }^{2}:=\sigma _{\theta }^{2}(\hat{F})=\frac{1}{n}%
\sum_{i=1}^{n}\psi ^{2}(X_{i};\theta ,\hat{F})  \label{eq:AVar-Index-Est}
\end{equation}%
where $\hat{F}$ denotes the EDF and $\psi $ is given in \eqref{eq:ALG}. \cite%
{Barrett2009} has validated the consistency of $\hat{\sigma }^{2}$; see
section \ref{sec:OS-AVar} for a more general scenario.


In the next two sections, we derive asymptotic inference methods for index
differences under dependent samples, when both sample sizes go to infinity.
In section \ref{sec: Intersection methods}, we consider generic dependent
samples, where we impose no restrictions on the dependence between the two
samples or the joint limiting distribution (if it exists). Our proposed
inference method is therefore applicable and robust to a wide range of
scenarios. Then, in section \ref{Sec: Overlapping samples}, we delve
consider a more specific data structure called overlapping samples, where
the dependence arises from matched pairs.

We use the following notation. For distribution $F_{k}$, $k=1,2$, we
estimate the index $\theta _{k}=\theta _{k}(F_{k})$ by $\hat{\theta }%
_{k}=\theta (\hat{F}_{k})$, where $\hat{F}_{k}$ is the EDF for $F_{k} $
based on $\{X_{ki}\}_{i=1}^{n_{k}}$. We denote $\Delta \hat{\theta }=\hat{%
\theta }_{1}-\hat{\theta }_{2}$ the estimate of the indices difference $%
\Delta \theta =\theta _{1}-\theta _{2}$. We wish to make inference on $%
\Delta \theta $. This problem is a nonparametric version of the Behrens\emph{%
-}Fisher problem, which involves comparing means of potentially distinct
distributions [\cite[p 846]{Lehmann2022}]. We will also consider the
possibility of dependence between the two samples.


\FloatBarrier

\newpageSlides

\section{\sectitlesize Intersection methods for generic dependent samples 
\label{sec: Intersection methods}}

\label{sec:DS} 

We start with the generic dependent samples, where the observations in one
sample may not be independent of those of the other sample. For simplicity,
the observations inside each sample are assumed to be independent and
identically distributed (iid). However, it is straightforward to extend
results to clustered data using techniques by \cite{Zheng2002}, \cite%
{Bhattacharya2007}, and \cite{Ibragimov2025}.

\begin{assumption}
\label{assump:DS}%
\captionassumption{\assumptionname}{Generic dependent samples}
Assume that $\{X_{1i}\}_{i=1}^{n_{1}} \overset{iid}{\sim} F_{1}$ and $%
\{X_{2i}\}_{i=1}^{n_{2}} \overset{iid}{\sim} F_{2}$.
\end{assumption}

Assumption \ref{assump:DS} only focuses on individual samples $%
\{X_{1i}\}_{i=1}^{n_{1}}$ and $\{X_{2i}\}_{i=1}^{n_{2}}$, allowing arbitrary
forms of dependence between the two samples. The generality of this
assumption will be more apparent when we introduce overlapping samples in
the subsequent section.


\subsection{Asymptotic intersection method \label{sec: Asymptotic
intersection method}}

\label{sec:Asym-IM}


As mentioned in \eqref{eq:AVar-Index-Est}, $\hat{\sigma }_{k}$ is a
consistent estimator for the asymptotic variance of the ALG functional $\hat{%
\theta }_{k}$, where 
\begin{equation}
\hat{\sigma }_{k}^{2}:=\sigma _{\theta }^{2}(\hat{F}_{k})=\frac{1}{n_{k}}%
\sum_{i=1}^{n_{k}}\psi ^{2}(X_{ki};\theta ,\hat{F}_{k}),\quad k=1,2.
\label{eq:AVar-Index-Est-k}
\end{equation}%
Let $q_{p}$ be the $p$-th quantile of the standard Gaussian distribution. We
then have the following asymptotic confidence interval (CI) for $\theta _{k}$
at level $(1-\alpha _{k})$, denoted by $\left[ L_{n_{k}},U_{n_{k}}\right] $,
where 
\begin{equation}
L_{n_{k}}=\hat{\theta }_{k}-q_{1-\frac{\alpha _{k}}{2}}\frac{\hat{\sigma }%
_{k}}{\sqrt{n_{k}}},\quad U_{n_{k}}=\hat{\theta }_{k}+q_{1-\frac{\alpha _{k}%
}{2}}\frac{\hat{\sigma }_{k}}{\sqrt{n_{k}}},\quad k=1,2.
\label{eq:AsymCI_Index}
\end{equation}%
The limiting confidence coefficient is defined as $\lim_{n_{k}\rightarrow
\infty }\inf_{\theta _{k}\in \Theta _{k}}\Pr_{\theta _{k}}\left(
L_{n_{k}}\leq \theta _{k}\leq U_{n_{k}}\right) $, provided the limit exists (%
\cite[p 526]{Lehmann2022}).

We propose an asymptotic version of the intersection method (IM) for $%
\Delta\theta$, adapted from \cite{Dufour2024d}, who extends results in \cite%
{Dufour1998} to finite-sample inference.

\begin{proposition}
\label{prop:AsymIM} 
\captiontheorem{Proposition}{Asymptotic intersection
method} Let $\theta (F_{k})$ be an asymptotically linear Gaussian
functional, and $[L_{n_{k}},U_{n_{k}}]$ in \eqref{eq:AsymCI_Index} an
asymptotic CI for $\theta _{k}$ at the level of $(1-\alpha _{k})$ for $k=1,2$%
. Then, under Assumption \ref{assump:DS}, the following properties hold$:$%
\newline
$(1)$ $\left[ L_{n_{1}}-U_{n_{2}},\ U_{n_{1}}-L_{n_{2}}\right] $ is an
asymptotic CI for $\Delta \theta $ with level $(1-\alpha )$, where $\alpha
_{1}$ and $\alpha _{2}$ are such that $\alpha \geq \alpha _{1}+\alpha _{2}$ $%
;$ \newline
$(2)$ if the two samples are independent and $[L_{n_{k}},U_{n_{k}}]$ for $%
k=1,2$ have limiting confidence coefficients $(1-\alpha _{k})$, then $\left[
L_{n_{1}}-U_{n_{2}},\ U_{n_{1}}-L_{n_{2}}\right] $ is an asymptotic CI for $%
\Delta \theta $ at the level of $(1-\alpha )$, where $\alpha _{1}$ and $%
\alpha _{2}$ are such that $\alpha =\alpha _{1}+\alpha _{2}-\alpha
_{1}\alpha _{2}$.
\end{proposition}


\subsection{Bootstrap intersection method \label{sec: Bootstrap intersection
method}}

\label{sec:Boot-IM} 

There are several ways to construct a bootstrap CI for $\theta _{k}$, $k=1,2$%
. One good choice, at least in theory, is the Studentized bootstrap, or
percentile-t, CI [\cite{Hall1992}]. The procedure is as follows [\cite%
{Davidson2004}].

\begin{enumerate}[$(1)$]%

\item
Given the original samples $\{X_{ki}\}_{i=1}^{n_{k}}$, obtain estimates $%
\hat{\theta}_{k}$ and $\hat{\sigma}_{k}$ for $k=1,2$.

\item
Construct the bootstrap samples $\{X^{*}_{ki}\}_{i=1}^{n_{k}}$ by resampling
with replacement from $\{X_{ki}\}_{i=1}^{n_{k}}$. Next, compute the
estimates $\hat{\theta}^{*}_{k}$ and $\hat{\sigma}^{*}_{k}$, and calculate
statistic $T^{*} = \big(\hat{\theta}^{*}_{k} - \hat{\theta}_{k}\big) / \hat{%
\sigma}^{*}_{k}$.

\item
Repeat the previous step $B$ times to obtain the statistics $%
\{T^{*}_{j}\}_{j = 1}^{B}$. Then, compute $q^{*}_{\frac{\alpha_{k}}{2}}$ and 
$q^{*}_{1 - \frac{\alpha_{k}}{2}}$, where $q^{*}_{p}$ is the $\lceil pB
\rceil$ order statistics of $\{T^{\ast}_{j}\}_{j = 1}^{B}$. Here, $\lceil
\cdot \rceil$ denotes the ceiling function.

\item
The bootstrap CI for $\theta _{k}$ at the level $(1-\alpha _{k})$ is $\left[
L_{n_{k}}^{B},U_{n_{k}}^{B}\right] $, where 
\begin{equation}
L_{n_{k}}^{B}=\hat{\theta }_{k}-\hat{\sigma }_{k}q_{1-\alpha _{k}/2}^{\ast
},\quad U_{n_{k}}^{B}=\hat{\theta }_{k}-\hat{\sigma }_{k}q_{\alpha
_{k}/2}^{\ast }.  \label{eq:BootCI_Index}
\end{equation}

\end{enumerate}%

Ideally, the number of bootstrap samples $B$ should be reasonably large and
satisfy the condition that $\alpha (B + 1)$ is an integer for any test level 
$\alpha$ [\cite{Davidson2000a}]. We then propose a bootstrap version of the
intersection method for $\Delta\theta$.

\begin{proposition}
\label{prop:BootIM} 
\captiontheorem{Proposition}{Bootstrap intersection
method} Let $\theta (F_{k})$ be an asymptotically linear Gaussian functional
and $[L_{n_{k}}^{B},U_{n_{k}}^{B}]$ in \eqref{eq:BootCI_Index} be a
bootstrap CI for $\theta _{k}$ at the level of $(1-\alpha _{k})$ for $k=1,2$%
. Then, Assumption \ref{assump:DS}, the following properties hold : \newline
$(1)$ $\left[ L_{n_{1}}^{B}-U_{n_{2}}^{B},\ U_{n_{1}}^{B}-L_{n_{2}}^{B}%
\right] $ is a bootstrap CI for $\Delta \theta $ with level $(1-\alpha )$,
where $\alpha _{1}$ and $\alpha _{2}$ are such that $\alpha \geq \alpha
_{1}+\alpha _{2}$ $;$\newline
$(2)$ if the two samples are independent and $[L_{n_{k}}^{B},U_{n_{k}}^{B}]$
for $k=1,2$ have limiting confidence coefficients $(1-\alpha _{k})$, then $%
\left[ L_{n_{1}}^{B}-U_{n_{2}}^{B},\ U_{n_{1}}^{B}-L_{n_{2}}^{B}\right] $ is
a bootstrap CI for $\Delta \theta $ at the level of $(1-\alpha )$, where $%
\alpha _{1}$ and $\alpha _{2}$ are such that $\alpha =\alpha _{1}+\alpha
_{2}-\alpha _{1}\alpha _{2}$.
\end{proposition}

Intersection methods (IMs) are asymptotically conservative, which is as
expected due to the trade-off between efficiency and robustness. Recall that
the IMs do not require any information on sample dependence or the limiting
joint distribution of estimators. The following section will demonstrate how
to achieve more efficient inference methods under additional assumptions.


\section{Overlapping samples \label{Sec: Overlapping samples}}

\label{sec:OS} 

In this section, we focus on a special type of sample known as overlapping
samples (OS), which accommodate sample dependence that stems exclusively
from the overlap (\emph{i.e.}, matched pairs). Such a data structure can be
employed for rotating or splitting panels in survey data. Several authors,
including \cite{Zheng2001}, \cite{Zheng2004}, and \cite{Biewen2002}, have
investigated overlapping samples in their analyses of inequality and poverty
and proposed inference methods for specific indices. We describe the OS as
follows.

\begin{assumption}
\label{assump:OS}\captionassumption{\assumptionname}{Overlapping samples}
Let $\{X_{1i}\}_{i=1}^{n_{1}}$ and $\{X_{2i}\}_{i=1}^{n_{2}}$ two samples
which satisfy Assumption \ref{assump:DS}.Further, the following conditions
are satisfied.

\begin{enumerate}
\item[(A1)] \label{assump:OS-MP-obs} The first $m$ observations are matched
pairs $\{(X_{1i},X_{2i})\}_{i=1}^{m}$, where $0\leq m\leq \min (n_{1},n_{2})$%
.

\item[(A2)] \label{assump:OS-MP-IID} The matched pairs are drawn from a
fixed joint distribution $\widetilde{F}$, whose marginal distributions are $%
F_{1}$ and $F_{2}$, \emph{i.e.} $\{(X_{1i},X_{2i})\}_{i=1}^{m}\overset{iid}{%
\sim }\widetilde{F}$.

\item[(A3)] \label{assump:OS-unMP} The observations that are unmatched in
each sample are independent of the observations in the other sample.

\begin{enumerate}
\item $\{X_{1i}\}_{i=m+1}^{n_{1}}$ is independent of $\{X_{2i}%
\}_{i=1}^{n_{2}}$.

\item $\{X_{2i}\}_{i=m+1}^{n_{2}}$ is independent of $\{X_{1i}%
\}_{i=1}^{n_{1}}$.
\end{enumerate}

\item[(A4)] \label{assump:OS-Asym} As $n_{1},n_{2}\rightarrow \infty $, we
require that $\frac{m}{n_{1}}\rightarrow \lambda _{1}$, $\frac{m}{n_{2}}%
\rightarrow \lambda _{2}$, $\frac{n_{1}}{n_{1}+n_{2}}\rightarrow \eta _{1}$
and $N:=\frac{n_{1}n_{2}}{n_{1}+n_{2}}\rightarrow \infty $, where $\lambda
_{1},\lambda _{2}\in \lbrack 0,1]$, and $\eta _{1}\in \lbrack 0,1]$.
\end{enumerate}
\end{assumption}

Condition {$($\textup{A1}$)$} assumes that the first $m$ observations in
each sample are matched pairs. Furthermore, it implicitly requires that we
know which two observations in each sample form a pair. For instance, when
comparing income difference between parents and children, the incomes of a
parent and a child from the same family make up a pair. If the pairing
information is lost, so that it is unclear whether or not the child and
parent come from the same household, then such samples will violate {$($%
\textup{A1}$)$}.

Condition {$($\textup{A2}$)$} makes the common assumption that pairs of data
are iid observations from a fixed unknown joint distribution $\widetilde{F}$%
. We may relax such a condition to independent but not identically
distributed.

Condition {$($\textup{A3}$)$} imposes independence between unmatched data
and entire observations in the other sample, indicating that the overlap is
the only source of dependence between the two samples.

Condition {$($\textup{A4}$)$} clarify what we mean by asymptotic in OS.
First, while both $n_{1}$ and $n_{2}$ go to infinity, the size of overlap $m$
can be zero or a fixed number, giving $\lambda _{1}=\lambda _{2}=0$. Second, 
$n_{1}$ and $n_{2}$ can increase to infinity at different orders of speed
such that $\eta _{1}=0$ or $\eta _{1}=1$ (and $\lambda _{1}=\lambda _{2}=0$%
). So, we do not view {$($\textup{A4}$)$} as a limitation; also see the
discussion following the Proposition \ref{prop:AsyN-dIndex}.

To complete Assumption \ref{assump:OS}, we make an asymptotic normality
assumption on matched pairs.

\begin{assumption}
\label{assump:JointNormal}%
\captionassumption{\assumptionname}{Asymptotically joint normality}
Let $\hat{\theta}_{k}$ be an asymptotically linear Gaussian functional with $%
\psi _{k}(x):=\psi (x;\theta ,F_{k})$ in \eqref{eq:ALG} for $k=1,2$. If $%
m\rightarrow \infty $, then 
\begin{equation}
\frac{1}{\sqrt{m}}\sum_{i=1}^{m}%
\begin{bmatrix}
\psi _{1}(X_{1i}) \\ 
\psi _{2}(X_{2i})%
\end{bmatrix}%
\underset{m\rightarrow \infty }{\overset{d}{\longrightarrow }}\mathrm{N}%
\left[ \left( 
\begin{array}{c}
0 \\ 
0%
\end{array}%
\right) ,\left( 
\begin{bmatrix}
\sigma _{1}^{2} & \sigma _{12} \\ 
\sigma _{12} & \sigma _{2}^{2}%
\end{bmatrix}%
\right) \right]  \label{eq:Pairs_MvtNormal}
\end{equation}%
where $\sigma _{k}^{2}=\mathrm{Var}[\psi _{k}(X_{ki})]$ and $\sigma _{12}=%
\mathrm{Cov}[\psi _{1}(X_{1i}),\psi _{2}(X_{2i})]$. Otherwise, we have $%
\lambda _{1}=\lambda _{2}=0$, where $\lambda _{1}$ and $\lambda _{2}$ are
given in Assumption \ref{assump:OS}.
\end{assumption}

Condition \eqref{eq:Pairs_MvtNormal} is technically necessary because
Gaussian marginal distributions do not imply a multivariate normal
distribution. In practice, however, this assumption is often satisfied for
many indices in Table \ref{tab:ALGIndex} under the conditions of Assumption %
\ref{assump:OS}. Specifically, $\left\{ (\psi _{1}(X_{1i}),\psi
_{2}(X_{2i}))\right\} _{i=1}^{m}$ are often iid random vectors. Then, by the
standard Central Limit Theorem, the sample average has bivariate normal
asymptotic distribution.


\subsection{Asymptotic distribution of index difference \label{sec:
Asymptotic distribution of index difference}}

\label{sec:OS-Asym} 

We first provide the asymptotic distribution of the estimator for the index
difference $\Delta \hat{\theta}$ with overlapping samples.

\begin{proposition}
\label{prop:AsyN-dIndex} 
\captionproposition{\propositionname}{Asymptotic method with overlapping samples}
If $\hat{\theta}_{1}$ and $\hat{\theta}_{2}$ are asymptotically linear
Gaussian functionals with $\psi _{k}(x):=\psi (x;\theta ,F_{k})$ in %
\eqref{eq:ALG} for $k=1,2$. Then, under the Assumptions \ref{assump:OS} and %
\ref{assump:JointNormal}, we have$:$ 
\begin{equation}
\sqrt{N}\big(\Delta \hat{\theta}-\Delta \theta \big)\underset{%
n_{1},n_{2}\rightarrow \infty }{\overset{d}{\longrightarrow }}\mathrm{N}%
[0,\sigma _{\Delta }^{2}]  \label{eq:AVar-IndexDiff}
\end{equation}%
where $\sigma _{\Delta }^{2}=(1-\eta _{1})\sigma _{1}^{2}+\eta _{1}\sigma
_{2}^{2}-2\sqrt{\eta _{1}(1-\eta _{1})\lambda _{1}\lambda _{2}}\sigma _{12}$%
, $\sigma _{k}^{2}=\mathrm{Var}[\psi _{k}(X_{k1})]$, $\sigma _{12}=\mathrm{%
Cov}[\psi _{1}(X_{11}),\psi _{2}(X_{21})]$ and $N\rightarrow \infty $, $\eta
_{1}$, $\lambda _{1}$, $\lambda _{2}\in \lbrack 0,1]$ are defined in
Assumption \ref{assump:OS}.
\end{proposition}

The last term of the asymptotic variance in \eqref{eq:AVar-IndexDiff}
indicates that the impact of sample dependency relies on both the overlap
portions (\emph{i.e.}, $\lambda _{1}\lambda _{2}$) and the covariance
between $\psi _{1}$ and $\psi _{2}$. Note that $\psi _{k}$ is often a
non-monotonic transformation; see Example \ref{Eg:Gini} and \ref{Eg:LC}. So
it is challenging to predict the value or even the sign of $\sigma _{12}$
based on $\mathrm{Cov}(X_{1},X_{2})$. We will revisit this issue in the
simulation experiments.

Assumption \ref{assump:OS} allows extremely unbalanced samples (where $\eta
_{1}=0$ or $1$). Take $\eta _{1}=1$ as an example. We then have $\lambda
_{1}=\lambda _{2}=0$, $N\approx n_{2}\rightarrow \infty $, and $\sigma
_{\Delta }^{2}=\sigma _{2}^{2}$, indicating $\hat{\theta}_{2}$ as the only
source of uncertainty. One way to interpret these results is that the sample
size $n_{1}$ is so much larger than $n_{2}$ that the uncertainty of $\hat{%
\theta}_{2}$ dominates that of $\hat{\theta}_{1}$. In other words, we can
treat $\{X_{1i}\}_{i=1}^{n_{1}}$ as the population and $\hat{\theta}_{1}$ as
the true parameter $\theta _{1}$ (without sampling errors).

We now apply the above proposition to obtain an asymptotic distribution for
estimates of differences in Gini indices and LC ordinates.

\begin{example}
\label{Eg:dGini} 
\captionresult{\examplename}{Gini index (continued)}
Under the conditions of Proposition \ref{prop:AsyN-dIndex}, we have$:$ 
\begin{equation}
\sqrt{N}\big(\Delta \hat{I}-\Delta I\big)\underset{n_{1},n_{2}\rightarrow
\infty }{\overset{d}{\longrightarrow }}\mathrm{N}\big(0,\sigma _{\Delta }^{2}%
\big)
\end{equation}%
where $\sigma _{\Delta }^{2}=(1-\eta _{1})\sigma _{1}^{2}+\eta _{1}\sigma
_{2}^{2}-2\sqrt{\eta _{1}(1-\eta _{1})\lambda _{1}\lambda _{2}}\sigma _{12}$%
, $\sigma _{k}^{2}=\mathrm{Var}[h(X_{k1};I,F_{k})]$ for $k=1,2$, and $\sigma
_{12}=\mathrm{Cov}[h(X_{11};I,F_{1}),h(X_{21};I,F_{2})]$ with $h(x;I,F)$
given in \eqref{eq:Gini-IF-aux}.
\end{example}

\begin{example}
\label{Eg:dLC} 
\captionresult{\examplename}{Lorenz curve ordinates (continued)}
By Proposition \ref{prop:AsyN-dIndex}, we have$:$ 
\begin{align}
& \sqrt{N}\big(\Delta \hat{L}(p)-\Delta L(p)\big)\underset{%
n_{1},n_{2}\rightarrow \infty }{\overset{d}{\longrightarrow }}\mathrm{N}\big(%
0,\sigma _{\Delta }^{2}(p)\big)\,, \\
\sigma _{\Delta }^{2}(p)& =(1-\eta _{1})\sigma _{1}^{2}(p)+\eta _{1}\sigma
_{2}^{2}(p)-2\sqrt{\eta _{1}(1-\eta _{1})\lambda _{1}\lambda _{2}}\sigma
_{12}(p)\,,
\end{align}%
where $\sigma _{k}^{2}(p)=\mathrm{Var}[h(X_{k1};L,F_{k},p)]$ for $k=1,2$,
and $\sigma _{12}(p)=\mathrm{Cov}[h(X_{11};L,F_{1},p),h(X_{21};L,F_{2},p)]$
with $h(x;L,F,p)$ given in \eqref{eq:LC-IF-aux}.

We can extend the results to a vector of Lorenz curve ordinates at $\mathbf{p%
}=(p_{1},\ldots ,\,p_{s})^{\prime }$. Let $h(X_{ki};L,F_{k},\mathbf{p}%
)=[h(X_{ki};L,F_{k},p_{1}),\ldots ,\,h(X_{ki};L,F_{k},p_{s})]^{\prime }$.
The asymptotic variance is 
\begin{equation}
\Sigma _{\Delta }(\mathbf{p})=(1-\eta _{1})\Sigma _{1}(\mathbf{p})+\eta
_{1}\Sigma _{2}(\mathbf{p})-2\sqrt{\eta _{1}(1-\eta _{1})\lambda _{1}\lambda
_{2}}\Sigma _{12}(\mathbf{p})  \label{eq:AVar-dLC}
\end{equation}%
where $\Sigma _{k}(\mathbf{p})=\mathrm{Var}[h(X_{ki};L,F_{k},\mathbf{p})]$
for $k=1,2$, and $\Sigma _{12}(\mathbf{p})=\mathrm{Cov}[h(X_{1i};L,F_{1},%
\mathbf{p}),h(X_{2i};L,F_{2},\mathbf{p})]$.
\end{example}


\subsection{Asymptotic variance estimation by influence functions \label%
{sec: Asymptotic variance estimation by influence functions}}

\label{sec:OS-AVar} 

To conduct inference on changes in indices, we need a consistent estimator
for the asymptotic variance. We introduce further notation to simplify our
discussion. Given distributions $F_{k}$ with $k=1,2$, we denote $\psi
_{k}(x):=\psi (x;\theta ,F_{k})$ and $\hat{\psi }_{n_{k}}(x):=\psi (x;\theta
,\hat{F}_{n_{k}})$ when no confusion arises. The subscript $n_{k}$ indicates
that the estimated functionals is based on sample $\{X_{i}\}_{i=1}^{n_{k}}$.
Similarly, we abbreviate $h_{k}$ and $\hat{h}_{n_{k}}$ as discussed in
section \ref{sec:index}. So, the estimate $\hat{\sigma }_{k}^{2}$ in %
\eqref{eq:AVar-Index-Est} will take the following form. 
\begin{equation}
\hat{\sigma }_{k}^{2}=\frac{1}{n_{k}}\sum_{i=1}^{n_{k}}\hat{\psi }%
_{n_{k}}^{2}(X_{ki}),\quad k=1,2.  \label{eq:AVar-Index-Est-Fk}
\end{equation}

Given the expression \eqref{eq:AVar-IndexDiff}, one can estimate the
asymptotic variance by 
\begin{equation}
\widetilde{\sigma }_{\Delta }^{2}=\frac{n_{2}}{n_{1}+n_{2}}\hat{\sigma}%
_{1}^{2}+\frac{n_{2}}{n_{1}+n_{2}}\hat{\sigma}_{2}^{2}-\frac{2m}{n_{1}+n_{2}}%
\hat{\sigma}_{12},  \label{eq:FalseAvar_Index}
\end{equation}%
where $\hat{\sigma}_{k}^{2}$ for $k=1,2$, is given in %
\eqref{eq:AVar-Index-Est-Fk}, and $\hat{\sigma}_{12}$ represents the sample
covariance of the matched pairs $\{(\hat{\psi}_{m}(X_{1i}),\hat{\psi}%
_{m}(X_{2i}))\}_{i=1}^{m}$. Indeed, several studies, including \cite%
{Zheng2001} and \cite{Zheng2004}, have proposed this type of estimator.
However, the estimator $\widetilde{\sigma }_{\Delta }^{2}$ in %
\eqref{eq:FalseAvar_Index} is not guaranteed to be positive (or positive
definite for a matrix). Consider an example where $n_{1}=n_{2}=2m$, then $%
\widetilde{\sigma }_{\Delta }^{2}=\frac{1}{2}(\hat{\sigma}_{1}^{2}+\hat{%
\sigma}_{2}^{2}-2\hat{\sigma}_{12})$. While the sample variances for
marginal distributions utilize the entire data, the estimate $\hat{\sigma}%
_{12}$ only uses $n/2$ observations, potentially leading to a negative $%
\widetilde{\sigma }_{\Delta }^{2}$.

We therefore consider an alternative estimator for the asymptotic variance.
On rewriting the covariance term in \eqref{eq:AVar-IndexDiff}, we get: 
\begin{equation}
\sigma _{\Delta }^{2}=(1-\eta _{1})\sigma _{1}^{2}+\eta _{1}\sigma _{2}^{2}-2%
\sqrt{\eta _{1}(1-\eta _{1})\lambda _{1}\lambda _{2}}\rho _{\theta }\sigma
_{1}\sigma _{2}  \label{eq:AVar-dIndex-rho}
\end{equation}%
where $\rho _{\theta }=\mathrm{corr}(\psi _{1}(X_{11}),\psi _{2}(X_{21}))$.
We consider the following estimate for the asymptotic variance $\sigma
_{\Delta }^{2}$ in \eqref{eq:AVar-IndexDiff}: 
\begin{equation}
\hat{\sigma }_{\Delta }^{2}=\frac{n_{2}}{n_{1}+n_{2}}\hat{\sigma }_{1}^{2}+%
\frac{n_{2}}{n_{1}+n_{2}}\hat{\sigma }_{2}^{2}-\frac{2m}{n_{1}+n_{2}}\hat{%
\rho }_{\theta }\hat{\sigma }_{1}\hat{\sigma }_{2}
\label{eq:AVar-dIndex-Est}
\end{equation}%
where $\hat{\sigma }_{k}^{2}$ for $k=1,2$, is given in %
\eqref{eq:AVar-Index-Est-Fk}, and $\hat{\rho }_{\theta }$ is the sample
correlation of the matched pairs $\left\{ (\hat{\psi }_{n_{1}}(X_{1i}),\hat{%
\psi }_{n_{2}}(X_{2i}))\right\} _{i=1}^{m}$. We provide justification for
the consistency of $\hat{\sigma }_{\Delta }^{2}$ in the appendix.

\begin{example}
\label{Eg:dGini-AVar-Est} 
\captionresult{\examplename}{Gini index (continued)}
In Example \ref{Eg:Gini}, we estimate the functional $h$ and asymptotic
variance using order statistics. However, sorting $\{X_{ki}\}_{i=1}^{n_{k}}$
will break the pairing relation in $\{(X_{1i},X_{2i})\}_{i=1}^{m}$, leading
to an inconsistent estimate for $\rho _{\theta }$ in %
\eqref{eq:AVar-dIndex-rho}. We instead consider estimates based on rank
statistics. Given order statistics $X_{k(1)}\leq \cdots \leq X_{k(n_{k})}$,
we define the rank of $X_{ki}$, denoted by $R_{ki}$, as $X_{i}=X_{k(R_{ki})}$%
. We estimate $h$ in \eqref{eq:Gini-IF-aux} as follows$:$ 
\begin{equation}
\hat{h}_{n_{k}}(X_{{ki}}):=h\big(X_{ki};I,\hat{F}_{n_{k}}\big)=\frac{\left( 
\frac{2R_{ki}}{n_{k}}\right) X_{ki}-\frac{2}{n}\sum_{j=1}^{R_{ki}}X_{k(j)}-%
\big(\hat{I}_{k}+1\big)X_{ki}}{\frac{1}{n}\sum_{i=1}^{n_{k}}X_{ki}}.
\end{equation}%
We then estimate the asymptotic variance by \eqref{eq:AVar-dIndex-Est} with $%
\hat{\psi }_{n_{k}}$ replaced by $\hat{h}_{n_{k}}:$ 
\begin{equation}
\hat{\sigma }_{\Delta }^{2}=\frac{n_{2}}{n_{1}+n_{2}}\frac{1}{n_{1}}\hat{%
\sigma }_{1}^{2}+\frac{n_{2}}{n_{1}+n_{2}}\hat{\sigma }_{2}^{2}-\frac{2m}{%
n_{1}+n_{2}}\hat{\rho }_{I}\hat{\sigma }_{1}\hat{\sigma }_{2}
\end{equation}%
where $\hat{\sigma }_{k}$ is the sample variance of $\left\{ \hat{h}%
_{n_{k}}^{2}(X_{ki})\right\} _{i=1}^{n_{k}}$ for $k=1,2$ and $\hat{\rho }%
_{I} $ is the sample correlation coefficient of $\{(\hat{h}_{n_{1}}(X_{1i}),%
\hat{h}_{n_{2}}(X_{2i}))\}_{i=1}^{m}$.
\end{example}

We can extend $\hat{\sigma}^{2}_{\Delta}$ in \eqref{eq:AVar-dIndex-Est} to
estimation for the asymptotic variance matrix for a vector of LC ordinates.

\begin{example}
\label{Eg:dLC-AVar-Est} 
\captionresult{\examplename}{Lorenz curve ordinates (continued)}
We can extend the Lorenz curve in Example \ref{Eg:dLC} to a vector of
ordinates at $\mathbf{p}$. Let $\hat{h}_{k}(X_{ki};\mathbf{p})=h(X_{ki};L,%
\hat{F}_{k},\mathbf{p})$ for $k=1,2$, and denote by $\big[\mathrm{diag}{(A)}%
\big]^{1/2}$ the diagonal matrix with elements being the square roots of
diagonal entries of $A$. So, we estimate $\Sigma _{\Delta }(\mathbf{p})$ in %
\eqref{eq:AVar-dLC} as follows: 
\begin{equation}
\begin{split}
\hat{\Sigma }_{\Delta }(\mathbf{p})=& \frac{n_{2}}{n_{1}+n_{2}}\hat{\Sigma }%
_{1}(\mathbf{p})+\frac{n_{2}}{n_{1}+n_{2}}\hat{\Sigma }_{2}(\mathbf{p}) \\
& -\frac{2m}{n_{1}+n_{2}}\Big[\mathrm{diag}{\big(\hat{\Sigma }_{1}(\mathbf{p}%
)\big)}\Big]^{1/2}\hat{\mathbf{\rho }}_{L}(\mathbf{p})\Big[\mathrm{diag}{%
\big(\hat{\Sigma }_{2}(\mathbf{p})\big)}\Big]^{1/2}
\end{split}%
\end{equation}%
where $\hat{\Sigma }_{k}(\mathbf{p})$ is sample variance matrix of $\big\{%
\hat{h}_{k}(X_{ki};\mathbf{p})\big\}_{i=1}^{n_{k}}$, and $\hat{\rho }_{L}(%
\mathbf{p})$ is the sample correlation coefficient matrix of $\big\{\big(%
\hat{h}_{1}(X_{1i};\mathbf{p}),\hat{h}_{2}(X_{2i};\mathbf{p})\big)\big\}%
_{i=1}^{m}$.
\end{example}

For testing $H_{0}:\Delta \theta =c$, we can construct a Student-T test
which, under the null hypothesis, converges in distribution to a standard
normal distribution: 
\begin{equation}
T=\dfrac{\sqrt{N}\big(\Delta \hat{\theta }-c\big)}{\hat{\sigma }_{\Delta }}%
\underset{n_{1},n_{2}\rightarrow \infty }{\overset{d}{\longrightarrow }}%
\mathrm{N}(0,1).  \label{eq:StudT-dIndex}
\end{equation}%
Inverting the test statistic yields an asymptotic confidence interval (CI)
of level $(1-\alpha )$ for $\Delta \theta $: 
\begin{equation}
\left[ \Delta \hat{\theta }-q_{1-\alpha /2}\frac{\hat{\sigma }_{\Delta }}{%
\sqrt{N}},\quad \Delta \hat{\theta }+q_{1-\alpha /2}\frac{\hat{\sigma }%
_{\Delta }}{\sqrt{N}}\right]  \label{eq:CI-dIndex-Asym}
\end{equation}%
where $q_{p}$ is the $p$-th quantile of the standard normal distribution.


\subsection{Bootstrap inference with overlapping samples \label{sec:
Bootstrap inference with overlapping samples}}

\label{sec:OS-Boot} 

We now examine the bootstrap method for $\Delta \theta $. Unlike the
percentile methods in \cite{Mills1997} and \cite{Biewen2002}, we will
explore the Studentized bootstrap, which can benefit from Beran's refinement
if the statistic bootstrapped is asymptotically pivotal; see \cite{Beran1988}
and \cite{Davidson2004} for instance. We consider the following procedure to
mimic the partial dependence under Assumption \ref{assump:OS}.

\begin{enumerate}
\item Given the original samples $\{X_{1i}\}_{i=1}^{n_{1}}$ and $%
\{X_{2i}\}_{i=1}^{n_{2}}$, obtain the estimates $\Delta \hat{\theta }$ and $%
\hat{\sigma }_{\Delta }^{2}$ in \eqref{eq:AVar-dIndex-Est}, and compute the
test statistic $T$ in \eqref{eq:StudT-dIndex}.

\item Build a bootstrap sample $\{X_{1i}^{\ast }\}_{i=1}^{n_{1}}$ and $%
\{X_{2i}^{\ast }\}_{i=1}^{n_{2}}$ as follows. First, draw in pair with
replacement from matched pairs $\{(X_{1i},X_{2i})\}_{i=1}^{m}$ to obtain $%
\{(X_{1i}^{\ast },X_{2i}^{\ast })\}_{i=1}^{m}$. Second, supplement $%
\{(X_{1i}^{\ast },X_{2i}^{\ast })\}_{i=1}^{m}$ by resampling with
replacement from the unmatched observations $\{X_{1i}\}_{i=m+1}^{n_{1}}$ and 
$\{X_{2i}\}_{i=m+1}^{n_{2}}$ to obtain $\{X_{1i}^{\ast }\}_{i=m+1}^{n_{1}}$
and $\{X_{2i}^{\ast }\}_{i=m+1}^{n_{2}}$, respectively.

\item For the bootstrap sample $\{X_{1i}^{\ast }\}_{i=1}^{n_{1}}$ and $%
\{X_{2i}^{\ast }\}_{i=1}^{n_{2}}$, compute the test statistic $T^{\ast }$
according to \eqref{eq:StudT-dIndex} with $c$ replaced by $\Delta \hat{%
\theta }$.

\item Repeat the above step $B$ times and get test statistics $\{T_{j}^{\ast
}\}_{j=1}^{B}$. Choose $B$ such that $\alpha (B+1)$ is a positive integer,
where $\alpha $ is the nominal level.

\item Compute the bootstrap p-value defined as the proportion of $%
T_{j}^{\ast }$ that is more extreme than $T$.

\item Reject the null at level $\alpha $ if the bootstrap p-value is less
than $\alpha $.
\end{enumerate}

We can also construct a two-sided bootstrap CI for $\Delta \theta $ at the
level of $(1-\alpha )$ using the empirical distribution of the bootstrap
statistics $T_{j}^{\ast }$: 
\begin{equation}
\left[ \Delta \hat{\theta }-q_{1-\alpha /2}^{\ast }\frac{\hat{\sigma }%
_{\Delta }}{\sqrt{N}},\quad \Delta \hat{\theta }-q_{\alpha /2}^{\ast }\frac{%
\hat{\sigma }_{\Delta }}{\sqrt{N}}\right]
\end{equation}%
where $q_{p}^{\ast }$ is the $\lceil pB\rceil $-th order statistics of $%
\{T_{j}^{\ast }\}_{j=1}^{B}$. Here, $\lceil \cdot \rceil $ denotes the
ceiling function. The following proposition establishes the validity of the
bootstrap method.

\begin{proposition}
\label{prop:Boot-dIndex} 
\captionproposition{\propositionname}{Bootstrap method with overlapping samples}
Let $\Delta \hat{\theta }$ and $\Delta \hat{\theta }^{\ast }$ be the
estimates based on original and bootstrap samples, respectively. Under
conditions of Proposition \ref{prop:AsyN-dIndex}, we have 
\begin{equation}
\sqrt{N}\big(\Delta \hat{\theta }^{\ast }-\Delta \hat{\theta }\big)\overset{d%
}{\longrightarrow }\mathrm{N}\big(0,\sigma _{\Delta }^{2}\big)
\end{equation}%
where $\sigma _{\Delta }^{2}$ is given in \eqref{eq:AVar-IndexDiff}.
\end{proposition}

Here are some comments on the above bootstrap method. First, while we focus
on the bootstrap method using the linearization technique, an alternative
approach is the bootstrap method based on U-statistic theory (\cite{Shi1986}%
). Second, we may view the partially dependent samples as samples with
missing observations. In particular, we can reformat the data as $\mathbf{Z}%
_{i} = (X_{1i}, \delta_{1i}, X_{2i}, \delta_{2i})^{\prime }$, where $%
\delta_{ki}$ is a variable that equals one if $X_{ki}$ is observed and zero
otherwise. Then we employ the pair bootstrap based on $\{\mathbf{Z}_{i}\}_{i
= 1}^{\bar{n}}$, where $\bar{n} = n_{1} + n_{2} - m$. The simulation
experiments (not reported here) suggest that the two procedures yield nearly
identical results. So, we will not report such types of CIs.


\newpageSlides

\section{\sectitlesize Monte Carlo simulations \label{sec:MC}}

\resetcountersSection

In this section, we report two simulation experiments on the impacts of
sample dependence and heavy tails on various confidence intervals for $%
\Delta \theta $. For simplicity, we assume that two samples share identical
sizes, \emph{i.e.}, $n_{1}=n_{2}=n$, unless stated otherwise. Then, we have $%
\eta _{1}=0.5$, $\lambda _{1}=\lambda _{2}=\lambda $ in Assumption \ref%
{assump:OS} (A3), and the asymptotic standard deviation in Proposition \ref%
{prop:AsyN-dIndex} reduces to the following quantity: 
\begin{equation}
\sigma _{\Delta }^{D}=\left( 0.5\sigma _{1}^{2}+0.5\sigma _{2}^{2}-\lambda
\rho _{\theta }\sigma _{1}\sigma _{2}\right) ^{1/2}
\label{eq:AVar-dIndex-MC}
\end{equation}%
where $\lambda $ is the overlap portion, $\sigma _{k}=[\mathrm{Var}(\psi
_{k}(X_{ki}))]^{1/2}$ for $k=1,2$, and $\rho _{\theta }=\mathrm{corr}[\psi
_{1}(X_{11}),\psi _{2}(X_{21})]$ in \eqref{eq:AVar-dIndex-rho}.


\subsection{Effects of sample dependence \label{sec: Effects of sample
dependence}}

\label{sec:MC-dep} 

We simulate data from the Singh-Maddala (SM) distribution, which has proven
effective in fitting the actual income distributions [\cite{McDonald1984}, 
\cite{Kleiber2003}]. The SM distribution [denoted $\text{SM}(b,a,q)$] has
distribution function: 
\begin{equation}
F(x)=1-\frac{1}{[1+(x/b)^{a}]^{q}},\quad a\geq 0,\,b>0\text{ and }q>1/a,
\label{eq:SM-CDF}
\end{equation}%
where $b$ is a scale parameter, and $a$ and $q$ are shape parameters. The
choice of scale parameter $b$ does not matter, as the Gini index and LC are
scale-invariant. We assume that the first sample contains data drawn from $%
\text{SM}(1,1.6971,8.3679)$, where the estimates are given by \cite%
{McDonald1984} for fitting the U.S. income distribution in 1980. For the
second sample, we follow \cite{Davidson2007a} and set $\text{SM}%
(0.4,2.8,1.7) $ in order to mimic the net income distribution of German
households.

We assume a Gaussian copula to capture the sample dependence for matched
pairs. 
\begin{equation}
C(u_{1},u_{2};\rho )=\Phi _{\rho }\big(\Phi ^{-1}(u_{1}),\Phi ^{-1}(u_{2})%
\big),\quad (u_{1},u_{2})\in \lbrack 0,1]\times \lbrack 0,1],\quad \rho \in
\lbrack -1,1],  \label{eq:DGP-GaussCopula}
\end{equation}%
where $\Phi $ and $\Phi _{\rho }$ denote the cumulative distribution
functions (CDFs) of the standard univariate and bivariate normal
distributions, respectively. The parameter $\rho $ indicates the degree of
sample dependence, with $0$ for independence and $1$ ($-1$) for positive
(negative) linear dependence.

We will explore various aspects of how sample dependence affects asymptotic
confidence intervals (CIs). Recall that it is challenging to predict the
value or sign of $\sigma _{12}$ in \eqref{eq:AVar-IndexDiff} for a given $%
\mathrm{Cov}(X_{1},X_{2})$. This difficulty arises because the functional $%
\psi (x)$ in \eqref{eq:ALG} is often not monotonic in $x$. To address this
issue, we will first evaluate the correlation between $\psi _{1}(X_{1i})$
and $\psi _{2}(X_{2i})$, namely $\rho _{\theta }$ in %
\eqref{eq:AVar-dIndex-MC}, against $\rho \in \{0,\pm 0.01,\ldots ,\,\pm 1\}$
in the Gaussian copula \eqref{eq:DGP-GaussCopula}. For simplicity, we
consider matched pairs of observations (i.e., $\lambda =1$). Since the
marginal distributions are known, we can directly compute $\psi_{k} $ for $k
= 1, 2$.

\begin{figure}%

\caption[Correlation between influence functions.]{} \label{fig:corIF-vs-rho}

\begin{center}
\begin{minipage}{\paperwidth}
\includegraphics[width=0.35\linewidth]{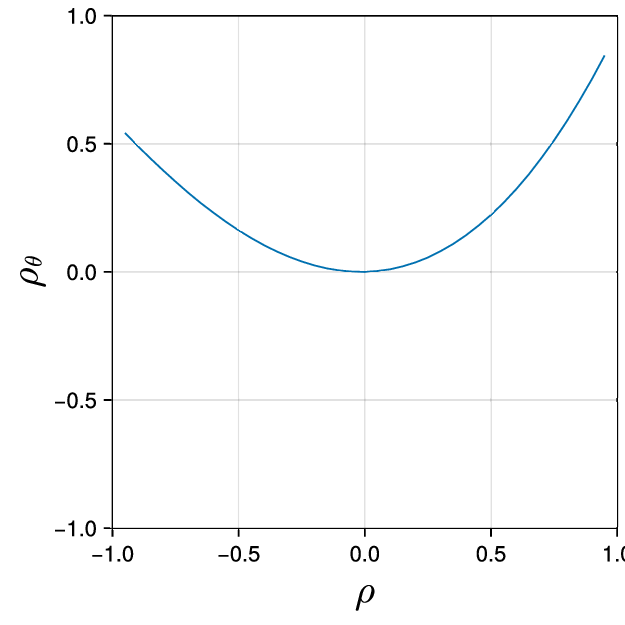} 
\includegraphics[width=0.35\linewidth]{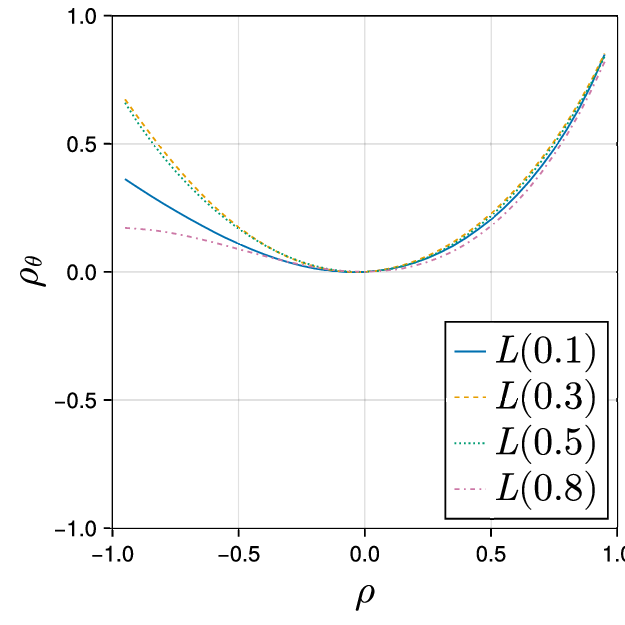}  
\end{minipage}
\end{center}

Figure \thefigure. $\rho_{\theta}$ in \eqref{eq:AVar-dIndex-MC} for Gini
index (left panel) and LC ordinates (right panel) against $\rho$ of Gaussian
copula in \eqref{eq:DGP-GaussCopula}.

{\footnotesize {The data are generated according to $F_{1}=\text{SM}%
(1,1.6971,8.3679)$ and $F_{2}=\text{SM}(0.4,2.8,1.7)$ with Gaussian copula %
\eqref{eq:DGP-GaussCopula}. The results are based on $5000$ replications of $%
5000$ observations according to DGP: $F_{1}=\text{SM}(1,1.6971,8.3679)$, $%
F_{2}=\text{SM}(0.4,2.8,1.7)$, $\rho \in \{0,\pm 0.01,\ldots ,\,\pm 1\}$ in %
\eqref{eq:DGP-GaussCopula}, overlap portion $\lambda =1$. } }

\end{figure}

Figure \ref{fig:corIF-vs-rho} plots the relationship between $\rho _{\theta
} $ and $\rho $ based on $5000$ simulations, each consisting of $5000$
observations. We find that both curves of the Gini index (left) and LC
ordinate (right) are $U$-shaped but asymmetric at zero, and that $\rho
_{\theta }$ is rarely negative. Therefore, by considering sample dependence,
we anticipate that the proposed inference method using overlapping samples
will yield more efficient results due to its smaller asymptotic variance.
However, the correlation $\rho _{\theta }$ can be monotonic in $\rho $. An
example is the population mean, where $\psi (x)=x-\mu $ and, thus, $\rho
_{\theta }=\rho $. We then expect that the conventional inference approach,
which ignores sample dependence, will underestimate the asymptotic variance
and thus lead to incorrect conclusions.

To quantify the effects of sample dependence, we take the asymptotic
standard deviation with independent samples as the benchmark, denoted by $%
\sigma _{\Delta }^{I}$, where 
\begin{equation}
\sigma _{\Delta }^{I}=\left( 0.5\sigma _{1}^{2}+0.5\sigma _{2}^{2}\right)
^{1/2}.  \label{eq:AVar-dIndex-IS-MC}
\end{equation}%
If we incorrectly specify $\lambda =0$, we end up with the asymptotic
variance $\sigma _{\Delta }^{I}$. We examine the increase in standard
deviations using the formula $(\sigma _{\Delta }^{I}-\sigma _{\Delta
}^{D})/\sigma _{\Delta }^{D}$, where $\sigma _{\Delta }^{D}$ is in %
\eqref{eq:AVar-dIndex-MC}. Note that $\sigma _{\Delta }^{D}$ is affected by
the correlation between $\psi _{1}$ and $\psi _{2}$ (\emph{i.e.}, $\rho
_{\theta }$) and overlap scope (\emph{i.e.}, $\lambda $).

We consider again the marginal distributions $\text{SM}(1, 1.6971, 8.3679)$
and $\text{SM}(0.4, 2.8, 1.7)$. The sample dependence of the matched pairs
is described by the Gaussian copula \eqref{eq:DGP-GaussCopula} with $\rho
\in \{0, \pm 0.1, \pm 0.5, \pm0.99\}$. The overlap portion of matched pairs
is $\lambda \in \{0.1, 0.5, 0.9\}$.

Table \ref{tab:rho-AStd} presents the results based on $5000$ replications
with $5000$ observations each. We use the true function $\psi_{k}$ for $k =
1, 2$, though typically unknown in practice. Note that the sample dependency
can substantially affect the asymptotic standard deviation if samples are
strongly correlated or share significant overlap. Take the change in
population mean $(\Delta \mu)$ as an example. We find that the increase
(decrease) in standard deviations resulting from neglecting sample
dependence can be as much as $180.9\%$ ($23.6\%$).

\begin{table}[tb]%

\caption{Increases in standard errors for estimated indices differences if
ignoring sample dependency.} \label{tab:rho-AStd}

\begin{center}
Table \thetable

Increases in standard errors for estimated indices differences if ignoring
sample dependency.
\end{center}

\hspace{-0.5\totalhormargin} \begin{minipage}{\paperwidth}
\centering%

\small%

\begin{center}
\begin{tabular}{ccccccc}
\hline
\rule[-1ex]{0cm}{4ex} $\rho$ & $\lambda$ & \multicolumn{5}{c}{Increases in
standard deviations $(\%)$} \\ \cline{3-7}
\rule[-1ex]{0cm}{4ex} &  & $\Delta I$ & $\Delta \mu$ & $\Delta L(0.1)$ & $%
\Delta L(0.5)$ & $\Delta L(0.8)$ \\ \hline
\rule[-1ex]{0cm}{4ex} -0.99 & 0.1 & 3.1 & -3.7 & 2.0 & 3.9 & 0.9 \\ 
& 0.5 & 18.8 & -15.4 & 11.3 & 25.3 & 4.6 \\ 
& 0.9 & 45.0 & -23.6 & 23.7 & 70.1 & 8.8 \\ 
\rule[-1ex]{0cm}{4ex} -0.5 & 0.1 & 0.8 & -2.0 & 0.6 & 0.9 & 0.5 \\ 
& 0.5 & 4.3 & -9.1 & 2.8 & 4.5 & 2.3 \\ 
& 0.9 & 8.2 & -14.8 & 5.3 & 8.5 & 4.3 \\ 
\rule[-1ex]{0cm}{4ex} -0.1 & 0.1 & 0.0 & -0.4 & 0.0 & 0.0 & 0.0 \\ 
& 0.5 & 0.1 & -2.1 & 0.0 & 0.1 & 0.1 \\ 
& 0.9 & 0.2 & -3.7 & 0.1 & 0.2 & 0.2 \\ 
\rule[-1ex]{0cm}{4ex} 0 & 0.1 & 0.0 & 0.0 & 0.0 & 0.0 & 0.0 \\ 
& 0.5 & 0.0 & 0.0 & 0.0 & 0.0 & 0.0 \\ 
& 0.9 & 0.0 & 0.0 & 0.0 & 0.0 & 0.0 \\ 
\rule[-1ex]{0cm}{4ex} 0.1 & 0.1 & 0.1 & 0.5 & 0.1 & 0.1 & 0.0 \\ 
& 0.5 & 0.3 & 2.3 & 0.3 & 0.3 & 0.2 \\ 
& 0.9 & 0.5 & 4.3 & 0.5 & 0.5 & 0.3 \\ 
\rule[-1ex]{0cm}{4ex} 0.5 & 0.1 & 1.1 & 2.4 & 1.0 & 1.1 & 0.9 \\ 
& 0.5 & 6.1 & 14.2 & 5.5 & 5.9 & 4.8 \\ 
& 0.9 & 11.8 & 31.2 & 10.6 & 11.5 & 9.2 \\ 
\rule[-1ex]{0cm}{4ex} 0.99 & 0.1 & 5.0 & 5.2 & 5.1 & 5.0 & 4.9 \\ 
& 0.5 & 36.2 & 39.4 & 37.8 & 36.3 & 35.7 \\ 
& 0.9 & 142.0 & 180.9 & 159.6 & 143.7 & 137.8 \\ \hline
\end{tabular}
\end{center}

\end{minipage}%

\medskip 
\noindent
\footnotesize%
Note -- The table provides the percentage increases in standard deviations
for estimators of several welfare index differences when ignoring sample
dependency, including the changes in Gini indices ($\Delta I$), population
mean ($\Delta \mu$) and LC ordinates ($\Delta L(p)$) at $p \in \{0.1, 0.5,
0.8\}$. The percentage increases in standard deviations are measured by $%
(\sigma^{I}_{\Delta} - \sigma^{D}_{\Delta}) / \sigma^{D}_{\Delta}$, where $%
\sigma^{I}_{\Delta}$ and $\sigma^{D}_{\Delta}$ are given in %
\eqref{eq:AVar-dIndex-IS-MC} and \eqref{eq:AVar-dIndex-MC}, respectively.
The results are based on $5000$ replications of $5000$ observations
according to the DGP: $F_{1} = \text{SM}(1, 1.6971, 8.3679)$, $F_{2} = \text{%
SM}(0.4, 2.8, 1.7)$, $\rho \in \{0, \pm 0.5, \pm0.99\}$ in %
\eqref{eq:DGP-GaussCopula}, overlap portion $\lambda \in \{0.1, 0.5, 0.9\}$.

\normalsize%

\end{table}%

We next examine the performance of $95\%$ confidence intervals (CIs) by
different approaches based on $1000$ replications of $1000$ observations
under the same DGP. Our goal is to investigate the impact of sample
dependence on the performance of inference. We set the conventional method
as the benchmark, which incorrectly assumes that two samples are
independent. By \eqref{eq:AVar-dIndex-Est}, we estimate the asymptotic
standard deviation as follows.%
\begin{equation}
\hat{\sigma }_{\Delta }^{D}=[0.5\hat{\sigma }_{1}^{2}+0.5\hat{\sigma }%
_{2}^{2}-(m/n)\hat{\rho }_{\theta }\hat{\sigma }_{1}\hat{\sigma }%
_{2}^{1/2}]^{1/2},  \label{eq:Est-AVar-dIndex-MC}
\end{equation}%
\begin{equation}
\hat{\sigma }_{\Delta }^{I}=[0.5\hat{\sigma }_{1}^{2}+0.5\hat{\sigma }%
_{2}^{2}]^{1/2},  \label{eq:Est-AVar-dIndex-IS-MC}
\end{equation}%
where for $k=1,2$, $\hat{\sigma }_{k}^{2}$ is the sample variance of $\hat{%
\psi }_{n_{k}}(X_{ki})$ in \eqref{eq:AVar-Index-Est-Fk}, and $\hat{\rho }%
_{\theta }$ is the sample correlation coefficient of $\left\{ (\hat{\psi }%
_{n_{1}}(X_{1i}),\hat{\psi }_{n_{2}}(X_{2i}))\right\} _{i=1}^{m}$.

We present simulation results in Table \ref{tab:CI-dI-DSvsIS} for the Gini
difference, Table \ref{tab:CI-dMean-DSvsIS} for the mean difference $\Delta
\mu$, and Table \ref{tab:CI-dLC-p05-DSvsIS} for LC ordinates differences at $%
p = 0.5$.

First, we find that conventional CIs, which fail to account for sample
dependence, may suffer from two problems. They can either be overly
conservative with overlarge widths (see columns $(9) - (12)$ in Table \ref%
{tab:CI-dI-DSvsIS} and \ref{tab:CI-dLC-p05-DSvsIS}), or they have coverage
rates much lower than the nominal level $95\%$ (see columns $(9) - (12)$ in
Table \ref{tab:CI-dMean-DSvsIS} when $\lambda = 0.9$ and $\rho = -0.99$).
Such findings are consistent with our expectations in Figure \ref%
{fig:corIF-vs-rho} and Table \ref{tab:rho-AStd}.

Second, our proposed asymptotic and bootstrap CIs with overlapping samples
can effectively control the level with reasonable widths; see columns $(5) -
(8)$ in Table \ref{tab:CI-dI-DSvsIS}, \ref{tab:CI-dMean-DSvsIS}, and Table %
\ref{tab:CI-dLC-p05-DSvsIS}.

Third, we find that the intersection methods (IMs) are valid, though
conservative, for all three indices differences, as indicated in $(1) - (4)$
in Table \ref{tab:CI-dI-DSvsIS}, \ref{tab:CI-dMean-DSvsIS} and \ref%
{tab:CI-dLC-p05-DSvsIS}. Note that the IMs are applicable in more general
scenarios and become the only valid procedure when overlapping samples fail
to hold; see section \ref{sec:MC-BeyondOS} for more details.

\begin{sidewaystable}%

\caption{Performance of proposed and conventional confidence intervals at level 
of $95\%$ for the difference in Gini index.} \label{tab:CI-dI-DSvsIS}

\begin{center}
Table \thetable

Performance of proposed and conventional confidence intervals at level of $%
95\%$ for the difference in Gini index.

\begin{adjustbox}{scale={0.9}{0.9}}%

\small%

\begin{tabular}{cc|cc|cc|cc|cc|cc|cc}
\hline
\rule[-1ex]{0cm}{4ex} &  & (1) & (2) & (3) & (4) & (5) & (6) & (7) & (8) & 
(9) & (10) & (11) & (12) \\ \cline{3-14}
\rule[-1ex]{0cm}{4ex} &  & \multicolumn{4}{c|}{Intersection method} & 
\multicolumn{4}{c|}{Overlapping samples} & \multicolumn{4}{c}{Independent
samples (incorrect)} \\ \cline{3-14}
\rule[-1ex]{0cm}{4ex} $\rho$ & $\lambda$ & \multicolumn{2}{c|}{Coverage $%
(\%) $} & \multicolumn{2}{c|}{Width} & \multicolumn{2}{c|}{Coverage $(\%)$}
& \multicolumn{2}{c|}{Width} & \multicolumn{2}{c|}{Coverage $(\%)$} & 
\multicolumn{2}{c}{Width} \\ \cline{3-14}
\rule[-1ex]{0cm}{4ex} &  & Asym & Boot & Asym & Boot & Asym & Boot & Asym & 
Boot & Asym & Boot & Asym & Boot \\ \hline
\rule[-1ex]{0cm}{4ex} -0.99 & 0.1 & 100 & 100 & 0.0666 & 0.0687 & 95.3 & 94.4
& 0.04 & 0.0405 & 95.7 & 95.1 & 0.0412 & 0.0418 \\ 
& 0.5 & 100 & 100 & 0.0664 & 0.0685 & 94.6 & 94 & 0.0345 & 0.035 & 98.4 & 
98.4 & 0.0411 & 0.0416 \\ 
& 0.9 & 100 & 100 & 0.0666 & 0.0687 & 94.1 & 93.1 & 0.0284 & 0.029 & \textbf{%
99.1} & \textbf{99} & 0.0413 & 0.042 \\ 
\rule[-1ex]{0cm}{4ex} -0.5 & 0.1 & 99.8 & 99.8 & 0.0664 & 0.0686 & 94.2 & 
93.9 & 0.0408 & 0.0413 & 94.9 & 95.5 & 0.0411 & 0.0416 \\ 
& 0.5 & 99.9 & 99.9 & 0.0665 & 0.0686 & 95.7 & 95.3 & 0.0395 & 0.04 & 96.1 & 
96.4 & 0.0412 & 0.0417 \\ 
& 0.9 & 100 & 100 & 0.0665 & 0.0685 & 95 & 95.4 & 0.0381 & 0.0386 & 96.8 & 
96.7 & 0.0412 & 0.0418 \\ 
\rule[-1ex]{0cm}{4ex} 0 & 0.1 & 99.9 & 99.9 & 0.0665 & 0.0684 & 94.8 & 94.6
& 0.0412 & 0.0419 & 95 & 95.8 & 0.0412 & 0.0418 \\ 
& 0.5 & 100 & 100 & 0.0666 & 0.0687 & 95.6 & 94.9 & 0.0413 & 0.0419 & 95.5 & 
95.4 & 0.0413 & 0.042 \\ 
& 0.9 & 100 & 99.9 & 0.0667 & 0.0689 & 95.3 & 95.3 & 0.0413 & 0.042 & 95.2 & 
95.3 & 0.0413 & 0.042 \\ 
\rule[-1ex]{0cm}{4ex} 0.5 & 0.1 & 99.8 & 99.9 & 0.0668 & 0.0689 & 95.3 & 94.6
& 0.0409 & 0.0416 & 95.7 & 94.9 & 0.0414 & 0.0421 \\ 
& 0.5 & 99.8 & 99.8 & 0.0664 & 0.0685 & 94 & 94.6 & 0.0388 & 0.0394 & 95.7 & 
95.5 & 0.0411 & 0.0418 \\ 
& 0.9 & 99.8 & 100 & 0.0665 & 0.0686 & 94.1 & 94.1 & 0.0368 & 0.0373 & 96.1
& 96 & 0.0412 & 0.0418 \\ 
\rule[-1ex]{0cm}{4ex} 0.99 & 0.1 & 99.9 & 99.8 & 0.0666 & 0.0687 & 94.2 & 
94.4 & 0.0393 & 0.0399 & 95.6 & 95.3 & 0.0412 & 0.0419 \\ 
& 0.5 & 100 & 100 & 0.0667 & 0.0688 & 95.6 & 95.9 & 0.0303 & 0.0308 & 
\textbf{99.5} & \textbf{99.4} & 0.0413 & 0.042 \\ 
& 0.9 & 100 & 100 & 0.0665 & 0.0688 & 96 & 95.4 & 0.017 & 0.0174 & \textbf{%
100} & \textbf{100} & 0.0412 & 0.0419 \\ \hline
\end{tabular}

\normalsize%

\end{adjustbox}%
\end{center}

\medskip 
\noindent
\footnotesize%
Note -- The table provides coverage rates and widths of $95\%$ confidence
intervals for the difference in Gini index. The data are generated according
to the DGP: $F_{1} = \text{SM}(1, 1.6971, 8.3679)$, $F_{2} = \text{SM}(0.4,
2.8, 1.7)$, $\rho \in \{0, \pm 0.5, \pm0.99\}$ in \eqref{eq:DGP-GaussCopula}%
, overlap portion $\lambda \in \{0.1, 0.5, 0.9\}$. The sample sizes are both 
$1000$. Columns $(1)$ - $(4)$ are performance by asymptotic intersection
method (Asym), bootstrap intersection method (Boot) in section \ref{sec:DS}.
Columns $(5)$ - $(8)$ are performance by asymptotic (Asym) and bootstrap
(Boot) methods with overlapping samples in section \ref{sec:OS}, using $\hat{%
\sigma}^{D}_{\Delta}$ in \eqref{eq:Est-AVar-dIndex-MC}. Columns $(9) $ - $%
(12)$ are performance by asymptotic (Asym) and bootstrap (Boot) methods if
incorrectly ignoring the sample dependence, using $\hat{\sigma}^{I}_{\Delta}$
in \eqref{eq:Est-AVar-dIndex-IS-MC}. The results are based on $1000$
replications with 399 bootstrap repetitions each. The bold font indicates
cases where coverage rates largely deviate from $95\%$.

\normalsize%

\end{sidewaystable}%

\begin{sidewaystable}%

\caption{Performance of proposed and conventional confidence intervals at 
level of $95\%$ for the mean difference.} \label{tab:CI-dMean-DSvsIS}

\begin{center}
Table \thetable

Performance of proposed and conventional confidence intervals at level of $%
95\%$ for the mean difference.

\begin{adjustbox}{scale={0.9}{0.9}}%

\small%

\begin{tabular}{cc|cc|cc|cc|cc|cc|cc}
\hline
\rule[-1ex]{0cm}{4ex} &  & (1) & (2) & (3) & (4) & (5) & (6) & (7) & (8) & 
(9) & (10) & (11) & (12) \\ \cline{3-14}
\rule[-1ex]{0cm}{4ex} &  & \multicolumn{4}{c|}{Intersection method} & 
\multicolumn{4}{c|}{Overlapping samples} & \multicolumn{4}{c}{Independent
samples (incorrect)} \\ \cline{3-14}
\rule[-1ex]{0cm}{4ex} $\rho$ & $\lambda$ & \multicolumn{2}{c|}{Coverage $%
(\%) $} & \multicolumn{2}{c|}{Width} & \multicolumn{2}{c|}{Coverage $(\%)$}
& \multicolumn{2}{c|}{Width} & \multicolumn{2}{c|}{Coverage $(\%)$} & 
\multicolumn{2}{c}{Width} \\ \cline{3-14}
\rule[-1ex]{0cm}{4ex} &  & Asym & Boot & Asym & Boot & Asym & Boot & Asym & 
Boot & Asym & Boot & Asym & Boot \\ \hline
\rule[-1ex]{0cm}{4ex} -0.99 & 0.1 & 99.8 & 99.9 & 0.0533 & 0.0544 & 95.1 & 95
& 0.0343 & 0.0346 & 94.2 & 94.2 & 0.033 & 0.0333 \\ 
& 0.5 & 99.5 & 99.7 & 0.0533 & 0.0544 & 96.3 & 95.6 & 0.039 & 0.0394 & 91.1
& 91.4 & 0.033 & 0.0333 \\ 
& 0.9 & 98.3 & 98.2 & 0.0534 & 0.0544 & 95.8 & 95.3 & 0.0433 & 0.0437 & 
\textbf{86.1} & \textbf{85.9} & 0.0331 & 0.0335 \\ 
\rule[-1ex]{0cm}{4ex} -0.5 & 0.1 & 99.8 & 99.8 & 0.0534 & 0.0545 & 94.9 & 95
& 0.0338 & 0.0341 & 94.6 & 94.1 & 0.0331 & 0.0335 \\ 
& 0.5 & 99.8 & 99.7 & 0.0535 & 0.0546 & 94.6 & 95 & 0.0365 & 0.0368 & 92.9 & 
92.7 & 0.0331 & 0.0334 \\ 
& 0.9 & 99.6 & 99.3 & 0.0535 & 0.0545 & 96.1 & 95.7 & 0.0389 & 0.0393 & 90.3
& 90.7 & 0.0331 & 0.0334 \\ 
\rule[-1ex]{0cm}{4ex} 0 & 0.1 & 99.9 & 99.9 & 0.0535 & 0.0546 & 95.5 & 95.5
& 0.0331 & 0.0334 & 95.5 & 95.2 & 0.0331 & 0.0334 \\ 
& 0.5 & 99.6 & 99.6 & 0.0534 & 0.0545 & 94 & 94.3 & 0.0331 & 0.0334 & 94.2 & 
94.2 & 0.0331 & 0.0335 \\ 
& 0.9 & 99.9 & 99.9 & 0.0535 & 0.0545 & 95.6 & 95.1 & 0.0332 & 0.0335 & 95.6
& 95.3 & 0.0331 & 0.0335 \\ 
\rule[-1ex]{0cm}{4ex} 0.5 & 0.1 & 99.8 & 99.9 & 0.0534 & 0.0545 & 94.1 & 94.1
& 0.0323 & 0.0327 & 95.1 & 95 & 0.0331 & 0.0334 \\ 
& 0.5 & 100 & 99.9 & 0.0534 & 0.0545 & 94.8 & 94.3 & 0.029 & 0.0293 & 96.2 & 
96.3 & 0.0331 & 0.0334 \\ 
& 0.9 & 100 & 100 & 0.0534 & 0.0545 & 93.9 & 93.9 & 0.0252 & 0.0255 & 98.4 & 
98.5 & 0.0331 & 0.0334 \\ 
\rule[-1ex]{0cm}{4ex} 0.99 & 0.1 & 100 & 100 & 0.0534 & 0.0545 & 94.3 & 94.3
& 0.0314 & 0.0318 & 95 & 95.6 & 0.0331 & 0.0334 \\ 
& 0.5 & 100 & 100 & 0.0534 & 0.0546 & 95.3 & 94.5 & 0.0237 & 0.0239 & 
\textbf{99.2} & \textbf{99.2} & 0.0331 & 0.0334 \\ 
& 0.9 & 100 & 100 & 0.0534 & 0.0545 & 94 & 93.9 & 0.0118 & 0.0119 & \textbf{%
100} & \textbf{100} & 0.0331 & 0.0334 \\ \hline
\end{tabular}

\normalsize%

\end{adjustbox}%
\end{center}

\medskip 
\noindent
\footnotesize%
Note -- The table provides coverage rates and widths of $95\%$ confidence
intervals for the mean difference. The data are generated according to DGP: $%
F_{1} = \text{SM}(1, 1.6971, 8.3679)$, $F_{2} = \text{SM}(0.4, 2.8, 1.7)$, $%
\rho \in \{0, \pm 0.5, \pm0.99\}$ in \eqref{eq:DGP-GaussCopula}, overlap
portion $\lambda \in \{0.1, 0.5, 0.9\}$. The sample sizes are both $1000$.
Columns $(1)$ - $(4)$ are performance by asymptotic intersection method
(Asym), bootstrap intersection method (Boot) in section \ref{sec:DS}.
Columns $(5)$ - $(8)$ are performance by asymptotic (Asym) and bootstrap
(Boot) methods with overlapping samples in section \ref{sec:OS}, using $\hat{%
\sigma}^{D}_{\Delta}$ in \eqref{eq:Est-AVar-dIndex-MC}. Columns $(9) $ - $%
(12)$ are performance by asymptotic (Asym) and bootstrap (Boot) methods if
incorrectly ignoring the sample dependence, using $\hat{\sigma}^{I}_{\Delta}$
in \eqref{eq:Est-AVar-dIndex-IS-MC}. The results are based on $1000$
replications with 399 bootstrap repetitions each. The bold font indicates
cases where coverage rates largely deviate from $95\%$.

\normalsize%

\end{sidewaystable}%

\begin{sidewaystable}%

\caption{Performance of proposed and conventional confidence intervals at level 
of $95\%$ for the difference in LC ordinate at percentile $0.5$.} \label%
{tab:CI-dLC-p05-DSvsIS}

\begin{center}
Table \thetable

Performance of proposed and conventional confidence intervals at level of $%
95\%$ for the difference in LC ordinate at percentile $0.5$.

\begin{adjustbox}{scale={0.9}{0.9}}%

\small%

\begin{tabular}{cc|cc|cc|cc|cc|cc|cc}
\hline
\rule[-1ex]{0cm}{4ex} &  & (1) & (2) & (3) & (4) & (5) & (6) & (7) & (8) & 
(9) & (10) & (11) & (12) \\ \cline{3-14}
\rule[-1ex]{0cm}{4ex} &  & \multicolumn{4}{c|}{Intersection method} & 
\multicolumn{4}{c|}{Overlapping samples} & \multicolumn{4}{c}{Independent
samples (incorrect)} \\ \cline{3-14}
\rule[-1ex]{0cm}{4ex} $\rho$ & $\lambda$ & \multicolumn{2}{c|}{Coverage $%
(\%) $} & \multicolumn{2}{c|}{Width} & \multicolumn{2}{c|}{Coverage $(\%)$}
& \multicolumn{2}{c|}{Width} & \multicolumn{2}{c|}{Coverage $(\%)$} & 
\multicolumn{2}{c}{Width} \\ \cline{3-14}
\rule[-1ex]{0cm}{4ex} &  & Asym & Boot & Asym & Boot & Asym & Boot & Asym & 
Boot & Asym & Boot & Asym & Boot \\ \hline
\rule[-1ex]{0cm}{4ex} -0.99 & 0.1 & 100 & 100 & 0.0451 & 0.0466 & 95.6 & 95
& 0.0269 & 0.0275 & 96 & 96.2 & 0.0279 & 0.0286 \\ 
& 0.5 & 100 & 100 & 0.045 & 0.0466 & 95.2 & 95.4 & 0.0222 & 0.0229 & \textbf{%
99.4} & \textbf{99.3} & 0.0279 & 0.0285 \\ 
& 0.9 & 100 & 100 & 0.0451 & 0.0467 & 94.4 & 95 & 0.0164 & 0.0172 & \textbf{%
100} & \textbf{99.8} & 0.0279 & 0.0285 \\ 
\rule[-1ex]{0cm}{4ex} -0.5 & 0.1 & 99.9 & 100 & 0.045 & 0.0466 & 95.2 & 95.6
& 0.0276 & 0.0283 & 95.3 & 95.6 & 0.0279 & 0.0286 \\ 
& 0.5 & 99.9 & 99.9 & 0.045 & 0.0466 & 94.9 & 95.1 & 0.0267 & 0.0274 & 95.8
& 96.1 & 0.0279 & 0.0285 \\ 
& 0.9 & 100 & 100 & 0.0451 & 0.0466 & 95 & 95.6 & 0.0257 & 0.0264 & 96.7 & 
97.2 & 0.0279 & 0.0286 \\ 
\rule[-1ex]{0cm}{4ex} 0 & 0.1 & 99.9 & 99.9 & 0.045 & 0.0468 & 95.2 & 95.6 & 
0.0279 & 0.0286 & 95 & 95.7 & 0.0279 & 0.0285 \\ 
& 0.5 & 99.8 & 100 & 0.0451 & 0.0467 & 95.3 & 95.8 & 0.0279 & 0.0286 & 95.5
& 95.1 & 0.0279 & 0.0286 \\ 
& 0.9 & 100 & 99.9 & 0.0451 & 0.0467 & 94.8 & 95.1 & 0.028 & 0.0286 & 94.8 & 
95 & 0.0279 & 0.0286 \\ 
\rule[-1ex]{0cm}{4ex} 0.5 & 0.1 & 99.7 & 100 & 0.0452 & 0.0468 & 95.4 & 95 & 
0.0277 & 0.0283 & 95.5 & 95.6 & 0.028 & 0.0286 \\ 
& 0.5 & 99.8 & 99.8 & 0.045 & 0.0466 & 94.2 & 94.8 & 0.0263 & 0.0269 & 95.6
& 95.7 & 0.0279 & 0.0285 \\ 
& 0.9 & 99.8 & 99.8 & 0.045 & 0.0466 & 95.6 & 95.5 & 0.025 & 0.0256 & 97.1 & 
97.7 & 0.0279 & 0.0285 \\ 
\rule[-1ex]{0cm}{4ex} 0.99 & 0.1 & 100 & 100 & 0.0451 & 0.0467 & 94.7 & 94.8
& 0.0266 & 0.0272 & 95.6 & 95.7 & 0.0279 & 0.0286 \\ 
& 0.5 & 100 & 100 & 0.0451 & 0.0468 & 96.2 & 96.3 & 0.0205 & 0.0211 & 
\textbf{99} & \textbf{99} & 0.0279 & 0.0287 \\ 
& 0.9 & 100 & 100 & 0.0451 & 0.0466 & 96.4 & 96.7 & 0.0115 & 0.0123 & 
\textbf{100} & \textbf{100} & 0.0279 & 0.0285 \\ \hline
\end{tabular}

\normalsize%

\end{adjustbox}%
\end{center}

\medskip 
\noindent
\footnotesize%
Note -- The table provides coverage rates and widths of $95\%$ confidence
intervals for the difference in LC ordinate at percentile $p=0.5$. The data
are generated according to DGP: $F_{1} = \text{SM}(1, 1.6971, 8.3679)$, $%
F_{2} = \text{SM}(0.4, 2.8, 1.7)$, $\rho \in \{0, \pm 0.5, \pm0.99\}$ in %
\eqref{eq:DGP-GaussCopula}, overlap portion $\lambda \in \{0.1, 0.5, 0.9\}$.
The sample sizes are both $1000$. Columns $(1)$ - $(4)$ are performance by
asymptotic intersection method (Asym), bootstrap intersection method (Boot)
in section \ref{sec:DS}. Columns $(5)$ - $(8)$ are performance by asymptotic
(Asym) and bootstrap (Boot) methods with overlapping samples in section \ref%
{sec:OS}, using $\hat{\sigma }_{\Delta }^{D}$ in %
\eqref{eq:Est-AVar-dIndex-MC}. Columns $(9)$ - $(12)$ are performance by
asymptotic (Asym) and bootstrap (Boot) methods if incorrectly ignoring the
sample dependence, using $\hat{\sigma }_{\Delta }^{I}$ in %
\eqref{eq:Est-AVar-dIndex-IS-MC}. The results are based on $1000$
replications with 399 bootstrap repetitions each. The bold font indicates
cases where coverage rates largely deviate from $95\%$.

\normalsize%

\end{sidewaystable}%

\FloatBarrier

\subsection{Effects of heavy tails \label{sec: Effects of heavy tails}}

\label{sec:MC-HeavyTail} 

Heavy-tailed distributions are notorious for causing problems for both
asymptotic and bootstrap inference. We will examine the impacts of heavy
tails on the performance of proposed inference methods. We still consider
the SM distribution $\text{SM}(b, a, q)$ in \eqref{eq:SM-CDF}. The upper
tail behaves similarly to the Pareto distribution with parameter $\beta = aq$%
, also known as the stability index. A smaller $\beta$ indicates a heavier
upper tail. For a variance to exist, $\beta$ must be no less than two.

To mimic the U.S. income distribution in 1980, we set the distribution of
the first sample to $\text{SM}(1, 1.6971, 8.3679$. For the second sample, we
explore three SM distributions, $\text{SM}(0.4, 2.8, 1.7)$, $\text{SM}(0.4,
2, 1.5)$, and $\text{SM}(0.4, 1.4, 1.5)$. The corresponding stability
indices are $4.76$, $3$, and $2.1$, indicating increasingly heavier upper
tails.

For simplicity, we assume that the two sample sizes are identical, with $n
\in \{100, 200, 500\}$. A portion $\lambda \in \{0.1, 0.5, 0.9\}$ of
observations is drawn in pairs, with dependence characterized by a Gaussian
copula in \eqref{eq:DGP-GaussCopula} with $\rho \in \{0, \pm 0.5, \pm0.99\}$%
. The performance of CIs is evaluated based on $1000$ replications, with $B
= 399$ bootstrap repetitions each.

We summarize the DGPs as follows.

\medskip \textbf{DGP I-A}: $F_{1} = \text{SM}(1, 1.6971, 8.3679)$, $F_{2} = 
\text{SM}(0.4, 2.8, 1.7)$, $\rho \in \{0, \pm 0.5, \pm0.99\}$ in %
\eqref{eq:DGP-GaussCopula}, stability index of $F_{2}$ is $\beta_{2} = 4.76$%
, overlap portion $\lambda \in \{0.1, 0.5, 0.9\}$.

\medskip

\textbf{DGP I-B}: $F_{1} = \text{SM}(1, 1.6971, 8.3679)$, $F_{2} = \text{SM}%
(0.4, 2, 1.5)$, $\rho \in \{0, \pm 0.5, \pm0.99\}$ in %
\eqref{eq:DGP-GaussCopula}, stability index of $F_{2}$ is $\beta_{2} = 3$,
overlap portion $\lambda \in \{0.1, 0.5, 0.9\}$.

\medskip

\textbf{DGP I-C}: $F_{1} = \text{SM}(1, 1.6971, 8.3679)$, $F_{2} = \text{SM}%
(0.4, 1.4, 1.5)$, $\rho \in \{0, \pm 0.5, \pm0.99\}$ in %
\eqref{eq:DGP-GaussCopula}, stability index of $F_{2}$ is $\beta_{2} = 2.1$,
overlap portion $\lambda \in \{0.1, 0.5, 0.9\}$.

\medskip

We consider changes in the Gini index, population mean and LC ordinate at
percentile of $0.5$. We assess the performance of four proposed confidence
intervals (CIs) at the level of $95\%$. These include the asymptotic
intersection method (IM-Asym), bootstrap intersection method (IM-Boot),
asymptotic inference with overlapping samples (OS-Asym), and bootstrap
inference with overlapping samples (OS-Boot).

For DGP I-A, we illustrate the performance of CIs for the difference in Gini
index, mean, and LC ordinate in Table \ref{tab:CI-dI-DGP1}, \ref%
{tab:CI-dmu-DGP1}, and \ref{tab:CI-dLC-DGP1}, respectively. As expected, the
asymptotic and bootstrap CIs with overlapping samples (OS-Asym and OS-Boot)
exhibit coverage rates close to the nominal level of $95\%$ due to the
absence of heavy tails. However, the asymptotic and bootstrap intersection
methods (IM-Asym and IM-Boot) are conservative, with consistently higher
coverage rates than the nominal level, yielding a trade-off between
efficiency and robustness.

For DGP I-B, we present the performance of CIs for three indices differences
in Table \ref{tab:CI-dI-DGP2}, \ref{tab:CI-dmu-DGP2}, and \ref%
{tab:CI-dLC-DGP2}. In this scenario, the second distribution has a heavy
upper tail. The coverage rates of asymptotic CIs with overlapping samples
are lower than $95\%$ when two samples are strongly positively related; see,
for example, column OS-Asym when $\rho = 0.99$ and $\lambda = 0.9$. The
bootstrap CI (OS-Boot), however, can alleviate this issue. The asymptotic
and bootstrap IMs (IM-Asym and IM-Boot) are still conservative.
Nevertheless, the widths are considered acceptable compared to the bootstrap
CIs.

For DGP I-C, we evaluate the performance of CIs for the three indices
differences in \ref{tab:CI-dI-DGP3}, \ref{tab:CI-dmu-DGP3}, and \ref%
{tab:CI-dLC-DGP3}. The second distribution now has an extremely heavy tail,
resulting in a considerable variance. Neither the asymptotic and bootstrap
CIs with overlapping samples (OS-Asym and OS-Boot) yield reliable results,
as the coverage falls below $95\%$. However, the bootstrap CI (OS-Boot)
suffers less coverage distortion compared to the asymptotic CI. The IMs
(IM-Asym and IM-Boot) remain valid and provide reasonable widths, sometimes
comparable to those of OS-Boot.

\begin{table}[tb]%

\caption{DGP I-A: coverage and width of confidence intervals for Gini indices
differences.} \label{tab:CI-dI-DGP1}

\begin{center}
Table \thetable

DGP I-A: coverage and width of confidence intervals for Gini indices
differences.
\end{center}

\begin{adjustbox}{scale={0.9}{0.9}}%

\hspace{-0.5\totalhormargin} \begin{minipage}{\paperwidth}
\centering%

\small%

\begin{center}
\begin{tabular}{ccc|cccc|cccc}
\hline
\rule[-1ex]{0cm}{4ex} $\rho$ & $\lambda$ & $n$ & \multicolumn{4}{c|}{
Coverage $(\%)$} & \multicolumn{4}{c}{Width} \\ \cline{4-11}
\rule[-1ex]{0cm}{4ex} &  &  & IM-Asym & IM-Boot & OS-Asym & OS-Boot & IM-Asym
& IM-Boot & OS-Asym & OS-Boot \\ \hline
\rule[-1ex]{0cm}{4ex} -0.99 & 0.1 & 100 & 99.9 & 99.9 & 94.9 & 94.3 & 0.2024
& 0.228 & 0.1231 & 0.1294 \\ 
&  & 200 & 99.7 & 99.8 & 95.5 & 95.4 & 0.1464 & 0.1579 & 0.0885 & 0.0916 \\ 
&  & 500 & 99.9 & 99.9 & 94.8 & 94.9 & 0.0937 & 0.0979 & 0.0564 & 0.0576 \\ 
& 0.5 & 100 & 99.9 & 99.9 & 94.3 & 93 & 0.2031 & 0.2293 & 0.1064 & 0.111 \\ 
&  & 200 & 100 & 100 & 94.4 & 93.4 & 0.1464 & 0.1581 & 0.0762 & 0.0782 \\ 
&  & 500 & 100 & 100 & 93.9 & 93.9 & 0.0938 & 0.0979 & 0.0488 & 0.0498 \\ 
& 0.9 & 100 & 100 & 100 & 94.3 & 93.4 & 0.2024 & 0.2274 & 0.0858 & 0.0889 \\ 
&  & 200 & 100 & 100 & 94.5 & 91.9 & 0.1453 & 0.1561 & 0.0617 & 0.0634 \\ 
&  & 500 & 100 & 100 & 94.8 & 93.3 & 0.0933 & 0.0975 & 0.0397 & 0.0404 \\ 
\rule[-1ex]{0cm}{4ex} -0.5 & 0.1 & 100 & 99.9 & 100 & 94.7 & 95.3 & 0.2026 & 
0.2267 & 0.1253 & 0.1317 \\ 
&  & 200 & 99.6 & 99.8 & 94.1 & 94.1 & 0.1452 & 0.1559 & 0.0897 & 0.0926 \\ 
&  & 500 & 100 & 100 & 93.9 & 94.3 & 0.0933 & 0.0975 & 0.0574 & 0.0585 \\ 
& 0.5 & 100 & 99.9 & 99.9 & 94.4 & 94.4 & 0.2031 & 0.228 & 0.1217 & 0.1281
\\ 
&  & 200 & 99.6 & 99.6 & 93.7 & 93.4 & 0.1457 & 0.1569 & 0.087 & 0.0899 \\ 
&  & 500 & 99.9 & 99.9 & 95.2 & 94.2 & 0.0936 & 0.0978 & 0.0557 & 0.0569 \\ 
& 0.9 & 100 & 99.8 & 99.9 & 94.2 & 93.6 & 0.2031 & 0.2293 & 0.1175 & 0.1238
\\ 
&  & 200 & 100 & 100 & 94.6 & 94.7 & 0.1463 & 0.1578 & 0.0843 & 0.0872 \\ 
&  & 500 & 100 & 100 & 95.7 & 95.4 & 0.0936 & 0.0978 & 0.0536 & 0.0547 \\ 
\rule[-1ex]{0cm}{4ex} 0 & 0.1 & 100 & 99.9 & 99.9 & 94.5 & 94.2 & 0.2018 & 
0.2267 & 0.1256 & 0.1323 \\ 
&  & 200 & 99.8 & 99.8 & 94.9 & 95.1 & 0.1456 & 0.1562 & 0.0904 & 0.0937 \\ 
&  & 500 & 99.6 & 99.7 & 94.2 & 93.8 & 0.0938 & 0.0981 & 0.0581 & 0.0594 \\ 
& 0.5 & 100 & 99.5 & 99.7 & 93.8 & 93.7 & 0.2029 & 0.2279 & 0.1263 & 0.134
\\ 
&  & 200 & 100 & 100 & 95.5 & 95.1 & 0.1464 & 0.1586 & 0.091 & 0.0946 \\ 
&  & 500 & 99.6 & 99.8 & 95.4 & 94.9 & 0.0939 & 0.0983 & 0.0582 & 0.0596 \\ 
& 0.9 & 100 & 99.4 & 99.7 & 94.1 & 93.8 & 0.2007 & 0.2244 & 0.1249 & 0.1319
\\ 
&  & 200 & 99.6 & 99.6 & 94.7 & 94.8 & 0.1459 & 0.1568 & 0.0905 & 0.0939 \\ 
&  & 500 & 99.9 & 100 & 93.8 & 93.6 & 0.094 & 0.0985 & 0.0583 & 0.0597 \\ 
\rule[-1ex]{0cm}{4ex} 0.5 & 0.1 & 100 & 99.6 & 99.8 & 93.7 & 93.5 & 0.2033 & 
0.2281 & 0.1252 & 0.132 \\ 
&  & 200 & 99.6 & 99.7 & 94.6 & 94.4 & 0.1452 & 0.1569 & 0.0893 & 0.0925 \\ 
&  & 500 & 99.9 & 99.9 & 95.1 & 94.6 & 0.0939 & 0.0984 & 0.0576 & 0.0589 \\ 
& 0.5 & 100 & 100 & 100 & 95.3 & 94.7 & 0.2022 & 0.2273 & 0.1187 & 0.1257 \\ 
&  & 200 & 100 & 99.9 & 94.9 & 94.5 & 0.1465 & 0.1589 & 0.0858 & 0.0893 \\ 
&  & 500 & 100 & 100 & 95.1 & 94.4 & 0.0938 & 0.0982 & 0.0548 & 0.0559 \\ 
& 0.9 & 100 & 100 & 99.9 & 92.7 & 93.5 & 0.2031 & 0.2286 & 0.1125 & 0.1197
\\ 
&  & 200 & 99.9 & 100 & 95 & 94.7 & 0.1463 & 0.1583 & 0.0812 & 0.0848 \\ 
&  & 500 & 100 & 100 & 94.5 & 94.4 & 0.0936 & 0.0981 & 0.052 & 0.0533 \\ 
\rule[-1ex]{0cm}{4ex} 0.99 & 0.1 & 100 & 99.4 & 99.7 & 94.1 & 94 & 0.2029 & 
0.2291 & 0.1206 & 0.1286 \\ 
&  & 200 & 99.8 & 99.9 & 95 & 95.4 & 0.1457 & 0.1568 & 0.0862 & 0.0895 \\ 
&  & 500 & 100 & 100 & 95.6 & 95.2 & 0.0936 & 0.0977 & 0.0553 & 0.0568 \\ 
& 0.5 & 100 & 99.9 & 99.9 & 94.3 & 93.8 & 0.2032 & 0.2291 & 0.093 & 0.0964
\\ 
&  & 200 & 100 & 100 & 94.7 & 94 & 0.145 & 0.1561 & 0.0661 & 0.068 \\ 
&  & 500 & 100 & 100 & 93.9 & 94 & 0.0935 & 0.0979 & 0.0425 & 0.0435 \\ 
& 0.9 & 100 & 100 & 100 & 94.6 & 93 & 0.2014 & 0.2255 & 0.0512 & 0.0518 \\ 
&  & 200 & 100 & 100 & 95.7 & 95.1 & 0.1462 & 0.158 & 0.0371 & 0.0381 \\ 
&  & 500 & 100 & 100 & 95 & 94.2 & 0.0935 & 0.0976 & 0.0238 & 0.0244 \\ 
\hline
\end{tabular}
\end{center}

\end{minipage}%

\end{adjustbox}%

\medskip 
\noindent
\footnotesize%
Note -- The table provides coverage rates and widths of confidence intervals
of level $95\%$ for Gini indices differences by asymptotic intersection
method (IM-Asym), bootstrap intersection method (IM-Boot), asymptotic method
with overlapping samples (OS-Asym) and bootstrap method with overlapping
samples (OS-Boot). The data are generated according to DGP I-A for given
Gaussian copula $\rho$ in \eqref{eq:DGP-GaussCopula}, overlap portion $%
\lambda$ and sample size $n$. The numbers are based on $1000$ replications
with $399$ bootstrap repetitions each.

\normalsize%

\end{table}%

\begin{table}[tb]%

\caption{DGP I-A: coverage and width of confidence intervals for the mean
difference.} \label{tab:CI-dmu-DGP1}

\begin{center}
Table \thetable

DGP I-A: coverage and width of confidence intervals for the mean difference.
\end{center}

\begin{adjustbox}{scale={0.9}{0.9}}%

\hspace{-0.5\totalhormargin} \begin{minipage}{\paperwidth}
\centering%

\small%

\begin{center}
\begin{tabular}{ccc|cccc|cccc}
\hline
\rule[-1ex]{0cm}{4ex} $\rho$ & $\lambda$ & $n$ & \multicolumn{4}{c|}{
Coverage $(\%)$} & \multicolumn{4}{c}{Width} \\ \cline{4-11}
\rule[-1ex]{0cm}{4ex} &  &  & IM-Asym & IM-Boot & OS-Asym & OS-Boot & IM-Asym
& IM-Boot & OS-Asym & OS-Boot \\ \hline
\rule[-1ex]{0cm}{4ex} -0.99 & 0.1 & 100 & 99.4 & 99.5 & 94 & 92.6 & 0.1674 & 
0.1779 & 0.1086 & 0.1093 \\ 
&  & 200 & 99.6 & 99.3 & 94.6 & 94.2 & 0.1192 & 0.1239 & 0.0771 & 0.0777 \\ 
&  & 500 & 99.9 & 99.7 & 94.8 & 94 & 0.0754 & 0.0772 & 0.0486 & 0.0491 \\ 
& 0.5 & 100 & 99 & 99.1 & 95.5 & 94.8 & 0.1671 & 0.1772 & 0.1232 & 0.1246 \\ 
&  & 200 & 99.4 & 99.3 & 93.9 & 93.4 & 0.119 & 0.1233 & 0.0875 & 0.0884 \\ 
&  & 500 & 99.6 & 99 & 95.3 & 95.5 & 0.0756 & 0.0775 & 0.0554 & 0.056 \\ 
& 0.9 & 100 & 97.8 & 97.9 & 95.3 & 94.6 & 0.1669 & 0.1764 & 0.1361 & 0.1385
\\ 
&  & 200 & 98.9 & 98.7 & 96 & 96.1 & 0.1185 & 0.1231 & 0.0964 & 0.0977 \\ 
&  & 500 & 98.3 & 98.1 & 93.6 & 94.1 & 0.0754 & 0.0772 & 0.0612 & 0.0619 \\ 
\rule[-1ex]{0cm}{4ex} -0.5 & 0.1 & 100 & 99.8 & 99.8 & 95.8 & 95.8 & 0.1674
& 0.1768 & 0.1063 & 0.1075 \\ 
&  & 200 & 99.8 & 99.7 & 93.9 & 94 & 0.1184 & 0.1227 & 0.0751 & 0.0757 \\ 
&  & 500 & 99.7 & 99.6 & 95.2 & 94.7 & 0.0754 & 0.0772 & 0.0477 & 0.0482 \\ 
& 0.5 & 100 & 99.5 & 99.6 & 95.6 & 94.3 & 0.1679 & 0.1778 & 0.1151 & 0.1172
\\ 
&  & 200 & 99.5 & 99.5 & 95.3 & 95.1 & 0.1184 & 0.1228 & 0.0809 & 0.0819 \\ 
&  & 500 & 99.5 & 99.6 & 94.6 & 94.7 & 0.0755 & 0.0772 & 0.0515 & 0.0521 \\ 
& 0.9 & 100 & 99.1 & 98.9 & 94.4 & 93.1 & 0.1672 & 0.1772 & 0.1221 & 0.1245
\\ 
&  & 200 & 98.8 & 99.1 & 94 & 93.8 & 0.1191 & 0.1236 & 0.0868 & 0.0881 \\ 
&  & 500 & 99.2 & 99.1 & 95.4 & 95.7 & 0.0754 & 0.0773 & 0.0549 & 0.0555 \\ 
\rule[-1ex]{0cm}{4ex} 0 & 0.1 & 100 & 99.9 & 99.9 & 94.5 & 93.3 & 0.1671 & 
0.177 & 0.1039 & 0.105 \\ 
&  & 200 & 100 & 100 & 94.3 & 94.4 & 0.1188 & 0.1234 & 0.0737 & 0.0746 \\ 
&  & 500 & 100 & 100 & 95.9 & 95.3 & 0.0754 & 0.0773 & 0.0468 & 0.0472 \\ 
& 0.5 & 100 & 99.7 & 99.7 & 95 & 93.6 & 0.1676 & 0.1776 & 0.104 & 0.1054 \\ 
&  & 200 & 99.8 & 99.7 & 95.4 & 95.6 & 0.119 & 0.1237 & 0.074 & 0.075 \\ 
&  & 500 & 100 & 100 & 95.5 & 94.7 & 0.0755 & 0.0774 & 0.0468 & 0.0472 \\ 
& 0.9 & 100 & 99.7 & 99.7 & 95.3 & 93.9 & 0.1667 & 0.1762 & 0.1032 & 0.1052
\\ 
&  & 200 & 100 & 100 & 95.2 & 94.4 & 0.1189 & 0.1233 & 0.0738 & 0.0748 \\ 
&  & 500 & 99.7 & 99.8 & 96.4 & 96.2 & 0.0757 & 0.0775 & 0.047 & 0.0474 \\ 
\rule[-1ex]{0cm}{4ex} 0.5 & 0.1 & 100 & 99.9 & 99.9 & 95.9 & 95.5 & 0.1672 & 
0.1774 & 0.1016 & 0.1031 \\ 
&  & 200 & 99.8 & 99.8 & 94.8 & 94.4 & 0.1185 & 0.1228 & 0.0718 & 0.0727 \\ 
&  & 500 & 99.9 & 100 & 94.6 & 93.7 & 0.0756 & 0.0775 & 0.0458 & 0.0464 \\ 
& 0.5 & 100 & 100 & 100 & 95.4 & 94.3 & 0.1674 & 0.1778 & 0.0911 & 0.0925 \\ 
&  & 200 & 100 & 100 & 96.4 & 95.7 & 0.1192 & 0.1238 & 0.0648 & 0.0658 \\ 
&  & 500 & 100 & 100 & 94.2 & 94.9 & 0.0756 & 0.0776 & 0.041 & 0.0415 \\ 
& 0.9 & 100 & 100 & 100 & 95.9 & 94.7 & 0.167 & 0.1776 & 0.0787 & 0.08 \\ 
&  & 200 & 100 & 100 & 94.9 & 94.6 & 0.1186 & 0.1233 & 0.0562 & 0.057 \\ 
&  & 500 & 100 & 100 & 93.8 & 93.5 & 0.0754 & 0.0773 & 0.0356 & 0.036 \\ 
\rule[-1ex]{0cm}{4ex} 0.99 & 0.1 & 100 & 99.8 & 99.8 & 93.1 & 92.8 & 0.1674
& 0.1785 & 0.0991 & 0.1015 \\ 
&  & 200 & 99.8 & 99.8 & 94 & 93.4 & 0.1185 & 0.1229 & 0.0699 & 0.0706 \\ 
&  & 500 & 99.8 & 99.8 & 94.3 & 94.1 & 0.0756 & 0.0775 & 0.0445 & 0.045 \\ 
& 0.5 & 100 & 99.9 & 99.9 & 94.8 & 94.5 & 0.1674 & 0.1773 & 0.0746 & 0.075
\\ 
&  & 200 & 100 & 100 & 94 & 92.7 & 0.1185 & 0.1231 & 0.0527 & 0.0533 \\ 
&  & 500 & 100 & 100 & 95.2 & 94.9 & 0.0754 & 0.0774 & 0.0335 & 0.0339 \\ 
& 0.9 & 100 & 100 & 100 & 94.2 & 91.7 & 0.166 & 0.1757 & 0.0361 & 0.0349 \\ 
&  & 200 & 100 & 100 & 95.7 & 94.4 & 0.119 & 0.1234 & 0.026 & 0.0259 \\ 
&  & 500 & 100 & 100 & 95.1 & 94.4 & 0.0753 & 0.0771 & 0.0165 & 0.0167 \\ 
\hline
\end{tabular}
\end{center}

\end{minipage}%

\end{adjustbox}%

\medskip 
\noindent
\footnotesize%
Note -- The table provides coverage rates and widths of confidence intervals
of level $95\%$ for the mean difference by asymptotic intersection method
(IM-Asym), bootstrap intersection method (IM-Boot), asymptotic method with
overlapping samples (OS-Asym) and bootstrap method with overlapping samples
(OS-Boot). The data are generated according to DGP I-A for given Gaussian
copula $\rho$ in \eqref{eq:DGP-GaussCopula}, overlap portion $\lambda$ and
sample size $n$. The numbers are based on $1000$ replications with 399
bootstrap repetitions each.

\normalsize%

\end{table}%

\begin{table}[tb]%

\caption{DGP I-A: coverage and width of confidence intervals for the LC
ordinate difference at $p = 0.5$.} \label{tab:CI-dLC-DGP1}

\begin{center}
Table \thetable

DGP I-A: coverage and width of confidence intervals for the LC ordinate
difference at $p = 0.5$.
\end{center}

\begin{adjustbox}{scale={0.9}{0.9}}%

\hspace{-0.5\totalhormargin} \begin{minipage}{\paperwidth}
\centering%

\small%

\begin{center}
\begin{tabular}{ccc|cccc|cccc}
\hline
\rule[-1ex]{0cm}{4ex} $\rho$ & $\lambda$ & $n$ & \multicolumn{4}{c|}{
Coverage $(\%)$} & \multicolumn{4}{c}{Width} \\ \cline{4-11}
\rule[-1ex]{0cm}{4ex} &  &  & IM-Asym & IM-Boot & OS-Asym & OS-Boot & IM-Asym
& IM-Boot & OS-Asym & OS-Boot \\ \hline
\rule[-1ex]{0cm}{4ex} -0.99 & 0.1 & 100 & 100 & 100 & 95.5 & 97.4 & 0.142 & 
0.1704 & 0.0852 & 0.098 \\ 
&  & 200 & 99.8 & 99.8 & 95.6 & 96.8 & 0.1005 & 0.1116 & 0.0601 & 0.065 \\ 
&  & 500 & 99.9 & 99.9 & 95.4 & 95.7 & 0.0637 & 0.067 & 0.038 & 0.0395 \\ 
& 0.5 & 100 & 100 & 100 & 94.7 & 97.5 & 0.142 & 0.17 & 0.0705 & 0.0843 \\ 
&  & 200 & 100 & 100 & 94.2 & 96.9 & 0.1007 & 0.1121 & 0.0498 & 0.0551 \\ 
&  & 500 & 100 & 100 & 94.8 & 95.4 & 0.0637 & 0.0671 & 0.0314 & 0.033 \\ 
& 0.9 & 100 & 100 & 100 & 95.2 & 99.3 & 0.1418 & 0.1703 & 0.0513 & 0.0692 \\ 
&  & 200 & 100 & 100 & 95 & 97.7 & 0.1003 & 0.1114 & 0.0364 & 0.043 \\ 
&  & 500 & 100 & 100 & 94.9 & 95.8 & 0.0636 & 0.067 & 0.0231 & 0.0249 \\ 
\rule[-1ex]{0cm}{4ex} -0.5 & 0.1 & 100 & 99.8 & 100 & 96.7 & 98.3 & 0.1418 & 
0.17 & 0.0874 & 0.0997 \\ 
&  & 200 & 99.7 & 100 & 93.7 & 96.2 & 0.1003 & 0.1115 & 0.0618 & 0.0666 \\ 
&  & 500 & 99.9 & 99.9 & 94.5 & 95.4 & 0.0636 & 0.067 & 0.0391 & 0.0404 \\ 
& 0.5 & 100 & 100 & 100 & 95.6 & 97.9 & 0.142 & 0.1702 & 0.0845 & 0.0975 \\ 
&  & 200 & 99.9 & 99.9 & 94.1 & 95.6 & 0.1004 & 0.1116 & 0.0597 & 0.0645 \\ 
&  & 500 & 99.9 & 99.8 & 95.8 & 95.8 & 0.0636 & 0.0671 & 0.0377 & 0.0392 \\ 
& 0.9 & 100 & 100 & 100 & 95.1 & 97.6 & 0.142 & 0.1707 & 0.0815 & 0.0948 \\ 
&  & 200 & 99.9 & 100 & 94.1 & 96.5 & 0.1006 & 0.1119 & 0.0576 & 0.0626 \\ 
&  & 500 & 100 & 100 & 96.1 & 96.3 & 0.0637 & 0.0672 & 0.0364 & 0.0377 \\ 
\rule[-1ex]{0cm}{4ex} 0 & 0.1 & 100 & 99.8 & 100 & 95.6 & 97.5 & 0.1415 & 
0.1696 & 0.0878 & 0.1008 \\ 
&  & 200 & 99.9 & 100 & 95.1 & 96.8 & 0.1004 & 0.1114 & 0.0623 & 0.0671 \\ 
&  & 500 & 99.7 & 99.8 & 95 & 95 & 0.0637 & 0.0674 & 0.0395 & 0.0409 \\ 
& 0.5 & 100 & 100 & 100 & 94.3 & 97.1 & 0.1422 & 0.171 & 0.0883 & 0.1015 \\ 
&  & 200 & 100 & 100 & 95.8 & 97.7 & 0.1007 & 0.1121 & 0.0625 & 0.0676 \\ 
&  & 500 & 99.9 & 100 & 95.5 & 95.6 & 0.0637 & 0.0673 & 0.0395 & 0.041 \\ 
& 0.9 & 100 & 99.5 & 100 & 94.2 & 97.4 & 0.1411 & 0.1692 & 0.0876 & 0.1005
\\ 
&  & 200 & 99.8 & 99.9 & 95.3 & 96.8 & 0.1005 & 0.1115 & 0.0623 & 0.0672 \\ 
&  & 500 & 99.8 & 99.9 & 94.7 & 95 & 0.0638 & 0.0674 & 0.0396 & 0.0411 \\ 
\rule[-1ex]{0cm}{4ex} 0.5 & 0.1 & 100 & 99.7 & 100 & 94.8 & 97.2 & 0.1423 & 
0.171 & 0.0875 & 0.1005 \\ 
&  & 200 & 99.8 & 99.9 & 94.7 & 95.7 & 0.1002 & 0.1111 & 0.0616 & 0.0662 \\ 
&  & 500 & 100 & 100 & 95.3 & 96.2 & 0.0638 & 0.0672 & 0.0391 & 0.0405 \\ 
& 0.5 & 100 & 100 & 100 & 95.1 & 97.6 & 0.1416 & 0.1704 & 0.0832 & 0.0962 \\ 
&  & 200 & 99.9 & 100 & 95.7 & 96.6 & 0.1007 & 0.1117 & 0.059 & 0.0641 \\ 
&  & 500 & 99.9 & 99.9 & 94.9 & 95.9 & 0.0638 & 0.0672 & 0.0373 & 0.0388 \\ 
& 0.9 & 100 & 100 & 100 & 94.8 & 97.4 & 0.1421 & 0.1709 & 0.0791 & 0.0925 \\ 
&  & 200 & 99.9 & 100 & 95.2 & 96.6 & 0.1007 & 0.1121 & 0.056 & 0.0613 \\ 
&  & 500 & 100 & 100 & 94.9 & 95 & 0.0637 & 0.0672 & 0.0355 & 0.037 \\ 
\rule[-1ex]{0cm}{4ex} 0.99 & 0.1 & 100 & 99.6 & 100 & 93.6 & 96.6 & 0.1417 & 
0.1702 & 0.084 & 0.097 \\ 
&  & 200 & 99.8 & 99.9 & 96 & 96.3 & 0.1005 & 0.1116 & 0.0594 & 0.0644 \\ 
&  & 500 & 100 & 100 & 95.8 & 96.5 & 0.0637 & 0.0671 & 0.0376 & 0.039 \\ 
& 0.5 & 100 & 100 & 100 & 94.7 & 97.7 & 0.1419 & 0.1706 & 0.065 & 0.0781 \\ 
&  & 200 & 100 & 100 & 95.6 & 96.8 & 0.1001 & 0.1115 & 0.0456 & 0.0508 \\ 
&  & 500 & 100 & 100 & 94.2 & 94.7 & 0.0636 & 0.0672 & 0.0289 & 0.0305 \\ 
& 0.9 & 100 & 100 & 100 & 95.8 & 99.3 & 0.1416 & 0.1698 & 0.0362 & 0.0541 \\ 
&  & 200 & 100 & 100 & 94.8 & 98.5 & 0.1004 & 0.1114 & 0.0255 & 0.0326 \\ 
&  & 500 & 100 & 100 & 94.7 & 97.2 & 0.0637 & 0.067 & 0.0162 & 0.0182 \\ 
\hline
\end{tabular}
\end{center}

\end{minipage}%

\end{adjustbox}%

\medskip 
\noindent
\footnotesize%
Note -- The table provides coverage rates and widths of confidence intervals
of level $95\%$ for the Lorenz curve (LC) ordinate difference at percentile $%
p = 0.5$ by asymptotic intersection method (IM-Asym), bootstrap intersection
method (IM-Boot), asymptotic method with overlapping samples (OS-Asym) and
bootstrap method with overlapping samples (OS-Boot). The data are generated
according to DGP I-A for given Gaussian copula $\rho$ in %
\eqref{eq:DGP-GaussCopula}, overlap portion $\lambda$ and sample size $n$.
The numbers are based on $1000$ replications with 399 bootstrap repetitions
each.

\normalsize%

\end{table}%

\begin{table}[tb]%

\caption{DGP I-B: coverage and width of confidence intervals for Gini indices
differences.} \label{tab:CI-dI-DGP2}

\begin{center}
Table \thetable

DGP I-B: coverage and width of confidence intervals for Gini indices
differences.
\end{center}

\begin{adjustbox}{scale={0.9}{0.9}}%

\hspace{-0.5\totalhormargin} \begin{minipage}{\paperwidth}
\centering%

\small%

\begin{center}
\begin{tabular}{ccc|cccc|cccc}
\hline
\rule[-1ex]{0cm}{4ex} $\rho$ & $\lambda$ & $n$ & \multicolumn{4}{c|}{
Coverage $(\%)$} & \multicolumn{4}{c}{Width} \\ \cline{4-11}
\rule[-1ex]{0cm}{4ex} &  &  & IM-Asym & IM-Boot & OS-Asym & OS-Boot & IM-Asym
& IM-Boot & OS-Asym & OS-Boot \\ \hline
\rule[-1ex]{0cm}{4ex} -0.99 & 0.1 & 100 & 99.6 & 99.6 & 93.8 & 94.5 & 0.2499
& 0.3099 & 0.156 & 0.18 \\ 
&  & 200 & 99.2 & 99.5 & 92 & 92.9 & 0.1852 & 0.2156 & 0.1156 & 0.1281 \\ 
&  & 500 & 99.8 & 99.7 & 94.4 & 94.9 & 0.1226 & 0.1364 & 0.0768 & 0.0831 \\ 
& 0.5 & 100 & 99.8 & 100 & 91.9 & 92.2 & 0.2508 & 0.3122 & 0.1407 & 0.1649
\\ 
&  & 200 & 99.6 & 99.8 & 92.4 & 92.9 & 0.1873 & 0.2222 & 0.106 & 0.122 \\ 
&  & 500 & 99.8 & 99.9 & 93.5 & 93.5 & 0.122 & 0.1362 & 0.0694 & 0.0768 \\ 
& 0.9 & 100 & 100 & 100 & 90.2 & 91 & 0.25 & 0.31 & 0.1217 & 0.1468 \\ 
&  & 200 & 99.8 & 99.9 & 93.3 & 93.7 & 0.1845 & 0.2165 & 0.0915 & 0.108 \\ 
&  & 500 & 99.9 & 100 & 93.9 & 93.9 & 0.1219 & 0.1383 & 0.0619 & 0.0727 \\ 
\rule[-1ex]{0cm}{4ex} -0.5 & 0.1 & 100 & 99.1 & 99.4 & 92.8 & 93.5 & 0.2498
& 0.3067 & 0.1579 & 0.1809 \\ 
&  & 200 & 99.6 & 99.8 & 93.8 & 94.4 & 0.1864 & 0.2221 & 0.1185 & 0.135 \\ 
&  & 500 & 99.6 & 99.8 & 94.8 & 94.9 & 0.1224 & 0.1379 & 0.078 & 0.0853 \\ 
& 0.5 & 100 & 99.6 & 99.8 & 92.7 & 93.3 & 0.2498 & 0.3092 & 0.154 & 0.1769
\\ 
&  & 200 & 99.1 & 99.3 & 93 & 92.8 & 0.1842 & 0.2167 & 0.1137 & 0.1276 \\ 
&  & 500 & 99.7 & 99.8 & 94.1 & 94.5 & 0.1216 & 0.1345 & 0.0752 & 0.081 \\ 
& 0.9 & 100 & 99.9 & 100 & 91.4 & 92.8 & 0.2489 & 0.3056 & 0.1485 & 0.1709
\\ 
&  & 200 & 99.9 & 100 & 94 & 93.8 & 0.1842 & 0.2162 & 0.1104 & 0.1238 \\ 
&  & 500 & 99.8 & 99.8 & 93.9 & 93.8 & 0.1226 & 0.1374 & 0.0739 & 0.0807 \\ 
\rule[-1ex]{0cm}{4ex} 0 & 0.1 & 100 & 99.3 & 99.7 & 92.8 & 93.8 & 0.2474 & 
0.3059 & 0.1569 & 0.1818 \\ 
&  & 200 & 99.5 & 99.6 & 92.4 & 92.5 & 0.1846 & 0.218 & 0.1177 & 0.1324 \\ 
&  & 500 & 99.6 & 99.9 & 94.2 & 94.1 & 0.123 & 0.1385 & 0.0788 & 0.0864 \\ 
& 0.5 & 100 & 99.6 & 99.8 & 92.7 & 94.5 & 0.2518 & 0.3136 & 0.1601 & 0.1866
\\ 
&  & 200 & 99.7 & 99.7 & 93 & 93.6 & 0.1865 & 0.2188 & 0.1191 & 0.1337 \\ 
&  & 500 & 99.5 & 99.6 & 93 & 92.5 & 0.1224 & 0.1365 & 0.0784 & 0.0852 \\ 
& 0.9 & 100 & 99.4 & 99.8 & 92.2 & 94 & 0.2527 & 0.319 & 0.1605 & 0.1913 \\ 
&  & 200 & 99.4 & 99.5 & 92.8 & 93.3 & 0.1855 & 0.2174 & 0.1183 & 0.1329 \\ 
&  & 500 & 99.7 & 99.8 & 93.6 & 93.1 & 0.1229 & 0.1382 & 0.0787 & 0.0862 \\ 
\rule[-1ex]{0cm}{4ex} 0.5 & 0.1 & 100 & 99.3 & 99.6 & 93.1 & 94.1 & 0.2484 & 
0.3042 & 0.156 & 0.1791 \\ 
&  & 200 & 99.6 & 99.8 & 93.8 & 94.1 & 0.1854 & 0.2171 & 0.117 & 0.1313 \\ 
&  & 500 & 99.7 & 99.8 & 93.9 & 93.6 & 0.1223 & 0.1359 & 0.0775 & 0.084 \\ 
& 0.5 & 100 & 99.4 & 99.6 & 91 & 93 & 0.249 & 0.3067 & 0.1501 & 0.1753 \\ 
&  & 200 & 99.6 & 99.8 & 92.8 & 93.2 & 0.1885 & 0.226 & 0.1152 & 0.1335 \\ 
&  & 500 & 99.8 & 99.8 & 92.9 & 92.5 & 0.1227 & 0.1371 & 0.075 & 0.0819 \\ 
& 0.9 & 100 & 99.5 & 99.5 & 90 & 91.5 & 0.248 & 0.3049 & 0.1424 & 0.1689 \\ 
&  & 200 & 99.6 & 99.7 & 91.2 & 92.6 & 0.1849 & 0.2161 & 0.1076 & 0.1225 \\ 
&  & 500 & 99.7 & 99.7 & 94.1 & 93.7 & 0.1215 & 0.1345 & 0.0711 & 0.0776 \\ 
\rule[-1ex]{0cm}{4ex} 0.99 & 0.1 & 100 & 99.4 & 99.8 & 92.1 & 92.9 & 0.2506
& 0.3102 & 0.1523 & 0.1767 \\ 
&  & 200 & 99.7 & 99.7 & 93.7 & 94.1 & 0.1877 & 0.2207 & 0.115 & 0.131 \\ 
&  & 500 & 99.4 & 99.5 & 92.5 & 92.3 & 0.1217 & 0.1346 & 0.0745 & 0.0806 \\ 
& 0.5 & 100 & 99.9 & 99.9 & 90.6 & 91.4 & 0.2501 & 0.3102 & 0.1213 & 0.146
\\ 
&  & 200 & 100 & 100 & 91.5 & 92.8 & 0.1835 & 0.2124 & 0.0899 & 0.1029 \\ 
&  & 500 & 100 & 100 & 93.8 & 93.9 & 0.1234 & 0.1387 & 0.0621 & 0.0709 \\ 
& 0.9 & 100 & 100 & 100 & 85.2 & 87.9 & 0.2491 & 0.3052 & 0.0785 & 0.11 \\ 
&  & 200 & 100 & 100 & 88.7 & 90.1 & 0.1865 & 0.2253 & 0.0633 & 0.0944 \\ 
&  & 500 & 100 & 100 & 90.2 & 92.7 & 0.1213 & 0.1356 & 0.043 & 0.0557 \\ 
\hline
\end{tabular}
\end{center}

\end{minipage}%

\end{adjustbox}%

\medskip 
\noindent
\footnotesize%
Note -- The table provides coverage rates and widths of confidence intervals
of level $95\%$ for Gini indices differences by asymptotic intersection
method (IM-Asym), bootstrap intersection method (IM-Boot), asymptotic method
with overlapping samples (OS-Asym) and bootstrap method with overlapping
samples (OS-Boot). The data are generated according to DGP I-B for given
Gaussian copula $\rho$ in \eqref{eq:DGP-GaussCopula}, overlap portion $%
\lambda$ and sample size $n$. The numbers are based on $1000$ replications
with 399 bootstrap repetitions each.

\normalsize%

\end{table}%

\begin{table}[tb]%

\caption{DGP I-B: coverage and width of confidence intervals for the mean
difference.} \label{tab:CI-dmu-DGP2}

\begin{center}
Table \thetable

DGP I-B: coverage and width of confidence intervals for the mean difference.
\end{center}

\begin{adjustbox}{scale={0.9}{0.9}}%

\hspace{-0.5\totalhormargin} \begin{minipage}{\paperwidth}
\centering%

\small%

\begin{center}
\begin{tabular}{ccc|cccc|cccc}
\hline
\rule[-1ex]{0cm}{4ex} $\rho$ & $\lambda$ & $n$ & \multicolumn{4}{c|}{
Coverage $(\%)$} & \multicolumn{4}{c}{Width} \\ \cline{4-11}
\rule[-1ex]{0cm}{4ex} &  &  & IM-Asym & IM-Boot & OS-Asym & OS-Boot & IM-Asym
& IM-Boot & OS-Asym & OS-Boot \\ \hline
\rule[-1ex]{0cm}{4ex} -0.99 & 0.1 & 100 & 99.2 & 99.6 & 95.1 & 93.6 & 0.2468
& 0.2858 & 0.1684 & 0.1839 \\ 
&  & 200 & 99.1 & 99.5 & 92.7 & 92 & 0.1752 & 0.1909 & 0.119 & 0.1249 \\ 
&  & 500 & 99.4 & 99.9 & 93.7 & 92.6 & 0.1127 & 0.1192 & 0.0765 & 0.0792 \\ 
& 0.5 & 100 & 97.7 & 98.5 & 94 & 93.7 & 0.2451 & 0.2814 & 0.1831 & 0.1943 \\ 
&  & 200 & 99 & 99.2 & 96 & 94.9 & 0.1776 & 0.1963 & 0.1321 & 0.1398 \\ 
&  & 500 & 99.3 & 99.3 & 94.8 & 94.3 & 0.1125 & 0.1191 & 0.0833 & 0.0859 \\ 
& 0.9 & 100 & 97.9 & 98 & 95.1 & 94.2 & 0.2448 & 0.2829 & 0.1967 & 0.2103 \\ 
&  & 200 & 97.8 & 98.5 & 93.8 & 94.2 & 0.1758 & 0.1975 & 0.1406 & 0.1498 \\ 
&  & 500 & 98 & 98.3 & 94.5 & 94.1 & 0.1128 & 0.1223 & 0.0899 & 0.0941 \\ 
\rule[-1ex]{0cm}{4ex} -0.5 & 0.1 & 100 & 99.1 & 99.3 & 93.6 & 93.6 & 0.2423
& 0.2753 & 0.1621 & 0.1742 \\ 
&  & 200 & 99.6 & 99.5 & 94.8 & 94.4 & 0.1765 & 0.1945 & 0.1185 & 0.1266 \\ 
&  & 500 & 99.7 & 99.7 & 93.8 & 93.4 & 0.1128 & 0.1204 & 0.0756 & 0.0796 \\ 
& 0.5 & 100 & 98.9 & 99 & 94.2 & 93.8 & 0.2443 & 0.2784 & 0.173 & 0.1865 \\ 
&  & 200 & 99.3 & 99.5 & 93.3 & 93.5 & 0.175 & 0.1927 & 0.1236 & 0.1315 \\ 
&  & 500 & 99.2 & 99.2 & 94.4 & 94 & 0.1121 & 0.1183 & 0.079 & 0.0815 \\ 
& 0.9 & 100 & 98.8 & 98.9 & 93.8 & 93 & 0.2446 & 0.2797 & 0.1815 & 0.1959 \\ 
&  & 200 & 98.5 & 99 & 93.6 & 93 & 0.175 & 0.1914 & 0.1295 & 0.136 \\ 
&  & 500 & 99 & 99.1 & 93.9 & 93.5 & 0.1129 & 0.12 & 0.0834 & 0.0868 \\ 
\rule[-1ex]{0cm}{4ex} 0 & 0.1 & 100 & 98.8 & 99 & 92.7 & 91.7 & 0.2443 & 
0.2841 & 0.1614 & 0.1793 \\ 
&  & 200 & 99.3 & 99.5 & 91.5 & 91 & 0.1753 & 0.1933 & 0.1157 & 0.1239 \\ 
&  & 500 & 99.7 & 99.6 & 94.1 & 93.5 & 0.1133 & 0.121 & 0.0749 & 0.0787 \\ 
& 0.5 & 100 & 99.5 & 99.7 & 93.7 & 93 & 0.2462 & 0.2842 & 0.1627 & 0.1813 \\ 
&  & 200 & 99.4 & 99.3 & 93.8 & 93.2 & 0.1756 & 0.1921 & 0.1159 & 0.1234 \\ 
&  & 500 & 99.5 & 99.6 & 94.5 & 94.1 & 0.1127 & 0.1196 & 0.0744 & 0.0779 \\ 
& 0.9 & 100 & 99.7 & 99.7 & 93.8 & 93.5 & 0.2501 & 0.2987 & 0.1657 & 0.1904
\\ 
&  & 200 & 99.1 & 99.1 & 93.8 & 93.6 & 0.1753 & 0.1917 & 0.1156 & 0.1237 \\ 
&  & 500 & 99.4 & 99.7 & 94.5 & 93.9 & 0.1132 & 0.1204 & 0.0748 & 0.0785 \\ 
\rule[-1ex]{0cm}{4ex} 0.5 & 0.1 & 100 & 99.7 & 99.8 & 93.2 & 92.8 & 0.2413 & 
0.2727 & 0.1558 & 0.1683 \\ 
&  & 200 & 99.6 & 99.8 & 95.5 & 95.2 & 0.1754 & 0.1914 & 0.1136 & 0.1212 \\ 
&  & 500 & 99.5 & 99.7 & 93.1 & 93 & 0.1128 & 0.1197 & 0.0732 & 0.0765 \\ 
& 0.5 & 100 & 99.6 & 99.6 & 93.1 & 92.3 & 0.2423 & 0.2749 & 0.1451 & 0.1599
\\ 
&  & 200 & 99.9 & 99.9 & 94.8 & 95.5 & 0.1793 & 0.2015 & 0.1088 & 0.1215 \\ 
&  & 500 & 99.9 & 99.7 & 94.5 & 93.9 & 0.1132 & 0.1198 & 0.0685 & 0.072 \\ 
& 0.9 & 100 & 99.7 & 99.8 & 91.7 & 91.9 & 0.2424 & 0.278 & 0.1326 & 0.1513
\\ 
&  & 200 & 99.9 & 99.9 & 92.6 & 92.1 & 0.1755 & 0.1923 & 0.0967 & 0.1062 \\ 
&  & 500 & 99.7 & 99.9 & 94.1 & 94.3 & 0.112 & 0.1181 & 0.062 & 0.0655 \\ 
\rule[-1ex]{0cm}{4ex} 0.99 & 0.1 & 100 & 99.4 & 99.8 & 92.9 & 92.7 & 0.2428
& 0.2762 & 0.1539 & 0.1682 \\ 
&  & 200 & 99.7 & 99.8 & 93.6 & 92.8 & 0.1773 & 0.1946 & 0.113 & 0.1213 \\ 
&  & 500 & 99.8 & 99.9 & 95.2 & 95.2 & 0.1122 & 0.1188 & 0.0713 & 0.0746 \\ 
& 0.5 & 100 & 100 & 100 & 92.4 & 92.5 & 0.2441 & 0.2782 & 0.1277 & 0.1466 \\ 
&  & 200 & 99.9 & 100 & 92.6 & 91.7 & 0.1737 & 0.1881 & 0.091 & 0.0989 \\ 
&  & 500 & 100 & 100 & 94.8 & 93.6 & 0.1136 & 0.1211 & 0.0606 & 0.0659 \\ 
& 0.9 & 100 & 100 & 100 & 89.5 & 90.6 & 0.243 & 0.273 & 0.0918 & 0.1181 \\ 
&  & 200 & 100 & 100 & 89.7 & 91.4 & 0.179 & 0.2077 & 0.0712 & 0.1054 \\ 
&  & 500 & 100 & 100 & 93 & 94.3 & 0.1121 & 0.1194 & 0.0445 & 0.0519 \\ 
\hline
\end{tabular}
\end{center}

\end{minipage}%

\end{adjustbox}%

\medskip 
\noindent
\footnotesize%
Note -- The table provides coverage rates and widths of confidence intervals
of level $95\%$ for the mean difference by asymptotic intersection method
(IM-Asym), bootstrap intersection method (IM-Boot), asymptotic method with
overlapping samples (OS-Asym) and bootstrap method with overlapping samples
(OS-Boot). The data are generated according to DGP I-B for given Gaussian
copula $\rho$ in \eqref{eq:DGP-GaussCopula}, overlap portion $\lambda$ and
sample size $n$. The numbers are based on $1000$ replications with 399
bootstrap repetitions each.

\normalsize%

\end{table}%

\begin{table}[tb]%

\caption{DGP I-B: coverage and width of confidence intervals for the LC
ordinate difference at $p = 0.5$.} \label{tab:CI-dLC-DGP2}

\begin{center}
Table \thetable

DGP I-B: coverage and width of confidence intervals for the LC ordinate
difference at $p = 0.5$.
\end{center}

\begin{adjustbox}{scale={0.9}{0.9}}%

\hspace{-0.5\totalhormargin} \begin{minipage}{\paperwidth}
\centering%

\small%

\begin{center}
\begin{tabular}{ccc|cccc|cccc}
\hline
\rule[-1ex]{0cm}{4ex} $\rho$ & $\lambda$ & $n$ & \multicolumn{4}{c|}{
Coverage $(\%)$} & \multicolumn{4}{c}{Width} \\ \cline{4-11}
\rule[-1ex]{0cm}{4ex} &  &  & IM-Asym & IM-Boot & OS-Asym & OS-Boot & IM-Asym
& IM-Boot & OS-Asym & OS-Boot \\ \hline
\rule[-1ex]{0cm}{4ex} -0.99 & 0.1 & 100 & 99.8 & 99.8 & 95.4 & 96.6 & 0.1614
& 0.1919 & 0.0973 & 0.1114 \\ 
&  & 200 & 99.8 & 100 & 94.9 & 95.5 & 0.1157 & 0.129 & 0.0696 & 0.0756 \\ 
&  & 500 & 99.9 & 99.9 & 94.7 & 95.5 & 0.0744 & 0.0796 & 0.0449 & 0.0471 \\ 
& 0.5 & 100 & 100 & 100 & 93 & 96.3 & 0.1615 & 0.1924 & 0.0822 & 0.0979 \\ 
&  & 200 & 100 & 100 & 94.1 & 95.8 & 0.1164 & 0.1305 & 0.0595 & 0.0664 \\ 
&  & 500 & 99.9 & 100 & 95 & 95.4 & 0.0742 & 0.0791 & 0.038 & 0.0405 \\ 
& 0.9 & 100 & 100 & 100 & 93.8 & 97.4 & 0.1616 & 0.1928 & 0.0631 & 0.0822 \\ 
&  & 200 & 100 & 100 & 94.8 & 97.6 & 0.1156 & 0.1291 & 0.0459 & 0.0543 \\ 
&  & 500 & 100 & 100 & 95.5 & 95.5 & 0.0742 & 0.079 & 0.0301 & 0.0339 \\ 
\rule[-1ex]{0cm}{4ex} -0.5 & 0.1 & 100 & 99.8 & 99.9 & 94.9 & 96.7 & 0.1615
& 0.1922 & 0.0999 & 0.1134 \\ 
&  & 200 & 99.8 & 100 & 96.2 & 97.2 & 0.1162 & 0.1303 & 0.072 & 0.0782 \\ 
&  & 500 & 100 & 100 & 95.8 & 96.2 & 0.0743 & 0.0792 & 0.046 & 0.0484 \\ 
& 0.5 & 100 & 99.8 & 100 & 94.3 & 96.6 & 0.1614 & 0.192 & 0.0966 & 0.1103 \\ 
&  & 200 & 99.7 & 99.9 & 93.2 & 94.7 & 0.1156 & 0.1289 & 0.0692 & 0.0754 \\ 
&  & 500 & 99.8 & 99.9 & 95.5 & 95.7 & 0.0741 & 0.0788 & 0.0443 & 0.0464 \\ 
& 0.9 & 100 & 100 & 100 & 94.5 & 97.2 & 0.1614 & 0.192 & 0.0931 & 0.1074 \\ 
&  & 200 & 99.9 & 100 & 94.9 & 96 & 0.1156 & 0.1291 & 0.0667 & 0.0727 \\ 
&  & 500 & 99.9 & 99.9 & 95.1 & 95 & 0.0745 & 0.0794 & 0.043 & 0.0453 \\ 
\rule[-1ex]{0cm}{4ex} 0 & 0.1 & 100 & 99.6 & 100 & 94.1 & 95.9 & 0.1607 & 
0.1913 & 0.1001 & 0.1142 \\ 
&  & 200 & 99.9 & 99.9 & 94.7 & 95.8 & 0.1156 & 0.1289 & 0.0721 & 0.0784 \\ 
&  & 500 & 100 & 100 & 95.3 & 95.5 & 0.0746 & 0.0797 & 0.0465 & 0.0488 \\ 
& 0.5 & 100 & 99.9 & 100 & 93.9 & 96.1 & 0.1622 & 0.1933 & 0.1011 & 0.1156
\\ 
&  & 200 & 99.8 & 99.9 & 95 & 96.4 & 0.1163 & 0.13 & 0.0726 & 0.0787 \\ 
&  & 500 & 99.9 & 99.8 & 93.2 & 94 & 0.0744 & 0.0793 & 0.0464 & 0.0486 \\ 
& 0.9 & 100 & 99.8 & 100 & 95 & 96.9 & 0.1626 & 0.1934 & 0.1013 & 0.1165 \\ 
&  & 200 & 99.4 & 99.5 & 93.7 & 95.1 & 0.1158 & 0.1292 & 0.0723 & 0.0786 \\ 
&  & 500 & 99.8 & 99.9 & 94.2 & 94.9 & 0.0746 & 0.08 & 0.0465 & 0.0488 \\ 
\rule[-1ex]{0cm}{4ex} 0.5 & 0.1 & 100 & 99.6 & 99.7 & 95.3 & 97 & 0.1608 & 
0.1913 & 0.0993 & 0.1128 \\ 
&  & 200 & 100 & 100 & 95.4 & 96.5 & 0.1158 & 0.129 & 0.0714 & 0.0776 \\ 
&  & 500 & 99.9 & 99.9 & 95.4 & 95.2 & 0.0743 & 0.0793 & 0.0459 & 0.0479 \\ 
& 0.5 & 100 & 99.8 & 100 & 93.7 & 96.1 & 0.1611 & 0.192 & 0.0951 & 0.109 \\ 
&  & 200 & 99.9 & 100 & 94.2 & 95.3 & 0.1169 & 0.1313 & 0.0692 & 0.0761 \\ 
&  & 500 & 99.9 & 100 & 94.1 & 94.5 & 0.0745 & 0.0796 & 0.0441 & 0.0463 \\ 
& 0.9 & 100 & 99.7 & 100 & 93.5 & 96.1 & 0.161 & 0.1914 & 0.0905 & 0.1054 \\ 
&  & 200 & 100 & 100 & 93.5 & 95 & 0.1156 & 0.1294 & 0.0652 & 0.0715 \\ 
&  & 500 & 99.9 & 99.9 & 95.4 & 95.2 & 0.0741 & 0.079 & 0.0418 & 0.0439 \\ 
\rule[-1ex]{0cm}{4ex} 0.99 & 0.1 & 100 & 99.7 & 99.9 & 94.6 & 96.1 & 0.1619
& 0.1929 & 0.0963 & 0.1102 \\ 
&  & 200 & 99.9 & 99.9 & 95.1 & 95.8 & 0.1165 & 0.1311 & 0.0694 & 0.0757 \\ 
&  & 500 & 99.9 & 99.9 & 94 & 94.1 & 0.0741 & 0.079 & 0.0441 & 0.0462 \\ 
& 0.5 & 100 & 100 & 100 & 94.3 & 96.9 & 0.1619 & 0.193 & 0.0753 & 0.09 \\ 
&  & 200 & 100 & 100 & 93.5 & 95.1 & 0.1154 & 0.1287 & 0.0538 & 0.0599 \\ 
&  & 500 & 100 & 100 & 95.3 & 94.1 & 0.0747 & 0.0799 & 0.0352 & 0.0377 \\ 
& 0.9 & 100 & 100 & 100 & 90.6 & 97.6 & 0.1612 & 0.192 & 0.0447 & 0.0639 \\ 
&  & 200 & 100 & 100 & 92.9 & 96.1 & 0.116 & 0.1294 & 0.0336 & 0.0445 \\ 
&  & 500 & 100 & 100 & 93.8 & 96.3 & 0.0741 & 0.0787 & 0.022 & 0.0257 \\ 
\hline
\end{tabular}
\end{center}

\end{minipage}%

\end{adjustbox}%

\medskip 
\noindent
\footnotesize%
Note -- The table provides coverage rates and widths of confidence intervals
of level $95\%$ for the Lorenz curve (LC) ordinate difference at percentile $%
p = 0.5$ by asymptotic intersection method (IM-Asym), bootstrap intersection
method (IM-Boot), asymptotic method with overlapping samples (OS-Asym) and
bootstrap method with overlapping samples (OS-Boot). The data are generated
according to DGP I-B for given Gaussian copula $\rho$ in %
\eqref{eq:DGP-GaussCopula}, overlap portion $\lambda$ and sample size $n$.
The numbers are based on $1000$ replications with 399 bootstrap repetitions
each.

\normalsize%

\end{table}%

\begin{table}[tb]%

\caption{DGP I-C: coverage and width of confidence intervals for Gini indices
differences.} \label{tab:CI-dI-DGP3}

\begin{center}
Table \thetable

DGP I-C: coverage and width of confidence intervals for Gini indices
differences.
\end{center}

\begin{adjustbox}{scale={0.9}{0.9}}%

\hspace{-0.5\totalhormargin} \begin{minipage}{\paperwidth}
\centering%

\small%

\begin{center}
\begin{tabular}{ccc|cccc|cccc}
\hline
\rule[-1ex]{0cm}{4ex} $\rho$ & $\lambda$ & $n$ & \multicolumn{4}{c|}{
Coverage $(\%)$} & \multicolumn{4}{c}{Width} \\ \cline{4-11}
\rule[-1ex]{0cm}{4ex} &  &  & IM-Asym & IM-Boot & OS-Asym & OS-Boot & IM-Asym
& IM-Boot & OS-Asym & OS-Boot \\ \hline
\rule[-1ex]{0cm}{4ex} -0.99 & 0.1 & 100 & 98.1 & 98.8 & 86.1 & 90.4 & 0.2961
& 0.4293 & 0.1923 & 0.2594 \\ 
&  & 200 & 97.8 & 98.3 & 88.5 & 91.6 & 0.2316 & 0.3181 & 0.1526 & 0.2025 \\ 
&  & 500 & 98.3 & 98.9 & 91 & 93.4 & 0.1615 & 0.2072 & 0.1081 & 0.1367 \\ 
& 0.5 & 100 & 98.5 & 99 & 83.7 & 86.9 & 0.2925 & 0.4187 & 0.1776 & 0.2409 \\ 
&  & 200 & 99 & 99.4 & 88.5 & 90.9 & 0.2317 & 0.3169 & 0.1443 & 0.1946 \\ 
&  & 500 & 98 & 99.1 & 87.4 & 91 & 0.1606 & 0.2101 & 0.1022 & 0.135 \\ 
& 0.9 & 100 & 99.2 & 99.4 & 83.2 & 87.1 & 0.2912 & 0.405 & 0.1629 & 0.2228
\\ 
&  & 200 & 98.5 & 99.5 & 85.9 & 88.3 & 0.2302 & 0.3205 & 0.1339 & 0.1934 \\ 
&  & 500 & 98.8 & 99.7 & 88.2 & 90.8 & 0.1601 & 0.2034 & 0.0966 & 0.1277 \\ 
\rule[-1ex]{0cm}{4ex} -0.5 & 0.1 & 100 & 96.5 & 98.7 & 85.2 & 88.1 & 0.2915
& 0.4101 & 0.1896 & 0.2503 \\ 
&  & 200 & 98.3 & 98.9 & 86.1 & 89.7 & 0.2321 & 0.3254 & 0.1541 & 0.2079 \\ 
&  & 500 & 97.9 & 98.9 & 88.9 & 90.8 & 0.1633 & 0.2155 & 0.1104 & 0.1427 \\ 
& 0.5 & 100 & 97.7 & 98.3 & 85.8 & 88.8 & 0.2927 & 0.409 & 0.1867 & 0.2437
\\ 
&  & 200 & 98.1 & 99.1 & 86.9 & 89.8 & 0.2285 & 0.3091 & 0.1483 & 0.1935 \\ 
&  & 500 & 98.2 & 98.8 & 88.7 & 90.7 & 0.1625 & 0.2107 & 0.1079 & 0.1385 \\ 
& 0.9 & 100 & 97.7 & 98.8 & 87.1 & 89.6 & 0.2972 & 0.429 & 0.1862 & 0.2523
\\ 
&  & 200 & 97.8 & 98.5 & 88.1 & 91.1 & 0.2322 & 0.3184 & 0.1483 & 0.1967 \\ 
&  & 500 & 98.6 & 99.4 & 88.4 & 90.2 & 0.1631 & 0.2106 & 0.1065 & 0.1361 \\ 
\rule[-1ex]{0cm}{4ex} 0 & 0.1 & 100 & 96.9 & 98.1 & 84.8 & 89.7 & 0.2954 & 
0.426 & 0.1928 & 0.26 \\ 
&  & 200 & 97.3 & 98.4 & 85.9 & 89.1 & 0.2261 & 0.3036 & 0.1497 & 0.1938 \\ 
&  & 500 & 97.7 & 98.6 & 87.6 & 91 & 0.1636 & 0.2151 & 0.111 & 0.144 \\ 
& 0.5 & 100 & 96.7 & 98.4 & 84.8 & 89.9 & 0.2935 & 0.4203 & 0.1914 & 0.259
\\ 
&  & 200 & 97.7 & 98.7 & 87.1 & 91.2 & 0.2321 & 0.3204 & 0.1543 & 0.2058 \\ 
&  & 500 & 98.3 & 99.4 & 90.8 & 93 & 0.1613 & 0.2099 & 0.1091 & 0.1397 \\ 
& 0.9 & 100 & 97 & 98.3 & 85.1 & 89.1 & 0.2932 & 0.4255 & 0.1913 & 0.2624 \\ 
&  & 200 & 97.9 & 98.7 & 89.1 & 91.6 & 0.2292 & 0.3099 & 0.152 & 0.1983 \\ 
&  & 500 & 97.9 & 98.7 & 88.8 & 91 & 0.1607 & 0.2058 & 0.1085 & 0.1362 \\ 
\rule[-1ex]{0cm}{4ex} 0.5 & 0.1 & 100 & 97.1 & 98.5 & 86.8 & 90.6 & 0.2953 & 
0.4228 & 0.1912 & 0.2579 \\ 
&  & 200 & 97.2 & 98.3 & 86.1 & 88.7 & 0.2307 & 0.3195 & 0.1521 & 0.2039 \\ 
&  & 500 & 98 & 98.8 & 90.3 & 92.5 & 0.1623 & 0.2164 & 0.1092 & 0.1438 \\ 
& 0.5 & 100 & 97.6 & 98.9 & 84.7 & 88.8 & 0.2913 & 0.409 & 0.1814 & 0.2449
\\ 
&  & 200 & 97.8 & 98.6 & 85.8 & 89.5 & 0.225 & 0.296 & 0.1429 & 0.1842 \\ 
&  & 500 & 98.9 & 99.3 & 89.2 & 91.2 & 0.1619 & 0.2145 & 0.1062 & 0.1417 \\ 
& 0.9 & 100 & 97.8 & 98.9 & 84.1 & 89.6 & 0.291 & 0.411 & 0.1751 & 0.2443 \\ 
&  & 200 & 98.7 & 99.4 & 84.5 & 88.5 & 0.2255 & 0.3034 & 0.1392 & 0.1881 \\ 
&  & 500 & 98.9 & 99.4 & 90 & 92.6 & 0.1632 & 0.2116 & 0.1046 & 0.137 \\ 
\rule[-1ex]{0cm}{4ex} 0.99 & 0.1 & 100 & 98.3 & 99 & 85.2 & 89.4 & 0.289 & 
0.4035 & 0.181 & 0.2393 \\ 
&  & 200 & 97.8 & 98.6 & 87.4 & 90.9 & 0.2267 & 0.3033 & 0.1448 & 0.1897 \\ 
&  & 500 & 99 & 99.7 & 89.1 & 92.7 & 0.1613 & 0.2082 & 0.1056 & 0.1354 \\ 
& 0.5 & 100 & 99.2 & 99.7 & 83.7 & 88.4 & 0.2913 & 0.4146 & 0.1529 & 0.2221
\\ 
&  & 200 & 99.1 & 99.5 & 82.1 & 87.1 & 0.2273 & 0.3088 & 0.1246 & 0.1791 \\ 
&  & 500 & 99.1 & 99.7 & 86.8 & 90.6 & 0.1645 & 0.2178 & 0.0957 & 0.1366 \\ 
& 0.9 & 100 & 99.9 & 100 & 72.7 & 82.8 & 0.2912 & 0.4099 & 0.1156 & 0.2151
\\ 
&  & 200 & 99.7 & 99.7 & 78.4 & 87.4 & 0.2288 & 0.3069 & 0.1011 & 0.1757 \\ 
&  & 500 & 99.9 & 100 & 81.3 & 88.6 & 0.1583 & 0.2007 & 0.0764 & 0.1206 \\ 
\hline
\end{tabular}
\end{center}

\end{minipage}%

\end{adjustbox}%

\medskip 
\noindent
\footnotesize%
Note -- The table provides coverage rates and widths of confidence intervals
of level $95\%$ for Gini indices differences by asymptotic intersection
method (IM-Asym), bootstrap intersection method (IM-Boot), asymptotic method
with overlapping samples (OS-Asym) and bootstrap method with overlapping
samples (OS-Boot). The data are generated according to DGP I-C for given
Gaussian copula $\rho$ in \eqref{eq:DGP-GaussCopula}, overlap portion $%
\lambda$ and sample size $n$. The numbers are based on $1000$ replications
with 399 bootstrap repetitions each.

\normalsize%

\end{table}%

\begin{table}[tb]%

\caption{DGP I-C: coverage and width of confidence intervals for the mean
difference.} \label{tab:CI-dmu-DGP3}

\begin{center}
Table \thetable

DGP I-C: coverage and width of confidence intervals for the mean difference.
\end{center}

\begin{adjustbox}{scale={0.9}{0.9}}%

\hspace{-0.5\totalhormargin} \begin{minipage}{\paperwidth}
\centering%

\small%

\begin{center}
\begin{tabular}{ccc|cccc|cccc}
\hline
\rule[-1ex]{0cm}{4ex} $\rho$ & $\lambda$ & $n$ & \multicolumn{4}{c|}{
Coverage $(\%)$} & \multicolumn{4}{c}{Width} \\ \cline{4-11}
\rule[-1ex]{0cm}{4ex} &  &  & IM-Asym & IM-Boot & OS-Asym & OS-Boot & IM-Asym
& IM-Boot & OS-Asym & OS-Boot \\ \hline
\rule[-1ex]{0cm}{4ex} -0.99 & 0.1 & 100 & 96.2 & 98.5 & 89.2 & 90.8 & 0.4457
& 0.7719 & 0.335 & 0.5667 \\ 
&  & 200 & 95.8 & 97.3 & 89.6 & 91.2 & 0.3259 & 0.4585 & 0.2453 & 0.3402 \\ 
&  & 500 & 97.4 & 98.3 & 92 & 92.6 & 0.2135 & 0.2762 & 0.161 & 0.2082 \\ 
& 0.5 & 100 & 95.5 & 98.2 & 91.3 & 92 & 0.4296 & 0.6582 & 0.3363 & 0.4708 \\ 
&  & 200 & 96.6 & 97.8 & 91.6 & 93.2 & 0.3177 & 0.4209 & 0.2482 & 0.3115 \\ 
&  & 500 & 96 & 97.5 & 89.5 & 90.7 & 0.2155 & 0.3291 & 0.1687 & 0.2505 \\ 
& 0.9 & 100 & 94.1 & 97.1 & 89.6 & 91.8 & 0.4215 & 0.6159 & 0.342 & 0.4493
\\ 
&  & 200 & 95.4 & 97.2 & 89.5 & 90.7 & 0.3336 & 0.5618 & 0.271 & 0.4244 \\ 
&  & 500 & 96.1 & 97.2 & 93 & 92.8 & 0.2134 & 0.3146 & 0.1723 & 0.244 \\ 
\rule[-1ex]{0cm}{4ex} -0.5 & 0.1 & 100 & 95.8 & 97.6 & 86.8 & 89.7 & 0.4289
& 0.7199 & 0.3184 & 0.5249 \\ 
&  & 200 & 96.7 & 98.1 & 88.7 & 90.4 & 0.3377 & 0.5376 & 0.2539 & 0.4052 \\ 
&  & 500 & 97.5 & 98.3 & 90.4 & 92.9 & 0.2167 & 0.2894 & 0.163 & 0.2199 \\ 
& 0.5 & 100 & 94.8 & 97.6 & 88.7 & 89.2 & 0.4201 & 0.7253 & 0.3193 & 0.5077
\\ 
&  & 200 & 95 & 97.4 & 89.1 & 90.5 & 0.3147 & 0.4253 & 0.2398 & 0.3152 \\ 
&  & 500 & 96.3 & 97.6 & 91.2 & 92 & 0.2147 & 0.2837 & 0.1646 & 0.2154 \\ 
& 0.9 & 100 & 94.5 & 96.5 & 87.3 & 89.7 & 0.445 & 1.0777 & 0.3483 & 0.7591
\\ 
&  & 200 & 97.4 & 98.7 & 92 & 92.8 & 0.3229 & 0.4807 & 0.2522 & 0.3532 \\ 
&  & 500 & 96.5 & 98.2 & 92 & 92 & 0.2124 & 0.2618 & 0.1658 & 0.1986 \\ 
\rule[-1ex]{0cm}{4ex} 0 & 0.1 & 100 & 96.1 & 98.2 & 86.5 & 90.3 & 0.4306 & 
0.7201 & 0.3167 & 0.5286 \\ 
&  & 200 & 97.6 & 99 & 90.2 & 92.4 & 0.3124 & 0.4469 & 0.2303 & 0.3322 \\ 
&  & 500 & 97.1 & 98.2 & 89.7 & 91.9 & 0.2169 & 0.2921 & 0.162 & 0.2181 \\ 
& 0.5 & 100 & 96.2 & 98.1 & 88.1 & 91 & 0.4337 & 0.6754 & 0.3191 & 0.4905 \\ 
&  & 200 & 96.8 & 98.5 & 89.1 & 89.8 & 0.3242 & 0.4515 & 0.2403 & 0.3344 \\ 
&  & 500 & 97.7 & 98.9 & 90.6 & 92.1 & 0.2143 & 0.3028 & 0.1597 & 0.2302 \\ 
& 0.9 & 100 & 96.7 & 98.1 & 88.4 & 91.7 & 0.4681 & 1.4358 & 0.3495 & 1.0698
\\ 
&  & 200 & 96.9 & 98.3 & 89.4 & 91.3 & 0.314 & 0.4209 & 0.2317 & 0.3086 \\ 
&  & 500 & 96.9 & 98.2 & 89.6 & 91.7 & 0.2091 & 0.2569 & 0.1552 & 0.1911 \\ 
\rule[-1ex]{0cm}{4ex} 0.5 & 0.1 & 100 & 96.3 & 98.2 & 87.5 & 91.1 & 0.4336 & 
0.6626 & 0.3164 & 0.4803 \\ 
&  & 200 & 96.2 & 98.4 & 86.9 & 88.9 & 0.3224 & 0.4536 & 0.2368 & 0.3373 \\ 
&  & 500 & 98.1 & 98.8 & 89.9 & 92.4 & 0.2217 & 0.3691 & 0.165 & 0.2939 \\ 
& 0.5 & 100 & 96.8 & 98.3 & 85.8 & 89.3 & 0.4232 & 0.6267 & 0.2963 & 0.4421
\\ 
&  & 200 & 97.6 & 98.9 & 89 & 90.2 & 0.3087 & 0.3954 & 0.2172 & 0.2835 \\ 
&  & 500 & 97.4 & 98.5 & 90.1 & 92.8 & 0.2183 & 0.311 & 0.1574 & 0.2369 \\ 
& 0.9 & 100 & 96.8 & 98.7 & 85.4 & 89.7 & 0.4224 & 0.653 & 0.2837 & 0.4829
\\ 
&  & 200 & 98.7 & 99.5 & 89.1 & 92 & 0.3118 & 0.4389 & 0.2118 & 0.3215 \\ 
&  & 500 & 97.8 & 98.6 & 90.7 & 92.9 & 0.2143 & 0.2666 & 0.1492 & 0.1953 \\ 
\rule[-1ex]{0cm}{4ex} 0.99 & 0.1 & 100 & 96.9 & 98.6 & 87.9 & 91.7 & 0.421 & 
0.6705 & 0.3021 & 0.4783 \\ 
&  & 200 & 97 & 98.4 & 89.4 & 90.9 & 0.3191 & 0.4711 & 0.2314 & 0.3507 \\ 
&  & 500 & 97.9 & 98.6 & 90.4 & 92.5 & 0.2107 & 0.2683 & 0.1538 & 0.2003 \\ 
& 0.5 & 100 & 96.8 & 98.3 & 84.9 & 90.1 & 0.4246 & 0.605 & 0.2792 & 0.4362
\\ 
&  & 200 & 98.3 & 98.7 & 87 & 90.6 & 0.3156 & 0.4265 & 0.2107 & 0.3083 \\ 
&  & 500 & 98.9 & 99.2 & 90.4 & 91.5 & 0.2206 & 0.2907 & 0.1518 & 0.2158 \\ 
& 0.9 & 100 & 98 & 99.2 & 81.9 & 90.2 & 0.4173 & 0.6589 & 0.2451 & 0.5257 \\ 
&  & 200 & 98.9 & 99.3 & 86.1 & 91.3 & 0.3149 & 0.4144 & 0.1912 & 0.2964 \\ 
&  & 500 & 99.2 & 99.5 & 87 & 92.3 & 0.2076 & 0.2586 & 0.1293 & 0.1849 \\ 
\hline
\end{tabular}
\end{center}

\end{minipage}%

\end{adjustbox}%

\medskip 
\noindent
\footnotesize%
Note -- The table provides coverage rates and widths of confidence intervals
of level $95\%$ for the mean difference by asymptotic intersection method
(IM-Asym), bootstrap intersection method (IM-Boot), asymptotic method with
overlapping samples (OS-Asym) and bootstrap method with overlapping samples
(OS-Boot). The data are generated according to DGP I-C for given Gaussian
copula $\rho$ in \eqref{eq:DGP-GaussCopula}, overlap portion $\lambda$ and
sample size $n$. The numbers are based on $1000$ replications with 399
bootstrap repetitions each.

\normalsize%

\end{table}%

\begin{table}[tb]%

\caption{DGP I-C: coverage and width of confidence intervals for the LC
ordinate difference at $p = 0.5$.} \label{tab:CI-dLC-DGP3}

\begin{center}
Table \thetable

DGP I-C: coverage and width of confidence intervals for the LC ordinate
difference at $p = 0.5$.
\end{center}

\begin{adjustbox}{scale={0.9}{0.9}}%

\hspace{-0.5\totalhormargin} \begin{minipage}{\paperwidth}
\centering%

\small%

\begin{center}
\begin{tabular}{ccc|cccc|cccc}
\hline
\rule[-1ex]{0cm}{4ex} $\rho$ & $\lambda$ & $n$ & \multicolumn{4}{c|}{
Coverage $(\%)$} & \multicolumn{4}{c}{Width} \\ \cline{4-11}
\rule[-1ex]{0cm}{4ex} &  &  & IM-Asym & IM-Boot & OS-Asym & OS-Boot & IM-Asym
& IM-Boot & OS-Asym & OS-Boot \\ \hline
\rule[-1ex]{0cm}{4ex} -0.99 & 0.1 & 100 & 99.3 & 99.8 & 92.4 & 95.3 & 0.1695
& 0.2033 & 0.1034 & 0.1217 \\ 
&  & 200 & 99.9 & 100 & 93 & 95 & 0.1244 & 0.1422 & 0.0761 & 0.0867 \\ 
&  & 500 & 99.8 & 99.9 & 94.3 & 94.9 & 0.0819 & 0.0907 & 0.0504 & 0.0555 \\ 
& 0.5 & 100 & 99.5 & 99.9 & 91.4 & 95.5 & 0.1685 & 0.201 & 0.0891 & 0.1083
\\ 
&  & 200 & 99.9 & 100 & 93.2 & 94.6 & 0.1245 & 0.1432 & 0.0669 & 0.0778 \\ 
&  & 500 & 99.4 & 99.6 & 92.1 & 92.8 & 0.0817 & 0.0903 & 0.0446 & 0.0511 \\ 
& 0.9 & 100 & 99.9 & 100 & 90.5 & 94.7 & 0.1679 & 0.2008 & 0.0724 & 0.0938
\\ 
&  & 200 & 99.9 & 100 & 89.8 & 93.5 & 0.1239 & 0.1416 & 0.0557 & 0.0702 \\ 
&  & 500 & 99.9 & 100 & 91 & 92.4 & 0.0814 & 0.0901 & 0.038 & 0.0446 \\ 
\rule[-1ex]{0cm}{4ex} -0.5 & 0.1 & 100 & 99.5 & 100 & 91.7 & 93.6 & 0.1684 & 
0.2007 & 0.1047 & 0.1217 \\ 
&  & 200 & 99.4 & 99.6 & 93.1 & 94.5 & 0.1244 & 0.1421 & 0.078 & 0.0887 \\ 
&  & 500 & 99.6 & 99.7 & 94.1 & 94 & 0.0824 & 0.0912 & 0.0519 & 0.0577 \\ 
& 0.5 & 100 & 99.4 & 99.6 & 92 & 95.8 & 0.1686 & 0.2015 & 0.1015 & 0.118 \\ 
&  & 200 & 99.8 & 99.8 & 93 & 93.6 & 0.1235 & 0.1408 & 0.0748 & 0.0844 \\ 
&  & 500 & 99.6 & 99.8 & 92.5 & 93.5 & 0.0822 & 0.0912 & 0.0503 & 0.0557 \\ 
& 0.9 & 100 & 99.7 & 99.9 & 92.3 & 95 & 0.1697 & 0.2034 & 0.0991 & 0.1171 \\ 
&  & 200 & 99.6 & 99.8 & 92.8 & 94.3 & 0.1245 & 0.1432 & 0.0732 & 0.0833 \\ 
&  & 500 & 100 & 99.8 & 92.4 & 93.7 & 0.0824 & 0.0916 & 0.0489 & 0.054 \\ 
\rule[-1ex]{0cm}{4ex} 0 & 0.1 & 100 & 99.1 & 100 & 90.9 & 93.3 & 0.1694 & 
0.2027 & 0.1061 & 0.1242 \\ 
&  & 200 & 99.5 & 99.8 & 90.2 & 92.8 & 0.1227 & 0.1399 & 0.0771 & 0.0864 \\ 
&  & 500 & 99.6 & 100 & 93 & 94 & 0.0825 & 0.0918 & 0.0523 & 0.0582 \\ 
& 0.5 & 100 & 99.4 & 99.8 & 91.6 & 95 & 0.1687 & 0.2012 & 0.1056 & 0.1236 \\ 
&  & 200 & 99.8 & 99.9 & 93.8 & 95.7 & 0.1246 & 0.1424 & 0.0783 & 0.0891 \\ 
&  & 500 & 99.7 & 99.8 & 93.7 & 94.6 & 0.0819 & 0.0907 & 0.0519 & 0.0574 \\ 
& 0.9 & 100 & 99 & 99.8 & 91.1 & 94.6 & 0.1687 & 0.2015 & 0.1056 & 0.1246 \\ 
&  & 200 & 99.9 & 99.9 & 93.4 & 94.3 & 0.1236 & 0.1409 & 0.0776 & 0.0875 \\ 
&  & 500 & 99.6 & 99.7 & 93.6 & 94.3 & 0.0817 & 0.0904 & 0.0517 & 0.0567 \\ 
\rule[-1ex]{0cm}{4ex} 0.5 & 0.1 & 100 & 99.5 & 99.9 & 92.6 & 94.2 & 0.1691 & 
0.2021 & 0.1049 & 0.1228 \\ 
&  & 200 & 99.7 & 99.9 & 91.7 & 94 & 0.1242 & 0.1418 & 0.0774 & 0.088 \\ 
&  & 500 & 99.6 & 99.6 & 93 & 94.5 & 0.0821 & 0.0912 & 0.0516 & 0.0577 \\ 
& 0.5 & 100 & 99.8 & 99.9 & 92.3 & 94.8 & 0.1685 & 0.2016 & 0.1001 & 0.1173
\\ 
&  & 200 & 99.8 & 99.9 & 92.7 & 94.5 & 0.1225 & 0.1391 & 0.0731 & 0.0819 \\ 
&  & 500 & 99.8 & 100 & 92.3 & 93.7 & 0.082 & 0.0908 & 0.0497 & 0.0561 \\ 
& 0.9 & 100 & 99.9 & 99.8 & 92 & 94.6 & 0.1683 & 0.2009 & 0.0959 & 0.1137 \\ 
&  & 200 & 99.9 & 99.9 & 92.3 & 93.8 & 0.1227 & 0.1391 & 0.0704 & 0.0804 \\ 
&  & 500 & 100 & 100 & 93.5 & 94.1 & 0.0823 & 0.0912 & 0.048 & 0.0536 \\ 
\rule[-1ex]{0cm}{4ex} 0.99 & 0.1 & 100 & 99.6 & 99.7 & 92.5 & 95 & 0.1676 & 
0.2001 & 0.1001 & 0.1165 \\ 
&  & 200 & 99.5 & 99.8 & 94.1 & 95.5 & 0.123 & 0.1401 & 0.0738 & 0.0836 \\ 
&  & 500 & 99.9 & 99.9 & 93.8 & 94.9 & 0.0819 & 0.0904 & 0.0496 & 0.055 \\ 
& 0.5 & 100 & 100 & 100 & 91.8 & 94.8 & 0.1677 & 0.2001 & 0.08 & 0.098 \\ 
&  & 200 & 100 & 100 & 89.7 & 92.1 & 0.123 & 0.1402 & 0.0597 & 0.0706 \\ 
&  & 500 & 100 & 100 & 92.8 & 94.1 & 0.0827 & 0.0921 & 0.0415 & 0.0489 \\ 
& 0.9 & 100 & 100 & 100 & 85 & 92.9 & 0.1682 & 0.2008 & 0.0522 & 0.0765 \\ 
&  & 200 & 100 & 100 & 87 & 91.1 & 0.1236 & 0.1411 & 0.0412 & 0.0564 \\ 
&  & 500 & 100 & 100 & 87.2 & 91.4 & 0.0811 & 0.0892 & 0.0291 & 0.0381 \\ 
\hline
\end{tabular}
\end{center}

\end{minipage}%

\end{adjustbox}%

\medskip 
\noindent
\footnotesize%
Note -- The table provides coverage rates and widths of confidence intervals
of level $95\%$ for the Lorenz curve (LC) ordinate difference at percentile $%
p = 0.5$ by asymptotic intersection method (IM-Asym), bootstrap intersection
method (IM-Boot), asymptotic method with overlapping samples (OS-Asym) and
bootstrap method with overlapping samples (OS-Boot). The data are generated
according to DGP I-C for given Gaussian copula $\rho$ in %
\eqref{eq:DGP-GaussCopula}, overlap portion $\lambda$ and sample size $n$.
The numbers are based on $1000$ replications with 399 bootstrap repetitions
each.

\normalsize%

\end{table}%

To sum up, the asymptotic CIs with overlapping samples (OS-Asym) perform
well, unless the tail is heavy. The bootstrap inference with overlapping
samples (OS-Boot) can effectively address the issue of OS-Asym, except that
the tail is too heavy. Such a finding is consistent with the literature;
see, for example, 
\mciteAYY{Davidson2009}{Davidson2012}%
. The intersection methods (IMs) tend to be conservative, provided the tail
is not too heavy so that the variance does not exist; see a numerical
exercise in the appendix. However, they can yield reliable results when both
OS-Asym and OS-Boot fail, particularly when overlapping samples (Assumption
3.2) or the limiting joint distribution (Assumption 3.3) are in question, as
we will illustrate below.

\FloatBarrier

\subsection{Beyond overlapping samples}

\label{sec:MC-BeyondOS} 

This section examines scenarios where the overlapping samples framework in
Assumption \ref{assump:OS} may not be applicable. We take the mean
difference $\Delta\mu = \mu_{1} - \mu_{2}$ as an example for two reasons.
First, $\psi$ in \eqref{eq:ALG} has a simple expression of $\psi(x) = x -
\mu $. As $\psi(x)$ is increasing in $x$, we can compute $\rho_{\theta}$ in %
\eqref{eq:AVar-dIndex-rho} as $\rho_{\theta} = \rho$. Second, the population
moments are usually easier to calculate than the welfare indices.

We demonstrate the robustness of intersection methods (IM) by comparing them
to approaches based on overlapping samples. Specifically, we aim to
investigate the performance of various CIs when Assumption \ref{assump:OS}
fails, while Assumption \ref{assump:DS} remains valid.

First, we consider samples that violate Assumption \ref{assump:OS} $(\mathbf{%
A2})$. Suppose that pairs of unobserved observations, denoted by $%
\{(U_{1i},U_{2i})\}_{i=1}^{m}$ are iid data from the joint distribution $%
(F_{1},F_{2},\rho )$, where $F_{1}=\text{SM}(1,1.6971,8.3679)$ and $F_{2}=%
\text{SM}(0.4,2.8,1.7)$. The correlation coefficient $\rho $ in %
\eqref{eq:DGP-GaussCopula} is either $\rho =\pm 0.95$ or selected randomly
from a uniform distribution over $(-1,1)$, \emph{i.e.}, $U(-1,1)$. Suppose
that we only observe the ordered values of $\{(U_{1i},U_{2i})\}_{i=1}^{m}$.
Let $X_{k,i}=U_{k(i)}$ for $k=1,2$ and $U_{k(1)}\leq \cdots ,\leq U_{k(n)}$.
The overlap of samples are then $(X_{1i},X_{2i})_{i=1}^{m}$. The condition $(%
\mathbf{A2})$ is violated in this case. The unmatched observations are iid
data from $F_{1}$ and $F_{2}$. The sample sizes are $n=100,200,500,1000$
with overlap portions $\lambda =0.5,1$. So, the DGP is as follows.

\medskip \textbf{DGP II-A}: (Violate $(\text{A}2)$) $\{X_{k, i}\}_{i=1}^{n}
= \{U_{k(i)}\}_{i=1}^{n}$ for $k = 1, 2$, where $U_{k(i)}$ is the $i$-th
order statistic. $\{U_{1, i}\}_{i=1}^{n} \overset{iid}{\sim} \text{SM}(1,
1.6971, 8.3679)$, $\{U_{2, i}\}_{i=1}^{n} \overset{iid}{\sim} \text{SM}(0.4,
2.8, 1.7)$, $\rho = 0.95$, $-0.95$ or $\rho \sim U(-1, 1)$. The sample sizes
are $n = 100, 200, 500, 1000$. The overlap portions are $\lambda = 0.5, 1$.
\medskip

Second, we conduct an experiment where condition Assumption $(\mathbf{A3})$
fails. We assume that the second sample consists of $2n$ iid observations
from $\text{SM}(1,1.6971,8.3679)$. Given the sample $\{X_{2,i}\}_{i=1}^{2n}$%
, define $X_{1,i}=X_{2,2i-1}-X_{2,2i}$ for $i=1,\ldots ,\,n$. Let $%
\{(X_{1,i},X_{2,2i-1})\}_{i=1}^{n}$ be the matched pairs. By construction, $%
\{X_{1,i}\}_{i=1}^{n}$ and $\{X_{2,2i-1}\}_{i=1}^{n}$ are dependent, which
violates condition $(\mathbf{A3})$. The DGP is summarized as below.

\medskip \textbf{DGP II-B}: (Violate $(\text{A}3)$) $\{X_{2, i}\}_{i=1}^{2n} 
\overset{iid}{\sim} \text{SM}(0.4, 2.8, 1.7)$ and $\{X_{1, i}\}_{i=1}^{n}$
where $X_{1, i} = X_{2, 2i - 1} - X_{2, 2i}$. $n = 100, 200, 500, 1000$.
\medskip

Table \ref{tab:CI-dMean-DGP-B1} provides simulation results for DGP II-A.
Table \ref{tab:CI-dMean-DGP-B2} provides simulation results for DGP II-B.
Both tables demonstrate that intersection methods are effective, but
approaches based on overlapping samples are invalid.

\begin{table}[tb]%

\caption{DGP II-A: coverage and width of confidence intervals for
mean difference.} \label{tab:CI-dMean-DGP-B1}

\begin{center}
Table \thetable

DGP II-A: coverage and width of confidence intervals for mean difference.
\end{center}

\begin{adjustbox}{scale={0.9}{0.9}}%

\hspace{-0.5\totalhormargin} \begin{minipage}{\paperwidth}
\centering%

\small%

\begin{center}
\begin{tabular}{ccc|cccc|cccc}
\hline
\rule[-1ex]{0cm}{4ex} $\rho $ & $\lambda $ & $n$ & \multicolumn{4}{c|}{
Coverage $(\%)$} & \multicolumn{4}{c}{Width} \\ \cline{4-11}
\rule[-1ex]{0cm}{4ex} &  &  & IM-Asym & IM-Boot & OS-Asym & OS-Boot & IM-Asym
& IM-Boot & OS-Asym & OS-Boot \\ \hline
\rule[-1ex]{0cm}{4ex} 0.95 & 0.5 & 100 & 100 & 100 & 69.6 & 66.7 & 0.1686 & 
0.1799 & 0.0228 & 0.0349 \\ 
&  & 200 & 100 & 100 & 67.6 & 65.3 & 0.1189 & 0.1235 & 0.015 & 0.0233 \\ 
&  & 500 & 100 & 100 & 67.1 & 68.4 & 0.0755 & 0.0775 & 0.0094 & 0.0144 \\ 
&  & 1000 & 100 & 100 & 63.6 & 64.9 & 0.0534 & 0.0545 & 0.0062 & 0.0082 \\ 
\rule[-1ex]{0cm}{4ex} & 1 & 100 & 100 & 100 & 65 & 63.6 & 0.167 & 0.1772 & 
0.0151 & 0.022 \\ 
&  & 200 & 100 & 100 & 61.8 & 60.3 & 0.1182 & 0.123 & 0.01 & 0.0159 \\ 
&  & 500 & 100 & 100 & 63.9 & 65.8 & 0.0754 & 0.0775 & 0.0064 & 0.009 \\ 
&  & 1000 & 100 & 100 & 66 & 67.9 & 0.0534 & 0.0545 & 0.0045 & 0.0058 \\ 
\rule[-1ex]{0cm}{4ex} -0.95 & 0.5 & 100 & 99.6 & 99.6 & 25.7 & 26.9 & 0.1664
& 0.1759 & 0.0365 & 0.0544 \\ 
&  & 200 & 98.9 & 99 & 22 & 24 & 0.1192 & 0.1239 & 0.0206 & 0.0292 \\ 
&  & 500 & 99.4 & 99.5 & 18.5 & 21.1 & 0.0752 & 0.0772 & 0.0112 & 0.0159 \\ 
&  & 1000 & 99.1 & 99.1 & 16.9 & 20.3 & 0.0535 & 0.0545 & 0.0073 & 0.0099 \\ 
\rule[-1ex]{0cm}{4ex} & 1 & 100 & 98.6 & 98.6 & 20.6 & 23.5 & 0.1676 & 0.1778
& 0.0216 & 0.034 \\ 
&  & 200 & 98.4 & 98.5 & 17.9 & 21 & 0.1186 & 0.1235 & 0.0126 & 0.0178 \\ 
&  & 500 & 98.7 & 98.8 & 18.2 & 22.3 & 0.0755 & 0.0774 & 0.0074 & 0.0102 \\ 
&  & 1000 & 98 & 98.3 & 16.5 & 18.9 & 0.0533 & 0.0545 & 0.0048 & 0.006 \\ 
$\rule[-1ex]{0cm}{4ex}U(-1,1)$ & 0.5 & 100 & 100 & 99.9 & 32.2 & 32.5 & 
0.1672 & 0.177 & 0.0348 & 0.0499 \\ 
&  & 200 & 99.8 & 99.8 & 27.8 & 27.6 & 0.1187 & 0.1233 & 0.0208 & 0.029 \\ 
&  & 500 & 99.6 & 99.7 & 25.8 & 27.9 & 0.0755 & 0.0776 & 0.0114 & 0.0162 \\ 
&  & 1000 & 99.8 & 99.8 & 22.9 & 25.7 & 0.0535 & 0.0546 & 0.0073 & 0.01 \\ 
\rule[-1ex]{0cm}{4ex} & 1 & 100 & 99.7 & 99.7 & 26.5 & 27.6 & 0.1672 & 0.1779
& 0.0209 & 0.0307 \\ 
&  & 200 & 100 & 100 & 22.2 & 24.8 & 0.1185 & 0.1229 & 0.0126 & 0.0171 \\ 
&  & 500 & 99.9 & 99.9 & 21.5 & 23 & 0.0755 & 0.0774 & 0.0071 & 0.0092 \\ 
&  & 1000 & 99.9 & 99.9 & 22.2 & 24.8 & 0.0535 & 0.0544 & 0.0049 & 0.0061 \\ 
\hline
\end{tabular}
\end{center}

\end{minipage}%

\end{adjustbox}%

\medskip 
\noindent
\footnotesize%
Note -- The table provides coverage rates and widths of confidence intervals
of level $95\%$ for mean difference by asymptotic intersection method
(IM-Asym), bootstrap intersection method (IM-Boot), asymptotic method with
overlapping samples (OS-Asym) and bootstrap method with overlapping samples
(OS-Boot). The data are generated according to DGP II-A for given sample
size $n$, overlap portion $\lambda$ and $\rho$ of Gaussian copula in %
\eqref{eq:DGP-GaussCopula}. The numbers are based on $1000$ replications
with 399 bootstrap repetitions each.

\normalsize%

\end{table}%

\begin{table}[tb]%

\caption{DGP II-B: coverage and width of confidence intervals for
mean difference.} \label{tab:CI-dMean-DGP-B2}

\begin{center}
Table \thetable

DGP II-B: coverage and width of confidence intervals for mean difference.
\end{center}

\begin{adjustbox}{scale={0.9}{0.9}}%

\hspace{-0.5\totalhormargin} \begin{minipage}{\paperwidth}
\centering%

\small%

\begin{center}
\begin{tabular}{c|cc|cc|cc|cc}
\hline
\rule[-1ex]{0cm}{4ex} n & \multicolumn{4}{c|}{Coverage $(\%)$} & 
\multicolumn{4}{c}{Width} \\ \cline{2-5}\cline{6-9}
\rule[-1ex]{0cm}{4ex} & IM-Asym & IM-Boot & OS-Asym & OS-Boot & IM-Asym & 
IM-Boot & OS-Asym & OS-Boot \\ \hline
\rule[-1ex]{0cm}{4ex} 100 & 99.9 & 99.9 & 86 & 76.6 & 0.1686 & 0.1738 & 
0.0852 & 0.0869 \\ 
200 & 99.7 & 99.8 & 88.8 & 79.4 & 0.1198 & 0.1227 & 0.0605 & 0.0614 \\ 
500 & 99.8 & 99.8 & 86.4 & 79.7 & 0.0758 & 0.0771 & 0.0382 & 0.0386 \\ 
1000 & 99.7 & 99.7 & 86.2 & 77.2 & 0.0536 & 0.0545 & 0.0271 & 0.0273 \\ 
\hline
\end{tabular}
\end{center}

\end{minipage}%

\end{adjustbox}%

\medskip 
\noindent
\footnotesize
Note -- The table provides coverage rates and widths of confidence intervals
of level $95\%$ for mean difference by asymptotic intersection method
(IM-Asym), bootstrap intersection method (IM-Boot), asymptotic method with
overlapping samples (OS-Asym) and bootstrap method with overlapping samples
(OS-Boot). The data are generated according to DGP II-B for given sample
size $n$. The numbers are based on $1000$ replications with 399 bootstrap
repetitions each.

\normalsize%

\end{table}%

\newpageSlides
\FloatBarrier

\section{\sectitlesize Application to the SHIW Data of the Bank of Italy 
\label{sec:App-SHIW}}

\resetcountersSection

This section examines the dynamic change in household financial inequality
over time in Italy. We utilize the data from the Survey of Household Income
and Wealth (SHIW) conducted by the Bank of Italy. This data has been used by
several studies, including \cite{Raffinetti2015}, \cite{Raffinetti2017}, 
\cite{Bouezmarni2024} and \cite{Dufour2024}.

For simplicity, we assume that each sample consists of iid observations. To
address the heterogeneity in household structures, we normalize data by the
equivalence scale proposed by \cite{Kakwani1998}: 
\begin{equation}
hhscale_{i}=(ad_{i}+0.2ch_{1,i}+0.4ch_{2,i}+0.7ch_{3,i})^{0.8}+0.1w_{i}
\end{equation}%
where, for family $i$, $ad_{i}$ is the number of adults within a family, $%
\{ch_{j,i}\}_{j=1}^{3}$ denote the number of children with age in $(0,5]$, $%
[6,14]$ and $[15,17]$, respectively, and $w_{i}$ is the number of employees
or self-employed individuals within the families. Table \ref%
{tab:SHIW-sum-stat} presents some summary statistics, and Figure \ref%
{fig:SHIW-12-14-hist-cdf} plots the estimated densities and distribution
functions.

\begin{table}[tb]%

\caption[Descriptive statistics for household financial income.]{} \label%
{tab:SHIW-sum-stat}

\begin{center}
Table \thetable

Descriptive statistics for household financial income.
\end{center}

\hspace{-0.5\totalhormargin} \begin{minipage}{\paperwidth}
\centering%

\small%

\begin{center}
\begin{tabular}{cccc}
\hline
\rule[-1ex]{0cm}{4ex} Statistic & 2012 &  & 2014 \\ \hline
\rule[-1ex]{0cm}{4ex} $n$ & 8151 &  & 8156 \\ 
$m$ & 4459 &  & 4459 \\ 
$\Pr (X<0)$ & 0.1002 &  & 0.0858 \\ 
$\Pr (X=0)$ & 0.187 &  & 0.1629 \\ 
$\Pr (X>0)$ & 0.7128 &  & 0.7512 \\ 
Min & -17123.2 &  & -10067 \\ 
$Q(0.25)$ & 0 &  & 0.3183 \\ 
Mean & 189.5418 &  & 108.1888 \\ 
Median & 29.6702 &  & 24.8268 \\ 
$Q(0.75)$ & 154.7083 &  & 125.8219 \\ 
Max & 73127.16 &  & 24280.84 \\ 
Skew & 15.703 &  & 7.6478 \\ 
Kurt & 528.6745 &  & 157.134 \\ 
$\mathbb{E}(X|X>0)$ & 458.921 &  & 285.0303 \\ 
$\mathbb{E}(X|X<0)$ & -1372.55 &  & -1234.28 \\ 
$\mathbb{E}(|X|)$ & 464.6487 &  & 320.0156 \\ \hline
\end{tabular}
\end{center}

\end{minipage}%

\medskip 
\noindent
\footnotesize%
Note -- $n$ is sample size. $m$ is the size of overlap. $Q(p)$ is p-th
quantile. Skew stands for skewness. Kurt stands for kurtosis. 

\normalsize%

\end{table}%

\begin{figure}%

\caption[2014 vs 2012: distributions of financial income.]{} \label%
{fig:SHIW-12-14-hist-cdf}

\begin{center}
\begin{minipage}{\paperwidth}
\includegraphics[width=0.35\linewidth]{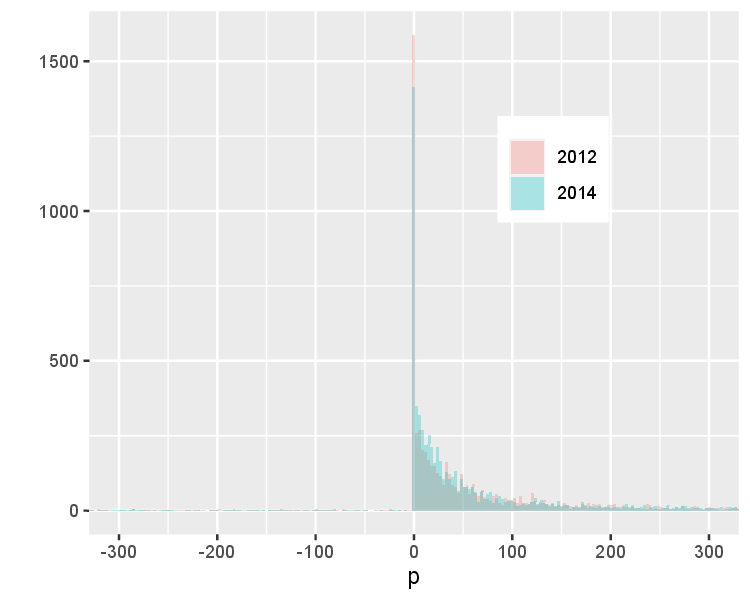} 
\includegraphics[width=0.35\linewidth]{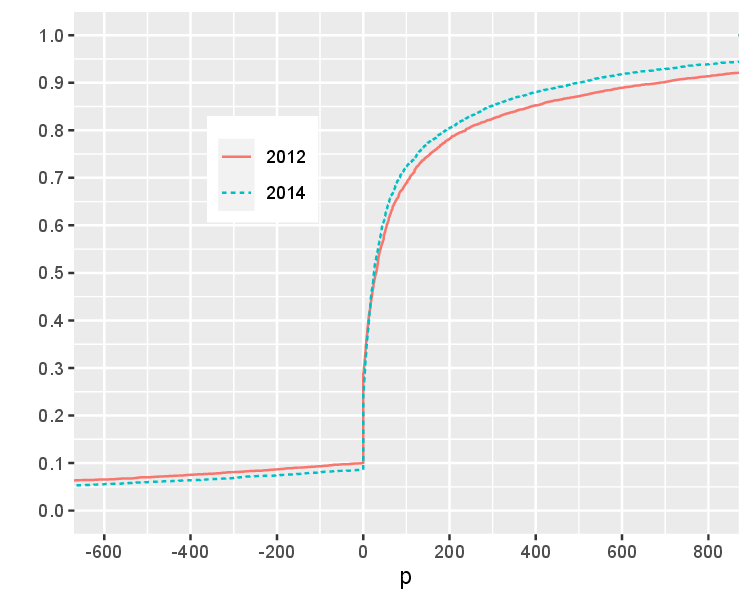}
\end{minipage}

Figure \thefigure. 2014 vs 2012: distributions of financial income.
\end{center}

{\footnotesize {\ } }

\end{figure}%

To analyze the changes in financial inequality in Italy from 2012 to 2014,
we must address two key issues. First, there is an overlap in the samples,
as nearly half of the observations appear in both waves. Therefore, we must
take the sample dependence into account. Second, financial income can be
negative, indicating that some households experienced losses. According to
Table \ref{tab:SHIW-sum-stat}, $10\%$ of households suffered losses in 2012,
while $8.5\%$ faced losses in 2014, respectively. We therefore consider a
family of extended inequality measures proposed by \cite{Bouezmarni2024} for
possibly negative variables. For convenience, we focus on two types of
specific measures.

\begin{definition}
\label{def:Ext-Ineq}%
\captiondefinition{\definitionname}{Extended Lorenz curve}%
\textbf{\ } Consider a general variable $X$ with distribution $F$. Let $\mu
(F)=\int x\,dF$ be the mean, $Q(p;F)=\inf \{x:F(x)\geq p\}$ be the p-th
quantile, and $\mu ^{p}(F)=\int |x|\,dF$ be the absolute mean. 
\begin{enumerate}[$(1)$]
\item
Let $F^{\oplus }(x)=\Pr (X\leq x\mid X>0)$ and $F^{\ominus }(x)=\Pr (X\leq
x\mid X<0)$. Then the sign-conditional Lorenz curves are 
\begin{equation}
L^{\oplus }(p;F^{\oplus })=\dfrac{\int_{0}^{p}Q(u;F^{\oplus })\,du}{\mu
(F^{\oplus })}\,,\quad L^{\ominus }(p;F^{\ominus })=\dfrac{%
\int_{0}^{p}Q(u;F^{\ominus })\,du}{\mu (F^{\ominus })}.
\label{eq:def-trun-LC}
\end{equation}

\item
The signed Lorenz curve is 
\begin{equation}
L^{s}(p;F)=\dfrac{\int_{0}^{p}Q(u;F)\,du}{\mu ^{p}(F)}.
\label{eq:def-sgn-LC}
\end{equation}

\item
Let $X_{1}^{\oplus }$ and $X_{2}^{\oplus }$ (resp. $X_{1}^{\ominus }$ and $%
X_{2}^{\ominus }$) be independent copies of variable $X^{\oplus }$ (resp. $%
X^{\ominus }$) with CDF $F^{\oplus }$ (resp. $F^{\ominus }$). Then the
sign-conditional Gini indices are 
\begin{equation}
I^{\oplus }(F^{\oplus })=\frac{\mathbb{E}(\lvert X_{1}^{\oplus
}-X_{2}^{\oplus }\rvert )}{2\mu (F^{\oplus })}\,,\quad I^{\ominus
}(F^{\ominus })=\frac{\mathbb{E}(\lvert X_{1}^{\ominus }-X_{2}^{\ominus
}\rvert )}{2\lvert \mu (F^{\ominus })\rvert }.  \label{eq:trun-Gini-def}
\end{equation}

\item
Let $X_{1}$ and $X_{2}$ are independent copies of variable $X$ with CDF $F$.
The positive Gini index is 
\begin{equation}
I^{p}(F)=\frac{\mathbb{E}(\lvert X_{1}-X_{2}\rvert )}{2\mu ^{p}(F)}.
\label{eq:pos-Gini-def}
\end{equation}%
\end{enumerate}%
\end{definition}

The sign-conditional LCs and Gini indices are tools for analyzing internal
inequality levels inside the group of winners or losers. The signed LC and
positive Gini index measure the overall level of inequality. See \cite%
{Bouezmarni2024} and \cite{Dufour2024} for more discussion. To simplify our
discussion, we will call $L^{\oplus}$ (resp. $L^{\ominus}$) the positive
(resp. negative) truncated LC. We call $I^{\oplus}$ and $I^{\ominus}$
similarly.

To facilitate our discussion, we use the superscript $\dagger$ (or $*$) to
simplify notations to indicate a specific object from a collection. For
example, $(L^{\dagger}, F^{\dagger})\in \{(L^{\oplus}, F^{\oplus}),
(L^{\ominus}, F^{\ominus})\}$ means that $(L^{\dagger}, F^{\dagger})$ can be
either the sign-conditional LCs with their corresponding distributions. We
will omit the collection if no confusion arises.

\begin{figure}%
\caption[2014 vs 2012: signed Lorenz curves.]{} \label{fig:SHIW-sLCs}

\begin{center}
\begin{minipage}{\paperwidth}
\includegraphics[width=0.35\linewidth]{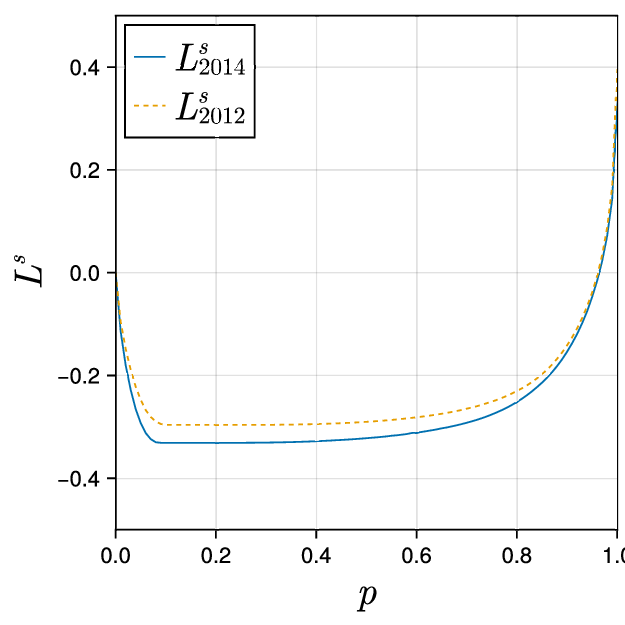} 
\includegraphics[width=0.35\linewidth]{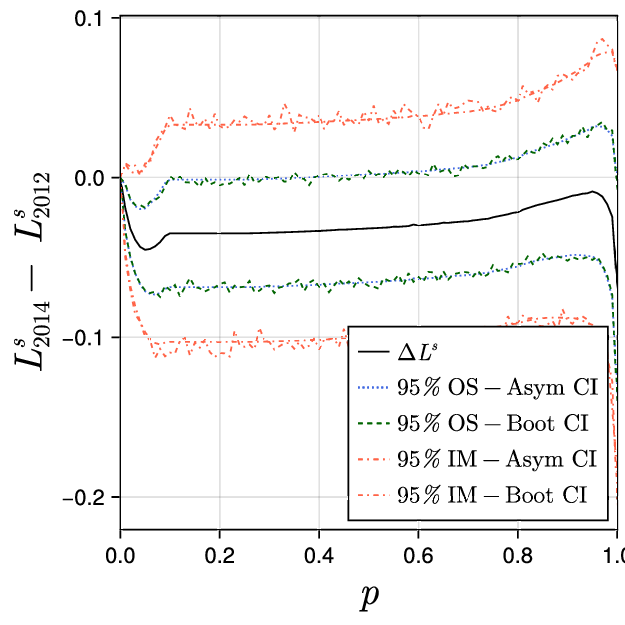}
\end{minipage}

Figure \thefigure. 2014 vs 2012: signed Lorenz curves
\end{center}

\end{figure}

We estimate the extended inequality measures by plugging the associated
EDFs. The estimated extended LCs are plotted in the left panels of Figure %
\ref{fig:SHIW-ppLCs}, \ref{fig:SHIW-nnLCs}, and \ref{fig:SHIW-sLCs}. The
estimates for extended Gini indices are displayed in Table \ref%
{tab:SHIW-dExtI-Est}. We will then conduct statistical inference for these
estimates. As shown in \cite{Dufour2024}, the estimators are ALG functionals
under specific conditions (outlined below).

\begin{lemma}
\label{lmm:ExtIneq-ALG} 
\captionlemma{\lemmaname}{Some properties of $H^{p}(x)$}
Suppose $\{X_{i}\}_{i=1}^{n}\overset{iid}{\sim }F$ where distribution
function $F$ has a finite variance. 
\begin{enumerate}[$(1)$]
\item
(Extended LC ordinate) Let $(L^{\dagger },F^{\dagger })\in \{(L^{\oplus
},F^{\oplus }),(L^{\ominus },F^{\ominus }),(L^{s},F)\}$. If $F^{\dagger }$
is differentiable at its $p$-th quantile, then $\hat{L^{\dagger }}(p)$
satisfies Definition \ref{def:ALG} with 
\begin{equation}
\psi ^{\dagger }(x;L^{\dagger },F^{\dagger },p)=W^{\dagger }(x;L^{\dagger
},F^{\dagger },p)-\mathbb{E}[W^{\dagger }(X^{\dagger };L^{\dagger
},F^{\dagger },p)]
\end{equation}%
where $(W^{\dagger },X^{\dagger })\in \{(W^{\oplus },X^{\oplus
}),(W^{\ominus },X^{\ominus }),(W^{s},X)\}$ 
\begin{equation}
\begin{split}
W^{\oplus }(x;L^{\oplus },F^{\oplus },p)& =\frac{\big[x-Q(p;F^{\oplus })\big]%
\mathbf{1}\big(x\leq Q(p;F^{\oplus })\big)-x\,L^{\oplus }(p;F^{\oplus })}{%
\mu (F^{\oplus })}, \\
W^{\ominus }(x;L^{\ominus },F^{\ominus },p)& =\frac{\big[x-Q(p;F^{\ominus })%
\big]\mathbf{1}\big(x\leq Q(p;F^{\ominus })\big)-x\,L^{\ominus
}(p;F^{\ominus })}{\mu (F^{\ominus })}, \\
W^{s}(x;L^{s},F,p)& =\frac{[x-Q(p;F)]\mathbf{1}(x\leq Q(p;F))-|x|\,L^{s}(p;F)%
}{\mu ^{p}(F)}.
\end{split}
\label{eq:ExtLC-ALG}
\end{equation}

\item
(Extended Gini index) Let $(I^{\dagger },F^{\dagger })\in \{(I^{\oplus
},F^{\oplus }),(I^{\ominus },I^{\ominus }),(I^{p},F)\}$. If $F^{\dagger }$
is continuous, then $\hat{I^{\dagger }}$ satisfies the Definition \ref%
{def:ALG} with 
\begin{equation}
\psi ^{\dagger }(x;I^{\dagger },F^{\dagger })=K^{\dagger }(x;I^{\dagger
},F^{\dagger })-\mathbb{E}[K^{\dagger }(X^{\dagger };I^{\dagger },F^{\dagger
})]
\end{equation}%
where $(K^{\dagger },X^{\dagger })\in \{(K^{\oplus },X^{\oplus
}),(K^{\ominus },X^{\ominus }),(K^{p},X)\}$ 
\begin{equation}
\begin{split}
K^{\oplus }(x;I^{\oplus },F^{\oplus })& =\frac{2xF^{\dagger
}(x)-2\int^{x}t\,dF^{\oplus }(t)-[I^{\oplus }(F^{\oplus })+1]x}{\mu
(F^{\oplus })}, \\
K^{\ominus }(x;I^{\ominus },F^{\ominus })& =\frac{2xF^{\ominus
}(x)-2\int^{x}t\,dF^{\ominus }(t)-[I^{\ominus }(F^{\ominus })+1]x}{\mu
(F^{\ominus })}, \\
K^{p}(x;I^{p},F)& =\frac{2xF(x)-2\int^{x}t\,dF(t)-I^{p}(F)\,|x|-x}{\mu
^{p}(F)}.
\end{split}
\label{eq:ExtI-ALG}
\end{equation}

\end{enumerate}%
\end{lemma}

Under Assumption \ref{assump:OS}, it is not difficult to show that the above
estimates fulfill Assumption \ref{assump:JointNormal} by multivariate
central limit theorem. So, we obtain the following corollary.

\begin{corollary}
\label{coro:AsyN-dExtIneq} 
\captionproposition{\propositionname}{Convergence in distribution for the extended inequality difference}
Suppose that the Assumption \ref{assump:OS} holds with finite variance
matrix. 
\begin{enumerate}[$(1)$]
\item
(Extended LC ordinate) If the distribution $F_{k}^{\dagger }\in
\{F_{k}^{\oplus },F_{k}^{\ominus },F_{k}\}$, $k=1,2$, is differentiable at
its $p$-th quantile, then 
\begin{equation}
\sqrt{N}\big(\Delta \hat{L^{\dagger }}(p)-\Delta L^{\dagger }(p)\big)\overset%
{d}{\longrightarrow }\mathrm{N}\big(0,\Sigma _{\Delta }^{\dagger }(p)\big)
\end{equation}%
where $\Delta \,L^{\dagger }\in \{\Delta \,L^{\oplus },\Delta \,L^{\ominus
},\Delta \,L^{s}\}$, and $\Sigma _{\Delta }^{\dagger }\in \{\Sigma _{\Delta
}^{\oplus },\Sigma _{\Delta }^{\ominus },\Sigma _{\Delta }^{s}\}$ is such
that 
\begin{equation}
\begin{split}
\Sigma _{\Delta }^{\dagger }(p)=& (1-\eta _{1})\mathrm{Var}\big[%
W_{1}^{\dagger }\big(X_{1}^{\dagger };p\big)\big]+\eta _{1}\mathrm{Var}\big[%
W_{2}^{\dagger }\big(X_{2}^{\dagger };p\big)\big] \\
& -2\sqrt{\eta _{1}(1-\eta _{1})\lambda _{1}\lambda _{2}}\mathrm{Cov}\big[%
W_{1}^{\dagger }\big(X_{1}^{\dagger };p\big),W_{2}^{\dagger }\big(%
X_{2}^{\dagger };p\big)\big],
\end{split}%
\end{equation}%
with $W_{k}^{\dagger }(x;p)=W^{\dagger }(x;L^{\dagger },F_{k}^{\dagger },p)$
given by \eqref{eq:ExtLC-ALG}.

\item
(Extended Gini index) If the distribution $F_{k}$, $k=1,2$, is continuous
with a possible jump at zero, then 
\begin{equation}
\sqrt{N}\big(\Delta \hat{I^{\dagger }}-\Delta I^{\dagger }\big)\overset{d}{%
\longrightarrow }\mathrm{N}\big(0,V_{\Delta }^{\dagger }\big)
\end{equation}%
where $\Delta \,I^{\dagger }\in \{\Delta \,I^{\oplus },\Delta \,I^{\ominus
},\Delta \,I^{p}\}$, and $V_{\Delta }^{\dagger }\in \{V_{\Delta }^{\oplus
},V_{\Delta }^{\ominus },V_{\Delta }^{p}\}$ is such that 
\begin{equation}
\begin{split}
V_{\Delta }^{\dagger }=& (1-\eta _{1})\mathrm{Var}\big[K_{1}^{\dagger }\big(%
X_{1}^{\dagger }\big)\big]+\eta _{1}\mathrm{Var}\big[K_{2}^{\dagger }\big(%
X_{2}^{\dagger }\big)\big] \\
& -2\sqrt{\eta _{1}(1-\eta _{1})\lambda _{1}\lambda _{2}}\mathrm{Cov}\big[%
K_{1}^{\dagger }\big(X_{1}^{\dagger }\big),K_{2}^{\dagger }\big(%
X_{2}^{\dagger }\big)\big]
\end{split}%
\end{equation}%
with $K_{k}^{\dagger }(x)=K^{\dagger }(x;I^{\dagger },F_{k}^{\dagger })$
given by \eqref{eq:ExtI-ALG}.

\end{enumerate}%
\end{corollary}

According to \eqref{eq:AVar-dIndex-Est}, we estimate the asymptotic variance
by the sample variance of the estimated influence function. We then obtain
CIs for the inequality changes at the level of $95\%$ using four proposed
methods, including the asymptotic intersection method (IM-Asym), bootstrap
intersection method (IM-Boot), asymptotic inference with overlapping samples
(OS-Asym), and bootstrap inference with overlapping samples (OS-Boot).

We first examine the internal inequality among winners. Figure \ref%
{fig:SHIW-ppLCs} plots the change in positive LC from 2012 to 2014, \emph{%
i.e.}, $\Delta L^{\oplus }=L_{2014}^{\oplus }-L_{2012}^{\oplus }$, and $95\%$
CIs. We find that $L_{2014}^{\oplus }$ is closer to the diagonal line, which
suggests a decrease in inequality level inside the group of winners.
Asymptotic and bootstrap CIs based on overlapping samples (OS-Asym and
OS-Boot) confirm this result, as zero falls outside the CIs. However, the
intersection methods (IM-Asym and IM-Boot) give an opposite conclusion.
Given the data structure, we believe that overlapping samples are a suitable
assumption and thus prefer the findings from OS-Asym and OS-Boot, as IM can
be conservative under such a scenario based on simulation results in the
previous section. We observe similar results for $\Delta I^{\oplus }$ in
Table \ref{tab:SHIW-dExtI-CI}. The estimate for $\Delta I^{\oplus }$ in
Table \ref{tab:SHIW-dExtI-Est} is only $-0.0246$, suggesting a relatively
small decrease in inequality level.

\begin{figure}%
\caption[2014 vs 2012: positively truncated Lorenz curves.]{} \label%
{fig:SHIW-ppLCs}

\begin{center}
\begin{minipage}{\paperwidth}
\includegraphics[width=0.35\linewidth]{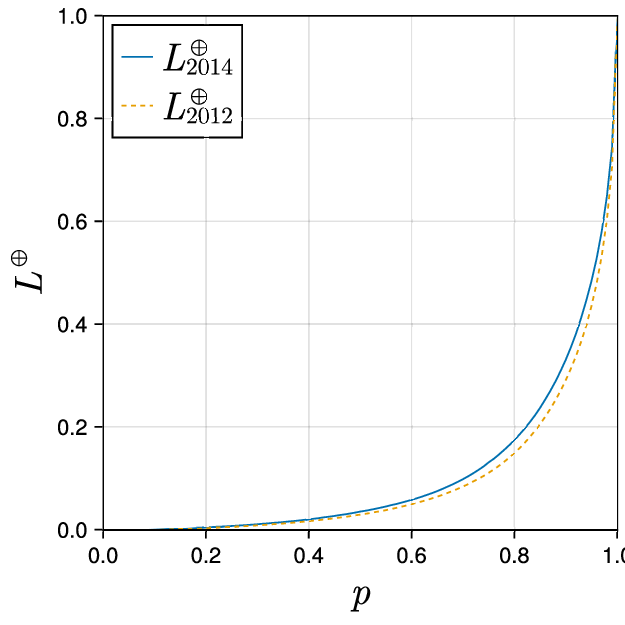} 
\includegraphics[width=0.35\linewidth]{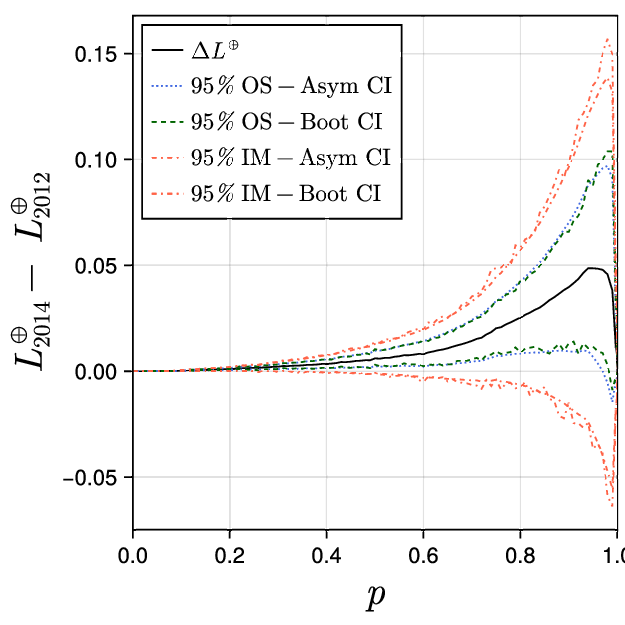}
\end{minipage}

Figure \thefigure. 2014 vs 2012: positively truncated Lorenz curves
\end{center}

\end{figure}%

We next study the internal inequality among the losers, as measured by $%
\Delta L^{\ominus}$ in Figure \ref{fig:SHIW-nnLCs} and $\Delta I^{\ominus}$
in Table \ref{tab:SHIW-dExtI-CI}. We find that only the OS-Asym CI for $%
\Delta I^{\ominus}$ excludes zero, while all the other CIs contain zero.
Given the simulation evidence in section \ref{sec:MC}, OS-Boot CI is
generally more reliable than OS-Asym CI. Therefore, we cannot conclude that
there is a significant change in internal inequality within the subgroup of
losers.

\begin{figure}%
\caption[2014 vs 2012: negatively truncated Lorenz curves.]{} \label%
{fig:SHIW-nnLCs}

\begin{center}
\begin{minipage}{\paperwidth}
\includegraphics[width=0.35\linewidth]{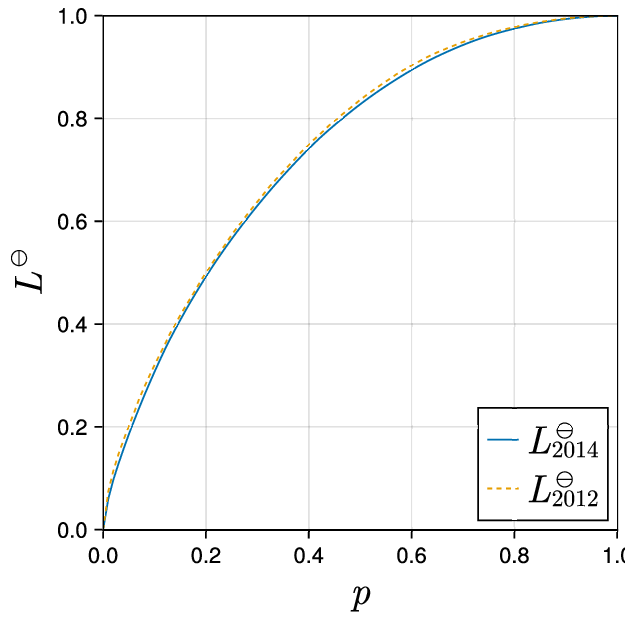} 
\includegraphics[width=0.35\linewidth]{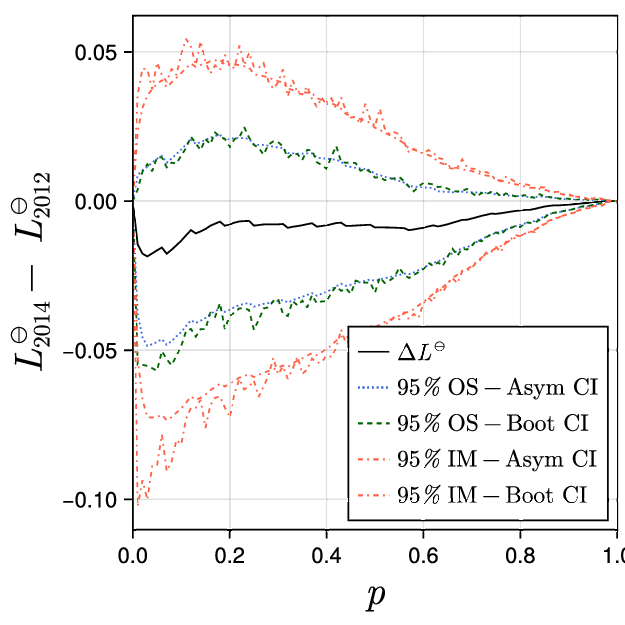}
\end{minipage}

Figure \thefigure. 2014 vs 2012: negatively truncated Lorenz curves
\end{center}

\end{figure}%

Finally, we analyze the overall inequality based on $\Delta L^{s}$ in Figure %
\ref{fig:SHIW-sLCs} and $\Delta I^{p}$ in Table \ref{tab:SHIW-dExtI-CI}. The
estimated signed LC displays a flat segment (in the left panel of Figure \ref%
{fig:SHIW-sLCs}), implying that the underlying distribution is not
differentiable at the corresponding quantile. So, we should be cautious when
interpreting inference results for the flat segments of $\Delta L^{s}$.

For $p < 0.1$, we note that asymptotic and bootstrap CIs with overlapping
samples (OS-Asym and OS-Boot) exclude zero, indicating a change over time.
We observe similar findings about $\Delta I^{p}$ by OS-Asym and OS-Boot CIs.
In contrast, the CIs based on IMs (IM-Asym and IM-Boot) include zero,
possibly due to the method's conservativeness. Note that, however, the
magnitude of the change in inequality is economically small.

\begin{table}[tb]%

\caption{2014 vs 2012: estimates for extended Gini indices.} \label%
{tab:SHIW-dExtI-Est}

\begin{center}
Table \thetable

2014 vs 2012: estimates for extended Gini indices
\end{center}

\hspace{-0.5\totalhormargin} \begin{minipage}{\paperwidth}
\centering%

\small%

\begin{center}
\begin{tabular}{cccc}
\hline
\rule[-1ex]{0cm}{4ex} Type $(I^{*})$ & $I^{*}_{2012}$ & $I^{*}_{2014}$ & $%
\Delta I^{*}$ \\ \hline
\rule[-1ex]{0cm}{4ex} $I^{\oplus}$ & 0.8110 & 0.7864 & -0.0246 \\ 
$I^{\ominus}$ & 0.4836 & 0.4699 & -0.0137 \\ 
$I^{+}$ & 0.8653 & 0.8395 & -0.0258 \\ 
$I^{-}$ & 0.9483 & 0.9546 & 0.0063 \\ 
$I^{p}$ & 0.8899 & 0.8776 & -0.0123 \\ \hline
\end{tabular}
\end{center}

\end{minipage}%

\medskip 
\noindent
\footnotesize%
Note -- $\Delta I^{*} = I^{*}_{2014} - I^{*}_{2012}$, where $I^{*}$ is one
of extended Gini indices $\{I^{\oplus}, I^{\ominus}, I^{p}\}$ in Definition %
\ref{def:Ext-Ineq}.

\normalsize%

\end{table}%

\begin{table}[tb]%

\caption{2014 vs 2012: $95\%$ CIs for difference in extended Gini index.} %
\label{tab:SHIW-dExtI-CI}

\begin{center}
Table \thetable

2014 vs 2012: $95\%$ CIs for difference in extended Gini index
\end{center}

\begin{adjustbox}{scale={0.75}{0.8}}%

\hspace{-0.5\totalhormargin} \begin{minipage}{\paperwidth}
\centering%

\small%

\begin{center}
\begin{tabular}{crlrlrlrl}
\hline
\rule[-1ex]{0cm}{4ex} Type $(I^{*})$ & \multicolumn{2}{c}{$95\%$ IM-Asym CI}
& \multicolumn{2}{c}{$95\%$ IM-Boot CI} & \multicolumn{2}{c}{$95\%$ OS-Asym
CI} & \multicolumn{2}{c}{$95\%$ OS-Boot CI} \\ \hline
\rule[-1ex]{0cm}{4ex} $I^{\oplus}$ & $[ -0.0584 $, & $0.0091]$ & $[ -0.0642 $%
, & $0.0096]$ & $[ -0.0411 $, & $-0.0082]$ & $[ -0.0409 $, & $-0.0092]$ \\ 
$I^{\ominus}$ & $[ -0.0685 $, & $0.0410]$ & $[ -0.0750 $, & $0.0456]$ & $[
-0.0231 $, & $-0.0044]$ & $[ -0.0429 $, & $0.0157]$ \\ 
$I^{p}$ & $[ -0.0271 $, & $0.0026]$ & $[ -0.0292 $, & $0.0037]$ & $[ -0.0201 
$, & $-0.0045]$ & $[ -0.0201 $, & $-0.0053]$ \\ \hline
\end{tabular}
\end{center}

\end{minipage}%

\end{adjustbox}%

\medskip 
\noindent
\footnotesize%
Note -- $\Delta I^{*} = I^{*}_{2014} - I^{*}_{2012}$, where $I^{*}$ is one
of extended Gini indices $\{I^{\oplus}, I^{\ominus}, I^{p}\}$ in Definition %
\ref{def:Ext-Ineq}.

\normalsize%

\end{table}%

\FloatBarrier
\newpageSlides

\section{\sectitlesize Extension to complex survey data \label{sec:clusters}}

\resetcountersSection

So far, we have assumed that researchers have a sample of micro-level data
drawn independently and identically distributed from a population
distribution. However, many large-scale household surveys by design involve
stratification and clustering. In this section, we briefly sketch the
potential extension of our methods to complex survey data.

In most surveys, the strata are chosen in advance and thus fixed. The
correction is straightforward; see \cite{Cochran1977} for a textbook-level
treatment. Therefore, we mainly discuss how to extend our methods,
especially the intersection approach, to clustered data. Cluster-robust
inference has become increasingly popular over the past decade and is now
employed routinely in many empirical microeconomics studies; see \cite%
{MacKinnon2023d} for a recent survey. There are at least two ways to
incorporate clusters into our framework.

\textit{First}, both samples in Assumption \ref{assump:DS} can be
generalized to clustered data. In general, the intra-cluster correlation is
positive, so when clusters are large, incorrectly assuming IID observations
can lead to smaller asymptotic variances and thus unreliable inference
(over-rejection of tests or insufficient coverage of CIs) [\cite%
{MacKinnon2016a}].

Two distinct types of assumptions are available for the asymptotic theory of
cluster-robust inference. The most common method is the "large number of
clusters" approach, where the number of clusters, often denoted by $G$,
tends to infinity [\cite{Djogbenou2019} and \cite{Hansen2019}]. The other is
the fixed-$G$ or "small number of large clusters" method [\cite%
{Ibragimov2010} and \cite{Ibragimov2016}]. The bootstrap methods, especially
the wild cluster restricted (WCR) bootstrap, often yield much more reliable
inferences than asymptotic procedures [\cite{MacKinnon2016} and \cite%
{MacKinnon2023c}]. The WCR bootstrap is usually based on the Rademacher
distribution [\cite{Davidson2008a} and \cite{Djogbenou2019}]. Nevertheless,
when the number of clusters is small, a six-point distribution is preferred [%
\cite{Webb2023}].

We may further extend to a multi-stage sampling design using techniques in 
\cite{Zheng2002} or \cite{Bhattacharya2005}.

\textit{Second}, the clustered data can generate dependence between the two
samples. To see it, suppose that there are $G$ groups, with group $g$
containing a sample of $n_{g}$ pairs of observations $(X_{1, gi}, X_{2, gi})$%
. One may conduct cluster-robust inference using either the "large number of
clusters" or the "small number of large clusters" approach. However, the
intersection methods (IMs) are also applicable. Furthermore, we can still
employ IMs even when asymptotic inference can fail.

One example is cluster heterogeneity, which we can demonstrate by the
simulation experiment under \textbf{DGP II-A}. We may view such a DGP for
the "large number of clusters" approach with $G = n$. That is, each group
contains only one pair of observations. Note that, due to sorting, the
correlation within each pair varies dramatically across groups, except for $%
\rho = 0.95$, the highly positively correlated case. As reported in Table %
\ref{tab:CI-dMean-DGP-B1}, the OS-based approach, which coincides trivially
with the cluster-robust inference, fails to provide reliable CIs. However,
the IM approach remains valid.

\FloatBarrier
\newpageSlides

\section{\sectitlesize Conclusion \label{sec:conclusion}}

\resetcountersSection

Traditional inference methods for comparing welfare indices typically assume
that samples are independent. However, this assumption can be problematic
because income samples are often dependent due to the overlap between
consecutive periods.

This paper considers a broad family of indices expressed as asymptotically
linear Gaussian functionals and proposes inference methods for index
differences with dependent samples. We first study generic dependent samples
and provide asymptotic and bootstrap intersection methods that are robust to
arbitrary sample dependence. We then explore the overlapping samples, a
specific type of dependent samples where the dependence arises solely from
the matched pairs, and propose asymptotic and bootstrap confidence intervals
for changes in indices. We additionally justify using influence functions to
estimate the asymptotic variance and covariance consistently.

We evaluate the performance of the proposed methods through a series of
simulation experiments. First, we find that sample dependence can
substantially impact asymptotic variance. Therefore, our proposed methods,
which utilize overlapping samples, can be more efficient than conventional
procedures based on independent samples. Next, we analyze the effects of
heavy tails and find that the asymptotic inference performs well for
realistic distributions but poorly for distributions with heavy tails. The
bootstrap method can alleviate this issue, except when the variance is
substantial or nonexistent. Finally, the intersection methods prove to be
reliable under generic dependent samples, particularly when the assumption
of overlapping samples is inappropriate. The intersection methods are also
reasonably efficient in certain instances, such as those involving extremely
heavy-tailed distributions.

To illustrate our methods, we applied them to analyze the changes in
financial inequality among Italian households from $2012$ to $2014$. The
application also highlights the practical importance of using extended
inequality measures in the presence of negative values.

We finally briefly discuss potential extension to clustered data and
multi-stage survey sampling. Moreover, we demonstrate that the intersection
methods are also robust to some cases where cluster-robust inference fails.

\newpage

\bibliographystyle{agsm}
\bibliography{HTYRefer}

\newpage

\appendix

\renewcommand{\thepage}{A--\arabic{page}}%

\renewcommand{\thesection}{\Alph{section}}%

\setcounter{page}{1}

\setcounter{section}{0}

\renewcommand{\thsec}{\thsection}

\renewcommand{\thseceq}{\thsectioneq}

\begin{center}
{\LARGE 
\vartitle%
}

by

Jean-Marie Dufour and Tianyu He

McGill University and Tianijn University

\today%

\quad

{\LARGE Technical appendix}
\end{center}

This appendix includes proofs of propositions and additional simulation
results.


\section{Proofs}



\subsection{Intersection methods in section \protect\ref{sec:DS}}


\cite{Dufour2024d} have proved the results for finite samples. We make
slight modifications to adapt to the asymptotic scenario. The following
proof works for both Proposition \ref{prop:AsymIM} and \ref{prop:BootIM}.

\begin{proof}[Proof of Proposition \protect\ref{prop:AsymIM}]
Take the asymptotic CI as an example, as the bootstrap CI can be similarly
derived. Under Assumption \ref{assump:DS}, we have the asymptotic CI for $%
\Delta \theta $ in \eqref{eq:CI-dIndex-Asym}. Then, we have 
\begin{align}
& \Pr \left( L_{n_{1}}-U_{n_{2}}\leq \Delta \theta \leq
U_{n_{1}}-L_{n_{2}}\right) \geq \Pr \left( \left\{ L_{n_{1}}\leq \theta
_{1}\leq U_{n_{1}}\right\} \bigcap \left\{ L_{n_{2}}\leq \theta _{2}\leq
U_{n_{2}}\right\} \right) \\
& \geq 1-\sum_{k=1}^{2}\Pr \left( \theta _{k}\notin \lbrack
L_{n_{k}},U_{n_{k}}]\right) \geq 1-(\alpha _{1}+\alpha _{2})\geq 1-\alpha
\end{align}%
where the second inequality is due to the Bonferroni inequality. When two
samples are independent, the Bonferroni inequality can be tight. As the
limiting confidence interval of $[L_{n_{k}},U_{n_{k}}]$ is $(1-\alpha _{k})$%
, we have the following result: 
\begin{equation}
\begin{split}
& \Pr \left( \left\{ L_{1}\leq \theta _{1}\leq U_{1}\right\} \bigcap \left\{
L_{2}\leq \theta _{2}\leq U_{2}\right\} \right) \\
=& \Pr \left( L_{1}\leq \theta _{1}\leq U_{1}\right) \Pr \left( L_{2}\leq
\theta _{2}\leq U_{2}\right) =(1-\alpha _{1})(1-\alpha _{2}) \\
=& 1-(\alpha _{1}+\alpha _{2})+\alpha _{1}\alpha _{2}=1-\alpha .
\end{split}%
\end{equation}
\end{proof}


\subsection{Asymptotic distribution of indices difference in section \protect
\ref{sec:OS-Asym}}


\begin{proof}[Proof of Proposition \protect\ref{prop:AsyN-dIndex}]
Under the Assumption \ref{assump:OS}, we have 
\begin{equation}
\begin{split}
\sqrt{N}(\hat{\theta }_{1}-\theta _{1})& =\sqrt{\frac{n_{2}}{n_{1}+n_{2}}}%
\sqrt{n_{1}}\sum_{i=1}^{n_{1}}\psi (X_{1i};\theta ,F_{1})+o_{p}(1)\overset{d}%
{\longrightarrow }\mathrm{N}\left( 0,\,(1-\eta _{1})\mathrm{Var}[\psi
(X_{1};\theta ,F_{1})]\right) \\
\sqrt{N}(\hat{\theta }_{2}-\theta _{2})& =\sqrt{\frac{n_{1}}{n_{1}+n_{2}}}%
\sqrt{n_{2}}\sum_{i=1}^{n_{2}}\psi (X_{2i};\theta ,F_{2})+o_{p}(1)\overset{d}%
{\longrightarrow }\mathrm{N}\left( 0,\,\eta _{1}\mathrm{Var}[\psi
(X_{2};\theta ,F_{2})]\right)
\end{split}%
\end{equation}%
Under the additional Assumption \ref{assump:JointNormal}, we have: 
\begin{equation}
\begin{split}
\sqrt{N}(\hat{\theta }_{1}-\theta _{1})& =\sqrt{\frac{n_{2}}{n_{1}+n_{2}}}%
\sqrt{\frac{m}{n_{1}}}\frac{1}{\sqrt{m}}\left[ \sum_{i=1}^{m}\psi
(X_{1i};\theta ,F_{1})+\sum_{i=m+1}^{n_{1}}\psi (X_{1i};\theta ,F_{1})\right]
+o_{p}(1)\,, \\
\sqrt{N}(\hat{\theta }_{2}-\theta _{2})& =\sqrt{\frac{n_{1}}{n_{1}+n_{2}}}%
\sqrt{\frac{m}{n_{2}}}\frac{1}{\sqrt{m}}\left[ \sum_{i=1}^{m}\psi
(X_{2i};\theta ,F_{2})+\sum_{i=m+1}^{n_{2}}\psi (X_{2i};\theta ,F_{2})\right]
+o_{p}(1)\,.
\end{split}%
\end{equation}%
By Assumption \ref{assump:OS}, only matched pairs contribute to the
covariance term. So we have asymptotic covariance $\sqrt{(1-\eta _{1})\eta
_{1}\lambda _{1}\lambda _{2}}\mathrm{Cov}(\psi (X_{1i};\theta ,F_{1}),\psi
(X_{2i};\theta ,F_{2}))$. Therefore, 
\begin{equation}
\begin{split}
& \sqrt{N}\left( 
\begin{bmatrix}
\hat{\theta }_{1} \\ 
\hat{\theta }_{2}%
\end{bmatrix}%
-%
\begin{bmatrix}
\theta _{1} \\ 
\theta _{2}%
\end{bmatrix}%
\right) =%
\begin{bmatrix}
\sqrt{\frac{n_{2}}{n_{1}+n_{2}}}\sqrt{\frac{m}{n_{1}}}\frac{1}{\sqrt{m}}%
\sum_{i=1}^{n_{1}}\psi (X_{1i};\theta ,F_{1}) \\ 
\sqrt{\frac{n_{1}}{n_{1}+n_{2}}}\sqrt{\frac{m}{n_{2}}}\frac{1}{\sqrt{m}}%
\sum_{i=1}^{n_{2}}\psi (X_{2i};\theta ,F_{2})%
\end{bmatrix}%
+%
\begin{bmatrix}
o_{p}(1) \\ 
o_{p}(1)%
\end{bmatrix}
\\
& \overset{d}{\longrightarrow }\mathrm{N}\left( 
\begin{bmatrix}
0 \\ 
0%
\end{bmatrix}%
,%
\begin{bmatrix}
(1-\eta _{1})\sigma _{1}^{2} & \sqrt{(1-\eta _{1})\eta _{1}\lambda
_{1}\lambda _{2}}\sigma _{12} \\ 
\sqrt{(1-\eta _{1})\eta _{1}\lambda _{1}\lambda _{2}}\sigma _{21} & \eta
_{1}\sigma _{2}^{2}%
\end{bmatrix}%
\right)
\end{split}%
\end{equation}%
where 
\begin{equation}
\sigma _{k}^{2}=\mathrm{Var}[\psi (X_{ki};\theta ,F_{k})],\quad \sigma
_{12}=\sigma _{21}=\mathrm{Cov}[\psi (X_{1i};\theta ,F_{1}),\psi
(X_{2i};\theta ,F_{2})]\,.
\end{equation}%
The desired result then follows by continuous mapping theorem.
\end{proof}


\subsection{Asymptotic variance estimation in section \protect\ref%
{sec:OS-AVar}}


The consistency proof often relies on specific conditions for given indices.
Nevertheless, we can gain insight into how to prove consistency based on
techniques presented in \cite{Barrett2009}, particularly Lemma 5 in their
appendix. We can also extend those results to an estimator for the
covariance between $\hat{\psi_{1}}$ and $\hat{\psi_{2}}$.

\begin{lemma}
\label{lmm:AVar-Est} 
\captionlemma{\lemmaname}{Consistent estimator for asymptotic variance}
Let $\hat{\psi }_{ki}$, $k=1,2$ be an estimate for $\psi _{ki}$ evaluated at 
$X_{ki}$ with $\psi _{k}$ in \eqref{eq:ALG} such that 
\begin{equation}
\left\vert \hat{\psi }_{ki}-\psi _{ki}\right\vert \leq \alpha _{k}\hat{A}%
_{k}+\rho _{ki}\hat{B}_{k}+\beta _{k}\hat{P}_{ki}+\omega _{ki}\hat{W}_{ki}.
\end{equation}%
Assume that the following conditions hold: 
\begin{enumerate}[$(1)$]
\item
$|\alpha _{k}|=O_{p}(1)$ and $\hat{A}_{k}=o_{p}(1)$ ,

\item
$\left|\frac{1}{n_{k}}\sum_{i=1}^{n_{k}} \rho^{2}_{ki}\right| = O_{p}(1)$
and $\hat{B}_{k} = o_{p}(1)$ ,

\item
$|\beta_{k}| = O_{p}(1)$ and $\frac{1}{n_{k}}\sum_{i=1}^{n_{k}} \hat{P}%
^{2}_{ki} = o_{p}(1)$ ,

\item
$|\frac{1}{n_{k}}\sum_{i=1}^{n_{k}} \omega^{4}_{ki}| = O_{p}(1)$ and $\frac{1%
}{n_{k}}\sum_{i=1}^{n_{k}} \hat{W}^{2}_{ki} = o_{p}(1)$ ,

\item
$\frac{1}{n_{k}} \sum_{i=1}^{n_{k}} \psi_{ki}^{2} = O_{p}(1)$ .

\end{enumerate}%

\begin{enumerate}[$(1)$]
\item
(\cite[Lemma $5$]{Barrett2009}) We then have 
\begin{equation}
\frac{1}{n_{k}}\sum_{i=1}^{n_{k}}\hat{\psi }_{ki}^{2}-\frac{1}{n_{k}}%
\sum_{i=1}^{n_{k}}\psi _{ki}^{2}=o_{p}(1).
\label{eq:AVar-Index-Est-Consistency}
\end{equation}%
\item
For matched pairs $\{(X_{1i},X_{2i})\}_{i=1}^{m}$, we have 
\begin{equation}
\frac{1}{m}\sum_{i=1}^{m}\hat{\psi }_{1i}\hat{\psi }_{2i}-\frac{1}{m}%
\sum_{i=1}^{m}\psi _{1i}\psi _{2i}=o_{p}(1).  \label{eq:ACov-Est-Consistency}
\end{equation}%
\end{enumerate}%
\end{lemma}

\begin{proofflexc}
\captionproofflexc{\lemmaname}{lmm:AVar-Est}
\cite[Lemma $5$]{Barrett2009} establish \eqref{eq:AVar-Index-Est-Consistency}%
. We include this result here for completeness. To prove %
\eqref{eq:ACov-Est-Consistency}, note that 
\begin{equation}
\begin{split}
\left\vert \frac{1}{m}\sum_{i=1}^{m}\hat{\psi }_{1i}\hat{\psi }_{2i}-\frac{1%
}{m}\sum_{i=1}^{m}\psi _{1i}\psi _{2i}\right\vert =& \Bigg|\frac{1}{m}%
\sum_{i=1}^{m}(\hat{\psi }_{1i}-\psi _{1i})(\hat{\psi }_{2i}-\psi _{2i}) \\
& +\frac{1}{m}\sum_{i=1}^{m}(\hat{\psi }_{1i}-\psi _{1i})\psi _{2i}+\frac{1}{%
m}\sum_{i=1}^{m}\psi _{1i}(\hat{\psi }_{2i}-\psi _{2i})\Bigg| \\
\leq & \left[ \frac{1}{m}\sum_{i=1}^{m}(\hat{\psi }_{1i}-\psi _{1i})^{2}%
\right] ^{1/2}\left[ \frac{1}{m}\sum_{i=1}^{m}(\hat{\psi }_{2i}-\psi
_{2i})^{2}\right] ^{1/2} \\
& +\left[ \frac{1}{m}\sum_{i=1}^{m}(\hat{\psi }_{1i}-\psi _{1i})^{2}\right]
^{1/2}\left( \frac{1}{m}\sum_{i=1}^{m}\psi _{2i}^{2}\right) ^{1/2} \\
& +\left( \frac{1}{m}\sum_{i=1}^{m}\psi _{1i}^{2}\right) ^{1/2}\left[ \frac{1%
}{m}\sum_{i=1}^{m}(\hat{\psi }_{2i}-\psi _{2i})^{2}\right] ^{1/2} \\
=& o_{p}(1),
\end{split}%
\end{equation}%
where the inequality is due to the Cauchy-Schwarz inequality and the last
equality is implied by conditions $(i)$ - $(v)$.
\end{proofflexc}


\subsection{Bootstrap inference in section \protect\ref{sec:OS-Boot}}


\begin{proofflexc}
\captionproofflexc{\lemmaname}{prop:Boot-dIndex}
By Assumption \ref{assump:OS} and the bootstrap algorithm, one can establish
the following asymptotic linear representation conditional on the original
data $\{X_{1i}\}_{i=1}^{n_{1}}$ and $\{X_{2i}\}_{i=1}^{n_{2}}$; see, for
example, \cite{Shao1995}. 
\begin{equation}
\sqrt{n_{k}}(\hat{\theta }_{k}^{\ast }-\hat{\theta }_{k})=\frac{1}{\sqrt{%
n_{k}}}\sum_{i=1}^{n_{k}}\psi (X_{ki}^{\ast };\theta ,\hat{F}_{k})+o_{p}(1)
\label{eq:Boot-AsymLinear}
\end{equation}%
where $\hat{\theta }_{k}^{\ast }$ denotes the estimate based on a bootstrap
sample $\{X_{ki}^{\ast }\}_{i=1}^{n_{k}}$ and $\hat{F}_{k}$ is the EDF for $%
F_{k}$. Conditional on the original data, we have, under additional
Assumption \ref{assump:JointNormal}, 
\begin{align}
\mathbb{E}\left[ \mathrm{IF}(X_{ki}^{\ast };\theta ,\hat{F}_{k})\mid P_{n}%
\right] & =\psi (X_{ki};\theta ,\hat{F}_{k})\overset{a.s.}{\longrightarrow }%
\mathbb{E}\left[ \psi (X_{ki};\theta ,F_{k})\right] =0\,, \\
\mathbb{E}\left[ \psi ^{2}(X_{ki}^{\ast };\theta ,\hat{F}_{k})\mid P_{n}%
\right] & =\frac{1}{n_{k}}\sum_{i=1}^{n_{k}}\psi ^{2}(X_{ki};\theta ,\hat{F}%
_{k})\overset{a.s.}{\longrightarrow }\mathbb{E}\left[ \psi
^{2}(X_{ki};\theta ,F_{k})\right] <\infty \,.
\end{align}%
Then by the central limit theorem, we have 
\begin{align}
\frac{1}{\sqrt{n_{k}}}\sum_{i=1}^{n_{k}}\psi (X_{ki}^{\ast };\theta ,\hat{F}%
_{k})& \overset{d}{\longrightarrow }\mathrm{N}(0,\Omega _{k}),\quad k=1,2,
\label{eq:Boot-AsymNormal} \\
\frac{1}{\sqrt{m}}\sum_{i=1}^{m}%
\begin{bmatrix}
\psi (X_{1i}^{\ast };\theta ,\hat{F}_{1}) \\ 
\psi (X_{2i}^{\ast };\theta ,\hat{F}_{2})%
\end{bmatrix}%
& \overset{d}{\longrightarrow }\mathrm{N}\left( 
\begin{bmatrix}
0 \\ 
0%
\end{bmatrix}%
,%
\begin{bmatrix}
\sigma _{1}^{2} & \sigma _{12} \\ 
\sigma _{21} & \sigma _{2}^{2}%
\end{bmatrix}%
\right) ,
\end{align}%
where $\sigma _{k}^{2}=\mathrm{Var}[\psi (X_{ki};\theta ,F_{k})]$ and $%
\sigma _{12}=\sigma _{21}=\mathrm{Cov}[\psi (X_{1i};\theta ,F_{1}),\psi
(X_{2i};\theta ,F_{2})]$. Similar to the proof of Proposition \ref%
{prop:AsyN-dIndex}, we have 
\begin{equation}
\sqrt{N}\left( 
\begin{bmatrix}
\hat{\theta }_{1}^{\ast } \\ 
\hat{\theta }_{2}^{\ast }%
\end{bmatrix}%
-%
\begin{bmatrix}
\hat{\theta }_{1} \\ 
\hat{\theta }_{2}%
\end{bmatrix}%
\right) \overset{d}{\longrightarrow }\mathrm{N}\left( 
\begin{bmatrix}
0 \\ 
0%
\end{bmatrix}%
,%
\begin{bmatrix}
(1-\eta _{1})\sigma _{1}^{2} & \sqrt{(1-\eta _{1})\eta _{1}\lambda
_{1}\lambda _{2}}\sigma _{12} \\ 
\sqrt{(1-\eta _{1})\eta _{1}\lambda _{1}\lambda _{2}}\sigma _{21} & \eta
_{1}\sigma _{2}^{2}%
\end{bmatrix}%
\right) \,.
\end{equation}%
By continuous mapping theorem, we show the validity of the bootstrap method.
\end{proofflexc}

\FloatBarrier

\subsection{Proof in section \protect\ref{sec:App-SHIW}}


\begin{proof}[Proof of Corollary \protect\ref{coro:AsyN-dExtIneq}]
\cite{Dufour2024} shows that the estimates for extended LCs and Gini indices
satisfy Definition \ref{def:ALG} with the influence function given in the
main text. Under the overlapping samples (\emph{i.e.}, Assumption \ref%
{assump:OS}), we can show that the Assumption \ref{assump:JointNormal} is
satisfied by the standard multivariate central limit theorem. We then can
apply Proposition \ref{prop:AsyN-dIndex} to conclude the results.
\end{proof}


\section{Additional simulation results}


This section provides a numerical exercise on the performance of the
proposed methods when the tail is extremely heavy, such that the variance
does not even exist. Under such a scenario, it is challenging to obtain
valid asymptotic or bootstrap CIs for $\theta_{k}$, $k = 1, 2$. So, we
expect asymptotic approaches to fail, including the intersection methods.
The DGP is as follows.

\medskip \textbf{DGP I-D}: $F_{1} = \text{SM}(1, 1.6971, 8.3679)$, $F_{2} = 
\text{SM}(0.4, 1.1, 1.1)$, $\rho \in \{0, \pm 0.5, \pm0.99\}$ in %
\eqref{eq:DGP-GaussCopula}, stability index of $F_{2}$ is $\beta_{2} = 1.21$%
, overlap portion $\lambda \in \{0.1, 0.5, 0.9\}$. \medskip

The simulation results are provided in Table \ref{tab:CI-dI-DGP4} for Gini
index difference, Table \ref{tab:CI-dmu-DGP4} for mean difference, and Table %
\ref{tab:CI-dLC-DGP4} for LC ordinate difference at $p = 0.5$.

\begin{table}[tb]%

\caption{DGP I-D: coverage and width of confidence intervals for Gini indices
differences.} \label{tab:CI-dI-DGP4}

\begin{center}
Table \thetable

DGP I-D: Coverage and width of confidence intervals for Gini indices
differences
\end{center}

\begin{adjustbox}{scale={0.9}{0.9}}%

\hspace{-0.5\totalhormargin} \begin{minipage}{\paperwidth}
\centering%

\small%

\begin{center}
\begin{tabular}{ccc|cccc|cccc}
\hline
\rule[-1ex]{0cm}{4ex} $\rho$ & $\lambda$ & $n$ & \multicolumn{4}{c|}{
Coverage $(\%)$} & \multicolumn{4}{c}{Width} \\ \cline{4-11}
\rule[-1ex]{0cm}{4ex} &  &  & IM-Asym & IM-Boot & OS-Asym & OS-Boot & IM-Asym
& IM-Boot & OS-Asym & OS-Boot \\ \hline
\rule[-1ex]{0cm}{4ex} -0.99 & 0.1 & 100 & 65.1 & 80.9 & 46.7 & 61.8 & 0.2972
& 0.5221 & 0.1953 & 0.3099 \\ 
&  & 200 & 67.4 & 82.2 & 47.1 & 63.4 & 0.2547 & 0.4648 & 0.1726 & 0.2875 \\ 
&  & 500 & 67.7 & 82.2 & 50.6 & 68.3 & 0.2054 & 0.3622 & 0.1453 & 0.2471 \\ 
& 0.5 & 100 & 65.9 & 79.3 & 42 & 58.9 & 0.2952 & 0.5143 & 0.1889 & 0.2985 \\ 
&  & 200 & 67.8 & 82.5 & 47.3 & 62.9 & 0.2533 & 0.4495 & 0.1671 & 0.2761 \\ 
&  & 500 & 67.5 & 81.2 & 48.9 & 66.3 & 0.2034 & 0.3483 & 0.1408 & 0.2385 \\ 
& 0.9 & 100 & 68 & 80.4 & 43.9 & 58.5 & 0.2964 & 0.5192 & 0.1818 & 0.2925 \\ 
&  & 200 & 70 & 84.4 & 46.9 & 64.9 & 0.2563 & 0.4642 & 0.1639 & 0.2852 \\ 
&  & 500 & 68.7 & 82.7 & 51.7 & 68.4 & 0.2085 & 0.3699 & 0.1423 & 0.254 \\ 
\rule[-1ex]{0cm}{4ex} -0.5 & 0.1 & 100 & 67.7 & 79.1 & 44.8 & 62.1 & 0.295 & 
0.5156 & 0.1932 & 0.3081 \\ 
&  & 200 & 70.4 & 82.4 & 50.5 & 66.9 & 0.2569 & 0.4737 & 0.1743 & 0.2923 \\ 
&  & 500 & 66.1 & 80.8 & 50.4 & 65.4 & 0.2039 & 0.3597 & 0.1442 & 0.245 \\ 
& 0.5 & 100 & 69.1 & 80.5 & 47.7 & 63.4 & 0.2955 & 0.5166 & 0.1918 & 0.3062
\\ 
&  & 200 & 68.1 & 83 & 48.1 & 63.7 & 0.2526 & 0.4638 & 0.1689 & 0.284 \\ 
&  & 500 & 68.4 & 82.7 & 50.9 & 67.5 & 0.2054 & 0.3639 & 0.1442 & 0.2477 \\ 
& 0.9 & 100 & 69.1 & 82.3 & 45.7 & 61.4 & 0.2992 & 0.5295 & 0.1911 & 0.3054
\\ 
&  & 200 & 70.9 & 84.9 & 48.6 & 66.7 & 0.2553 & 0.4549 & 0.1685 & 0.2776 \\ 
&  & 500 & 67.6 & 83.8 & 50.4 & 67.5 & 0.2059 & 0.362 & 0.1429 & 0.246 \\ 
\rule[-1ex]{0cm}{4ex} 0 & 0.1 & 100 & 66.6 & 80.1 & 45.4 & 62.1 & 0.2974 & 
0.5167 & 0.1946 & 0.3107 \\ 
&  & 200 & 69.1 & 83.9 & 48.8 & 66.4 & 0.254 & 0.4709 & 0.1716 & 0.2902 \\ 
&  & 500 & 68.5 & 82.4 & 53.1 & 68.8 & 0.2061 & 0.3653 & 0.1459 & 0.2482 \\ 
& 0.5 & 100 & 70.9 & 82.7 & 48 & 63.6 & 0.2986 & 0.5298 & 0.195 & 0.3174 \\ 
&  & 200 & 68.4 & 82 & 48.8 & 66.2 & 0.2521 & 0.4506 & 0.17 & 0.2804 \\ 
&  & 500 & 66.4 & 84 & 50.6 & 67.1 & 0.206 & 0.3635 & 0.1458 & 0.2496 \\ 
& 0.9 & 100 & 70 & 83.5 & 46.9 & 63.5 & 0.2981 & 0.5232 & 0.1948 & 0.3149 \\ 
&  & 200 & 69 & 84.5 & 50.5 & 66.6 & 0.2591 & 0.4833 & 0.1759 & 0.3029 \\ 
&  & 500 & 66 & 83 & 51.3 & 65.8 & 0.2068 & 0.3611 & 0.1464 & 0.2498 \\ 
\rule[-1ex]{0cm}{4ex} 0.5 & 0.1 & 100 & 66.6 & 79.8 & 44.3 & 59.6 & 0.296 & 
0.5193 & 0.1918 & 0.3101 \\ 
&  & 200 & 67.8 & 82 & 48 & 65.4 & 0.2535 & 0.4589 & 0.17 & 0.2852 \\ 
&  & 500 & 65.8 & 80.1 & 47.5 & 65 & 0.2037 & 0.3545 & 0.1431 & 0.244 \\ 
& 0.5 & 100 & 67.9 & 80.5 & 45 & 60.6 & 0.2969 & 0.5211 & 0.1869 & 0.3094 \\ 
&  & 200 & 68.8 & 82.9 & 46.8 & 66.1 & 0.2537 & 0.4524 & 0.1666 & 0.281 \\ 
&  & 500 & 67.1 & 82.6 & 49.2 & 65.2 & 0.2049 & 0.3549 & 0.1423 & 0.2453 \\ 
& 0.9 & 100 & 68 & 81.4 & 43.7 & 61 & 0.2954 & 0.5144 & 0.1817 & 0.3088 \\ 
&  & 200 & 68.3 & 83.1 & 46.2 & 64.1 & 0.255 & 0.4666 & 0.1645 & 0.2939 \\ 
&  & 500 & 66.8 & 83.9 & 49.3 & 66.7 & 0.2044 & 0.3611 & 0.1401 & 0.2446 \\ 
\rule[-1ex]{0cm}{4ex} 0.99 & 0.1 & 100 & 66.8 & 80.7 & 44.6 & 61.3 & 0.2962
& 0.517 & 0.188 & 0.3077 \\ 
&  & 200 & 68.3 & 82.2 & 46.2 & 64.5 & 0.2542 & 0.4629 & 0.1676 & 0.2844 \\ 
&  & 500 & 68.8 & 83.1 & 49.8 & 69.6 & 0.2046 & 0.3576 & 0.1419 & 0.2442 \\ 
& 0.5 & 100 & 69.2 & 81.1 & 40 & 60.2 & 0.2978 & 0.5268 & 0.1658 & 0.2964 \\ 
&  & 200 & 66.9 & 82 & 44.2 & 61.9 & 0.2544 & 0.4666 & 0.1511 & 0.283 \\ 
&  & 500 & 65.5 & 81.6 & 46.7 & 63.9 & 0.207 & 0.3707 & 0.1344 & 0.2553 \\ 
& 0.9 & 100 & 69.4 & 83.1 & 34.8 & 57.9 & 0.2998 & 0.529 & 0.1405 & 0.3089
\\ 
&  & 200 & 65.6 & 82 & 36 & 58.7 & 0.2502 & 0.4427 & 0.1309 & 0.2833 \\ 
&  & 500 & 69.6 & 85.1 & 46.7 & 69.2 & 0.2074 & 0.368 & 0.1262 & 0.2686 \\ 
\hline
\end{tabular}
\end{center}

\end{minipage}%

\end{adjustbox}%

\medskip 
\noindent
\footnotesize%
Note -- The table provides coverage rates and widths of confidence intervals
of level $95\%$ for Gini indices differences by asymptotic intersection
method (IM-Asym), bootstrap intersection method (IM-Boot), asymptotic method
with overlapping samples (OS-Asym) and bootstrap method with overlapping
samples (OS-Boot). The data are generated according to DGP I-D for given
Gaussian copula $\rho$ in \eqref{eq:DGP-GaussCopula}, overlap portion $%
\lambda$ and sample size $n$. The numbers are based on $1000$ replications
with 399 bootstrap repetitions each.

\normalsize%

\end{table}%

\begin{table}[tb]%

\caption{DGP I-D: coverage and width of confidence intervals for the mean
difference.} \label{tab:CI-dmu-DGP4}

\begin{center}
Table \thetable

DGP I-D: Coverage and width of confidence intervals for the mean difference
\end{center}

\begin{adjustbox}{scale={0.9}{0.9}}%

\hspace{-0.5\totalhormargin} \begin{minipage}{\paperwidth}
\centering%

\small%

\begin{center}
\begin{tabular}{ccc|cccc|cccc}
\hline
\rule[-1ex]{0cm}{4ex} $\rho$ & $\lambda$ & $n$ & \multicolumn{4}{c|}{
Coverage $(\%)$} & \multicolumn{4}{c}{Width} \\ \cline{4-11}
\rule[-1ex]{0cm}{4ex} &  &  & IM-Asym & IM-Boot & OS-Asym & OS-Boot & IM-Asym
& IM-Boot & OS-Asym & OS-Boot \\ \hline
\rule[-1ex]{0cm}{4ex} -0.99 & 0.1 & 100 & 50.3 & 71.2 & 45.7 & 65.1 & 
104.6436 & 6.06E+06 & 91.4418 & 1.98E+06 \\ 
&  & 200 & 50.8 & 70.1 & 45.3 & 64.8 & 3.1009 & 49.4768 & 2.6664 & 42.7888
\\ 
&  & 500 & 52.5 & 71 & 47 & 64.4 & 2.4276 & 19.6046 & 2.0936 & 16.8995 \\ 
& 0.5 & 100 & 46.9 & 70.2 & 41.4 & 61.3 & 3.3023 & 65.7548 & 2.8374 & 53.9223
\\ 
&  & 200 & 49.2 & 70 & 44.4 & 63.3 & 2.4144 & 15.0058 & 2.0734 & 11.6824 \\ 
&  & 500 & 50.7 & 70.1 & 45.2 & 64.4 & 2.9188 & 136.0929 & 2.5268 & 109.0324
\\ 
& 0.9 & 100 & 49.2 & 70 & 44.4 & 62 & 3.7567 & 86.6199 & 3.244 & 71.0627 \\ 
&  & 200 & 52.8 & 72.6 & 48 & 65.7 & 4.3442 & 443.3181 & 3.7662 & 359.0844
\\ 
&  & 500 & 54.2 & 72.6 & 49.4 & 68.1 & 2.6242 & 29.9872 & 2.2718 & 20.6227
\\ 
\rule[-1ex]{0cm}{4ex} -0.5 & 0.1 & 100 & 47.3 & 72 & 43 & 62.7 & 3.7911 & 
73.9171 & 3.2506 & 59.166 \\ 
&  & 200 & 52.1 & 71.9 & 48.2 & 65.1 & 3.3213 & 113.5945 & 2.8577 & 101.6268
\\ 
&  & 500 & 50.3 & 69.5 & 44.8 & 62.8 & 2.9738 & 81.4345 & 2.5705 & 73.8106
\\ 
& 0.5 & 100 & 50.5 & 71.9 & 46.1 & 66.3 & 4.6656 & 222.3083 & 4.022 & 
202.3445 \\ 
&  & 200 & 52.9 & 73.2 & 48.1 & 65.9 & 3.5026 & 54.1624 & 3.0205 & 47.2103
\\ 
&  & 500 & 52.9 & 72.1 & 47.4 & 65.8 & 3.3481 & 123.9841 & 2.9002 & 112.9123
\\ 
& 0.9 & 100 & 51.3 & 71.5 & 46.1 & 64.2 & 3.2174 & 37.1448 & 2.7618 & 29.1254
\\ 
&  & 200 & 51.9 & 72.8 & 47.2 & 65.3 & 2.7184 & 26.8982 & 2.3384 & 22.0257
\\ 
&  & 500 & 53.6 & 71.8 & 48.6 & 65.5 & 2.355 & 14.8571 & 2.0334 & 12.0181 \\ 
\rule[-1ex]{0cm}{4ex} 0 & 0.1 & 100 & 48 & 70.8 & 42.8 & 63.3 & 3.3135 & 
74.8295 & 2.8305 & 65.924 \\ 
&  & 200 & 53.7 & 74.8 & 48.8 & 68.4 & 3.1065 & 27.9355 & 2.6682 & 24.698 \\ 
&  & 500 & 52.1 & 72.1 & 47.8 & 67.1 & 2.491 & 21.7902 & 2.1476 & 17.7954 \\ 
& 0.5 & 100 & 51.7 & 73.7 & 46.8 & 66.8 & 3.6396 & 68.2653 & 3.1151 & 58.158
\\ 
&  & 200 & 50.4 & 69.7 & 46.4 & 63.9 & 3.1844 & 91.6626 & 2.7366 & 83.719 \\ 
&  & 500 & 51.3 & 72.1 & 47.3 & 63.2 & 3.0778 & 150.742 & 2.6606 & 134.0302
\\ 
& 0.9 & 100 & 50.6 & 73.7 & 44.1 & 65.9 & 4.1431 & 163.2247 & 3.5553 & 
139.7482 \\ 
&  & 200 & 51.6 & 73.5 & 46.8 & 67.2 & 3.4441 & 70.3926 & 2.9636 & 55.5171
\\ 
&  & 500 & 51.3 & 70.3 & 46.7 & 64.2 & 2.3801 & 21.565 & 2.0506 & 18.6042 \\ 
\rule[-1ex]{0cm}{4ex} 0.5 & 0.1 & 100 & 49.8 & 71.7 & 45.9 & 64.5 & 4.4097 & 
350.4052 & 3.7857 & 304.7753 \\ 
&  & 200 & 51.3 & 72.5 & 45.2 & 65.4 & 3.7808 & 335.106 & 3.2559 & 293.4675
\\ 
&  & 500 & 49.3 & 68.5 & 44.8 & 63.2 & 2.4709 & 24.4397 & 2.1288 & 20.8386
\\ 
& 0.5 & 100 & 49 & 70.9 & 43.3 & 63.3 & 4.2868 & 255.1002 & 3.6686 & 219.9049
\\ 
&  & 200 & 50.9 & 71.2 & 45.8 & 64.6 & 3.1769 & 76.8024 & 2.722 & 68.8833 \\ 
&  & 500 & 50.2 & 70.3 & 45.2 & 62.8 & 3.971 & 1282.26 & 3.4374 & 367.0488
\\ 
& 0.9 & 100 & 47.6 & 71.3 & 42.7 & 64.6 & 4.7578 & 226.987 & 4.0723 & 
211.8639 \\ 
&  & 200 & 49.8 & 71.5 & 44.4 & 64.1 & 3.6566 & 101.4453 & 3.1361 & 84.5179
\\ 
&  & 500 & 51.8 & 70.1 & 44.5 & 64.7 & 2.6412 & 27.3444 & 2.2719 & 22.6686
\\ 
\rule[-1ex]{0cm}{4ex} 0.99 & 0.1 & 100 & 50.2 & 70.4 & 45.7 & 63.4 & 4.62 & 
161.5719 & 3.9662 & 142.5533 \\ 
&  & 200 & 48.7 & 70.4 & 43 & 63.9 & 3.8731 & 404.2647 & 3.3346 & 404.5954
\\ 
&  & 500 & 52.3 & 72.6 & 48.3 & 66.2 & 19.6546 & 127460.4 & 17.1535 & 
116212.1 \\ 
& 0.5 & 100 & 52.2 & 72.3 & 45.9 & 66 & 5.1184 & 1230.286 & 4.3802 & 1177.597
\\ 
&  & 200 & 50.7 & 71 & 45.1 & 65.2 & 3.4308 & 50.8803 & 2.9338 & 45.1805 \\ 
&  & 500 & 51.8 & 69.9 & 45.9 & 64 & 2.4524 & 18.3954 & 2.1038 & 16.5795 \\ 
& 0.9 & 100 & 48.3 & 72.9 & 43.9 & 65.4 & 3.7557 & 141.8391 & 3.1699 & 
144.3589 \\ 
&  & 200 & 49.2 & 68.2 & 43.3 & 62.5 & 3.4507 & 96.5791 & 2.9393 & 95.9893
\\ 
&  & 500 & 52.7 & 73.4 & 47.9 & 67.4 & 2.9023 & 67.7659 & 2.4909 & 64.3352
\\ \hline
\end{tabular}
\end{center}

\end{minipage}%

\end{adjustbox}%

\medskip 
\noindent
\footnotesize%
Note -- The table provides coverage rates and widths of confidence intervals
of level $95\%$ for the mean difference by asymptotic intersection method
(IM-Asym), bootstrap intersection method (IM-Boot), asymptotic method with
overlapping samples (OS-Asym) and bootstrap method with overlapping samples
(OS-Boot). The data are generated according to DGP I-D for given Gaussian
copula $\rho$ in \eqref{eq:DGP-GaussCopula}, overlap portion $\lambda$ and
sample size $n$. The numbers are based on $1000$ replications with 399
bootstrap repetitions each.

\normalsize%

\end{table}%

\begin{table}[tb]%

\caption{DGP I-D: coverage and width of confidence intervals for the LC
ordinate difference at $p = 0.5$.} \label{tab:CI-dLC-DGP4}

\begin{center}
Table \thetable

DGP I-D: Coverage and width of confidence intervals for the LC ordinate
difference at $p=0.5$
\end{center}

\begin{adjustbox}{scale={0.9}{0.9}}%

\hspace{-0.5\totalhormargin} \begin{minipage}{\paperwidth}
\centering%

\small%

\begin{center}
\begin{tabular}{ccc|cccc|cccc}
\hline
\rule[-1ex]{0cm}{4ex} $\rho$ & $\lambda$ & $n$ & \multicolumn{4}{c|}{
Coverage $(\%)$} & \multicolumn{4}{c}{Width} \\ \cline{4-11}
\rule[-1ex]{0cm}{4ex} &  &  & IM-Asym & IM-Boot & OS-Asym & OS-Boot & IM-Asym
& IM-Boot & OS-Asym & OS-Boot \\ \hline
\rule[-1ex]{0cm}{4ex} -0.99 & 0.1 & 100 & 97.7 & 98.4 & 76.7 & 81.4 & 0.1459
& 0.1766 & 0.0899 & 0.1099 \\ 
&  & 200 & 95.5 & 97 & 72.7 & 76.7 & 0.1103 & 0.1292 & 0.0678 & 0.0826 \\ 
&  & 500 & 92.3 & 95 & 69.3 & 74.4 & 0.0777 & 0.0915 & 0.0482 & 0.0603 \\ 
& 0.5 & 100 & 97.4 & 97.5 & 71 & 76.7 & 0.1462 & 0.1769 & 0.0803 & 0.1012 \\ 
&  & 200 & 96.3 & 97.9 & 68.6 & 74.4 & 0.1103 & 0.13 & 0.0612 & 0.0762 \\ 
&  & 500 & 92.6 & 93.9 & 63 & 70 & 0.0773 & 0.0907 & 0.0442 & 0.0569 \\ 
& 0.9 & 100 & 99 & 99.4 & 66.8 & 73.4 & 0.146 & 0.1768 & 0.0697 & 0.0937 \\ 
&  & 200 & 97.3 & 97.7 & 64.8 & 71.9 & 0.1107 & 0.131 & 0.0549 & 0.0735 \\ 
&  & 500 & 93.8 & 95.3 & 60.8 & 70.4 & 0.0782 & 0.0926 & 0.0415 & 0.0577 \\ 
\rule[-1ex]{0cm}{4ex} -0.5 & 0.1 & 100 & 96.9 & 98.4 & 76.8 & 81.4 & 0.1459
& 0.1767 & 0.0915 & 0.112 \\ 
&  & 200 & 95.3 & 96.6 & 76.4 & 78.9 & 0.1107 & 0.1307 & 0.0692 & 0.0841 \\ 
&  & 500 & 91.3 & 93.2 & 68.4 & 72.6 & 0.0775 & 0.091 & 0.0489 & 0.061 \\ 
& 0.5 & 100 & 97.3 & 98.6 & 78.8 & 82.1 & 0.1454 & 0.1758 & 0.0887 & 0.1083
\\ 
&  & 200 & 95.2 & 96.6 & 73.5 & 77.5 & 0.1095 & 0.1291 & 0.0667 & 0.0813 \\ 
&  & 500 & 91.9 & 94.6 & 69.9 & 74.1 & 0.0776 & 0.0916 & 0.0478 & 0.0601 \\ 
& 0.9 & 100 & 96.8 & 98 & 78 & 81.1 & 0.1471 & 0.1775 & 0.0868 & 0.1072 \\ 
&  & 200 & 95.3 & 96.8 & 74 & 79.1 & 0.1103 & 0.1301 & 0.0651 & 0.0789 \\ 
&  & 500 & 92.6 & 94.5 & 66.1 & 72.7 & 0.0777 & 0.0915 & 0.0467 & 0.0589 \\ 
\rule[-1ex]{0cm}{4ex} 0 & 0.1 & 100 & 96.4 & 97.8 & 76.8 & 79.9 & 0.1466 & 
0.177 & 0.0919 & 0.1119 \\ 
&  & 200 & 94.3 & 96.4 & 75.4 & 79.6 & 0.1092 & 0.1289 & 0.0686 & 0.0835 \\ 
&  & 500 & 92.5 & 94.5 & 69.5 & 74.7 & 0.0777 & 0.0915 & 0.0492 & 0.0612 \\ 
& 0.5 & 100 & 97 & 98.1 & 78.6 & 83.1 & 0.1462 & 0.1771 & 0.0919 & 0.113 \\ 
&  & 200 & 94.6 & 96.4 & 73.9 & 77.2 & 0.1097 & 0.1289 & 0.0688 & 0.0824 \\ 
&  & 500 & 92.1 & 94.9 & 68.2 & 74.6 & 0.0776 & 0.0917 & 0.0492 & 0.0614 \\ 
& 0.9 & 100 & 96.6 & 98.2 & 79.9 & 83.9 & 0.1458 & 0.177 & 0.0916 & 0.1125
\\ 
&  & 200 & 94.8 & 96.4 & 76 & 80 & 0.1108 & 0.1312 & 0.0697 & 0.0856 \\ 
&  & 500 & 92.2 & 94.1 & 68.5 & 73.3 & 0.0781 & 0.0921 & 0.0494 & 0.0613 \\ 
\rule[-1ex]{0cm}{4ex} 0.5 & 0.1 & 100 & 96.9 & 97.8 & 79.7 & 83.9 & 0.1458 & 
0.1766 & 0.0908 & 0.111 \\ 
&  & 200 & 94.8 & 96 & 74.3 & 77.9 & 0.1101 & 0.13 & 0.0684 & 0.0828 \\ 
&  & 500 & 90 & 94.2 & 67.8 & 71.1 & 0.0775 & 0.0908 & 0.0486 & 0.0604 \\ 
& 0.5 & 100 & 97.2 & 98.3 & 76.8 & 81.8 & 0.1465 & 0.1777 & 0.0878 & 0.1085
\\ 
&  & 200 & 95.3 & 96.6 & 71.7 & 76.5 & 0.1101 & 0.1295 & 0.066 & 0.0803 \\ 
&  & 500 & 92.2 & 94.6 & 69.1 & 73.8 & 0.0777 & 0.0912 & 0.0474 & 0.0596 \\ 
& 0.9 & 100 & 97 & 98.9 & 75 & 79.8 & 0.1453 & 0.176 & 0.0842 & 0.1053 \\ 
&  & 200 & 95.8 & 96.6 & 71.1 & 75.6 & 0.1103 & 0.1301 & 0.0644 & 0.0803 \\ 
&  & 500 & 92.6 & 95.1 & 66.2 & 71.6 & 0.0774 & 0.0914 & 0.0461 & 0.0584 \\ 
\rule[-1ex]{0cm}{4ex} 0.99 & 0.1 & 100 & 96.1 & 98.1 & 78.6 & 82.1 & 0.1459
& 0.1766 & 0.0879 & 0.1086 \\ 
&  & 200 & 93.9 & 95 & 73.1 & 77.4 & 0.11 & 0.1295 & 0.0661 & 0.0811 \\ 
&  & 500 & 92.7 & 94.5 & 68.6 & 75.5 & 0.0773 & 0.0908 & 0.047 & 0.0593 \\ 
& 0.5 & 100 & 98.8 & 99.1 & 70 & 75.2 & 0.1462 & 0.177 & 0.0724 & 0.0937 \\ 
&  & 200 & 96.6 & 97.7 & 63.1 & 70 & 0.1102 & 0.1302 & 0.0555 & 0.0724 \\ 
&  & 500 & 94 & 95.9 & 59.7 & 67.4 & 0.0779 & 0.0918 & 0.041 & 0.0563 \\ 
& 0.9 & 100 & 99.8 & 99.9 & 53.3 & 65.6 & 0.1466 & 0.1777 & 0.0542 & 0.0821
\\ 
&  & 200 & 99.1 & 99.3 & 49.7 & 61.9 & 0.1098 & 0.1289 & 0.0432 & 0.0647 \\ 
&  & 500 & 96.3 & 97.5 & 53.4 & 66.9 & 0.0779 & 0.0919 & 0.0351 & 0.0549 \\ 
\hline
\end{tabular}
\end{center}

\end{minipage}%

\end{adjustbox}%

\medskip 
\noindent
\footnotesize%
Note -- The table provides coverage rates and widths of confidence intervals
of level $95\%$ for the Lorenz curve (LC) ordinate difference at percentile $%
p=0.5$ by asymptotic intersection method (IM-Asym), bootstrap intersection
method (IM-Boot), asymptotic method with overlapping samples (OS-Asym) and
bootstrap method with overlapping samples (OS-Boot). The data are generated
according to DGP I-D for given Gaussian copula $\rho $ in %
\eqref{eq:DGP-GaussCopula}, overlap portion $\lambda $ and sample size $n$.
The numbers are based on $1000$ replications with 399 bootstrap repetitions
each.

\normalsize%

\end{table}%

\end{document}

%% file: Dufour_arXivPreamble.tex
\usepackage[utf8]{inputenc}
\usepackage{amssymb}
\usepackage{amsfonts}
\usepackage{amsmath}
\usepackage{mathptmx}
\usepackage{eurosym}
\usepackage{graphicx}
\usepackage{rotating}
\usepackage[justification=centering]{caption}
\usepackage{comment}
\usepackage{booktabs}
\usepackage{placeins}

\input{DufourP0}
\SpecialFonts
\excludecomment{Abstract}
\excludecomment{DP}
\excludecomment{graphicpack}
\excludecomment{Plan}
\excludecomment{ps2pdf}
\includecomment{pdflatex}
\includecomment{Alternative}
\includecomment{Old}
\includecomment{Tentative}
\includecomment{Elementary}
\includecomment{Intermediate}
\includecomment{Advanced}
\includecomment{French}
\includecomment{reversemargin}
\includecomment{AbstractF}
\includecomment{TOCLists}
\includecomment{Calculations}
\includecomment{Article}
\includecomment{Book}
\includecomment{DP-W}
\includecomment{DP-CIR}
\includecomment{DP-SSRN}
\includecomment{DP-arXiv}
\includecomment{Journal}
\includecomment{Lectures}
\includecomment{Organization}
\excludecomment{SlidesIn} \includecomment{SlidesOut} \includecomment{Slides}
\includecomment{SlidesLandscapeOut} \includecomment{SlidesLandscape}
\input{DufourC0}

\input{DufourC1_SeparateThNumberingSecArt}

\providecommand{\vartitleadjust}{}
\renewcommand{\vartitleadjust}{}

\providecommand{\varthanks}{\ \ \standardthanks}
\renewcommand{\varthanks}{\ \ \standardthanks}
\renewcommand{\varthanks}{\ \ This work was supported by the William Dow Chair in Political
  Economy (McGill University), Tianjin University, the Bank of Canada (Research Fellowship), 
  the Toulouse School of Economics (Pierre-de-Fermat Chair of excellence), 
  the Universitad Carlos III de Madrid (Banco Santander de Madrid Chair of excellence), 
  the Natural Sciences and Engineering Research Council of Canada, and the Social Sciences and Humanities Research
  Council of Canada.}
\providecommand{\thanksvar}{\thanks{\varthanks}}
\renewcommand{\thanksvar}{\thinspace \thanks{\ \ The authors thank Md Nazmul Ahsan, 
Russell Davidson, John Galbraith, Silvia Goncalves, James MacKinnon, Russell Steele, Brennan Thompson, 
Victoria Zinde-Walsh, Haitian Xie, Jun Cai, and all seminar participants 
at McGill, the 22nd China Economic Annual Conference, IAAE2024, AMES2024, and the 2025 Canadian Econometric Study Group. \varthanks}}

\providecommand{\vartitle}{\vartitleadjust Nonparametric methods for comparing distribution functionals for dependent samples with application to welfare}
\renewcommand{\vartitle}{\vartitleadjust Nonparametric methods for comparing distribution functionals for dependent samples with application to welfare}
\renewcommand{\vartitle}{\vartitleadjust Nonparametric methods for comparing distribution functionals for dependent samples with application to inequality measures}

\providecommand{\TianyuAddress}{Ma Yinchu School of Economics, Tianjin University;
  e-mail: he\_tianyu@tju.edu.cn}
\renewcommand{\TianyuAddress}{Ma Yinchu School of Economics, Tianjin University; 
  e-mail: he\_tianyu@tju.edu.cn}

\providecommand{\varAuthors}{Jean-Marie Dufour \thanks{\ \ \DufourAddress} \\
 McGill University}
\renewcommand{\varAuthors}{Jean-Marie Dufour \thanks{\ \ \DufourAddress} \\
 McGill University \and Tianyu He\thanks{\ \ \TianyuAddress} \\
  Tianjin University}

\providecommand{\vardate}{\today, \texttime}
\providecommand{\vardate}{\dateformat{\August 2019}{\revised{\April 2020, \September 2021}}{\thisversion{\January 2023}}}
\renewcommand{\vardate}{\firstversion{\August 2019} \\ \revised{\April 2020, \September 2021,       \September 2022, \January 2023, \December 2024} 
     \\ \thisversion{\December 2025} \\ \compiled{\today, \texttime}}
 \renewcommand{\vardate}{\firstversion{\October 2023} \\ \revised{\December 2024, \September 2025, \December 2025} 
\\ \thisversion{\December 2025} \\ \compiled{\today, \texttime}}
\renewcommand{\vardate}{\December 2025}

\providecommand{\forthcoming}{\vfill \noindent This paper has been published 
    in \emph{Econometrica}.}
\renewcommand{\forthcoming}{}

\providecommand{\listttheoremnameb}{List of Definitions, Assumptions, Propositions 
     and Theorems}
\renewcommand{\listttheoremnameb}{List of Definitions, Assumptions, Propositions 
     and Theorems}

\renewcommand{\thefootnote}{\alph{footnote}}

\begin{French}
\end{French}

\begin{graphicpack}
\end{graphicpack}

\begin{DP}
\end{DP}

\begin{DP-W}
\end{DP-W}

\begin{DP-CIR}
\end{DP-CIR}

\begin{DP-SSRN}
\end{DP-SSRN}

\begin{DP-arXiv}
\end{DP-arXiv}

\begin{Lectures} 
\end{Lectures}

\begin{Organization}
\end{Organization}

\begin{reversemargin}
\end{reversemargin}

\begin{SlidesIn}
\input{DufourC1Slides}
\end{SlidesIn}

\begin{SlidesOut}
\end{SlidesOut}

\begin{Slides}
\end{Slides}

\begin{SlidesLandscapeOut}
\end{SlidesLandscapeOut}

\begin{Article} 
\end{Article}

\begin{Journal}
\end{Journal}

\input{tcilatex}

%% file: DufourP0.tex
\input{DufourP2_Base}

%% file: DufourP2_Base.tex
\usepackage{ifthen} 
\usepackage{xifthen} 

\usepackage{calc}  
\usepackage{time}  
\usepackage{clock}  

\usepackage{harvard}

\providecommand{\cite}{\citeasnoun}
\renewcommand{\cite}{\citeasnoun}

\usepackage{bibunits}

\usepackage{fancybox}
  
\usepackage{adjustbox} 

\usepackage[justification=centering]{caption} 

\usepackage{section}
\usepackage{theorem}
\usepackage{verbatim}
\providecommand{\bTentative}{}

\usepackage{eurosym} 
\usepackage{fix-cm}  
\usepackage{textcomp} 
\usepackage{pifont}  
\usepackage[T1]{fontenc}
    
\providecommand{\SpecialFonts}{
    \usepackage{calligra} 
	
    \usepackage[varumlaut]{yfonts}  
    }

\usepackage{pdfpages} 

\usepackage{gloss}
\usepackage{makeidx} 

\usepackage{enumerate}  
\usepackage{etaremune}  

\usepackage{amssymb}  
\usepackage{amsfonts}  
\usepackage{amsmath}  

\usepackage{array}
\usepackage{bbold} 
\usepackage{dsfont}  
\usepackage{fixmath} 
\usepackage{mathtools}  
\usepackage{mathrsfs} 

\usepackage{endnotes}

\usepackage{afterpage} 
\usepackage{layout} 
\usepackage{lscape} 

\usepackage{setspace} 

\usepackage{url}  

\usepackage{longtable} 

\usepackage{subfigure}

\providecommand{\paraNumbering}{\theoremstyle{change}}

%% file: DufourC0.tex
\input{DufourC1_Base}

%% file: DufourC1_Base.tex
\providecommand{\sectitlesize}{}

\newcommand{\mciteA}{\citename}
\newcommand{\mciteY}{\citeyear*}

\newcommand{\mciteAYY}[3][]{\mciteA{#2} (\mciteY{#2}, \mciteY{#3}#1)}

\newcommand{\cbu}{,\xspace}
\newcommand{\dbu}{.\xspace}
\newcommand{\bbibuniteconometrica}{\begin{bibunit}[econometrica] 
    \renewcommand{\cite}{\nocite*}
    \renewcommand\refname{} \renewcommand{\cbu}{} 
    \renewcommand{\dbu}{} \renewcommand{\newline}{}
    \vspace{-3.5\baselineskip}}
\newcommand{\bbibunitagsm}{\begin{bibunit}[agsm] \renewcommand{\cite}{\nocite*}
    \renewcommand\refname{} \renewcommand{\cbu}{} \renewcommand{\dbu}{} 
    \renewcommand{\newline}{}
    \vspace{-3.5\baselineskip}}
\newcommand{\ebibunit}{\putbib[refer1] \end{bibunit}}

\newcommand{\bbibunitunsrt}{\begin{bibunit}[unsrt] \renewcommand{\cite}{\nocite*}
    \renewcommand\refname{} \renewcommand{\cbu}{} \renewcommand{\dbu}{} \renewcommand{\newline}{}
    \vspace{-3.5\baselineskip}}
\newcommand{\ebibunitunsrt}{\putbib[refer1] \end{bibunit}}

\makeindex  
\makeglossary  
\makegloss

\newcommand{\NumericNumberedLists}{
\def\labelenumi{\arabic{enumi}.}
\def\theenumi{\arabic{enumi}}
\def\labelenumii{\arabic{enumii}.}
\def\theenumii{\arabic{enumii}}
\def\p@enumii{\theenumi.}
\def\labelenumiii{\arabic{enumiii}.}
\def\theenumiii{\arabic{enumiii}}
\def\p@enumiii{\theenumi.\theenumii.}
\def\labelenumiv{\arabic{enumiv}.}
\def\theenumiv{\arabic{enumiv}}
\def\p@enumiv{\p@enumiii.\theenumiii}
}

\allowdisplaybreaks

\providecommand{\newpageSlides}{}
\providecommand{\sectitlesize}{}

\newlength{\totalhormargin}
\newlength{\totalvermargin}

\providecommand{\newpageSlides}{}
\providecommand{\slideFontSize}{}

\newcommand{\December}{December }

\def\todayMY{\ifcase\month\or
  January\or February\or March\or April\or May\or June\or
  July\or August\or September\or October\or November\or
  December\fi\ \number\year}

\providecommand{\vartitleadjust}{}
\providecommand{\vartitle}{}

\providecommand{\varAuthors}{Jean-Marie Dufour \thanks{\ \ \DufourAddress} \\
  McGill University}

\providecommand{\vardate}{\today, \texttime}

\providecommand{\forthcoming}{}

\providecommand{\varthanks}{\ \ \standardthanksIFM}
\providecommand{\thanksvar}{\thanks{\varthanks}}

\providecommand{\DufourAddress}{William Dow Professor of Economics, McGill University,
  Centre interuniversitaire de recherche en analyse des
  organisations (CIRANO), and Centre interuniversitaire de recherche en
  \'{e}conomie quantitative (CIREQ). Mailing address:
  Department of Economics, McGill University, Leacock Building, Room 414,
  855 Sherbrooke Street West, Montr\'{e}al, Qu\'{e}bec H3A 2T7, Canada.
  TEL: (1) 514 398 6071; FAX: (1) 514 398 4800; e-mail: 
  \protect\url=jean-marie.dufour@mcgill.ca=\thinspace. Web page:
  \protect\url{http://www.jeanmariedufour.com} }

\newcommand{\sptha}{\hspace{-0.01em}}
\newcommand{\spthb}{\hspace{-0.01em}}

\newcommand{\sppr}{}
\newcommand{\theoremname}{Theorem}
\newcommand{\acknowledgementname}{Acknowledgement}
\newcommand{\algorithmname}{Algorithm}
\newcommand{\assumptionname}{Assumption}
\newcommand{\axiomname}{Axiom}
\newcommand{\casename}{Case}
\newcommand{\claimname}{Claim}
\newcommand{\conclusionname}{Conclusion}
\newcommand{\conditionname}{Condition}
\newcommand{\conjecturename}{Conjecture}
\newcommand{\corollaryname}{Corollary}
\newcommand{\criterionname}{Criterion}
\newcommand{\definitionname}{Definition}
\newcommand{\examplename}{Example}
\newcommand{\exercisename}{Exercise}
\newcommand{\lemmaname}{Lemma}
\newcommand{\notationname}{Notation}
\newcommand{\problemname}{Problem}
\newcommand{\proofname}{\sppr Proof}

\newcommand{\propertyname}{\sppr Property}
\newcommand{\propositionname}{Proposition}
\newcommand{\reflistname}{References}
\newcommand{\remarkname}{Remark}
\newcommand{\resultname}{Result}
\newcommand{\solutionname}{Solution}

\newcommand{\summaryname}{Summary}

\providecommand{\proofof}{Proof of }
\renewcommand{\proofof}{Proof of }

\newenvironment{proof}[1][\sptha \proofname]{\par
  \normalfont
  \trivlist
  \item[\hskip\labelsep\scshape
    #1{.}]\ignorespaces}
    {\qed\endtrivlist\vspace{\baselineskip}}

\newenvironment{proofnoname}[1][\sptha \proofname]{\par
  \noindent
  \normalfont}%
  {\qed\vspace{\baselineskip}}

\newenvironment{proofflex}[1][\spthb \proofname]{\par
  \normalfont
  \trivlist
  \item[\hskip\labelsep\scshape
    #1{\ }]\ignorespaces}
    {\qed\endtrivlist\vspace{\baselineskip}}

\newenvironment{proofflexb}[1][\spthb \proofname]{\par
  \normalfont
  \trivlist
  \item[\hskip\labelsep\scshape
    #1{\ }]\ignorespaces}
    {\qed\endtrivlist\vspace{\baselineskip}}

\newenvironment{proofflexc}[1][\spthb \proofname]{\par
  \noindent
  \normalfont}
  {\qed\vspace{\baselineskip}}

\newenvironment{proofflexwc}[1][\spthb \proofname]{\par
  \normalfont
  \trivlist
  \item[\hskip\labelsep\scshape
    #1{\ }]\ignorespaces}

\newenvironment{proofflexd}[1][\spthb \proofname]{\par
  \noindent
  \normalfont}
  {\vspace{-0.5\baselineskip}}

\newenvironment{sec2subsec}{\renewcommand{\section}{\subsection}}{}
\newenvironment{reflist}{\quad \newline \noindent {\Large \bf \reflistname}
  \newline \quad \newline
  \begin{list}{}{\itemindent -.15in \leftmargin .15in \parsep 0in
                 \itemsep 0in} \item \vspace{-.35in} }{\end{list}}

\newenvironment{npar}{\noindent \bf{\thesection.\thenpar}}{}
\newenvironment{subnpar}{\noindent \bf{\thesubsection.\thesubnpar}}{}

\newcounter{paran}

\newcounter{npar}[section]
\newcounter{nparr}[section]

\newcounter{subnpar}[subsection]
\newcounter{subnparr}[subsection]

\DeclareRobustCommand{\qed}{%
  \ifmmode 
  \else \leavevmode\unskip\penalty9999 \hbox{}\nobreak\hfill
  \fi
  \quad\hbox{\qedsymbol}}
\newcommand{\openbox}{\leavevmode
  \hbox to.77778em{%
  \hfil\vrule
  \vbox to.675em{\hrule width.6em\vfil\hrule}%
  \vrule\hfil}}
\DeclareRobustCommand{\qeddirect}{%
  \ifmmode 
  \else \leavevmode\unskip\penalty9999 \hbox{}\nobreak\hfill
  \fi
  \quad\hbox{\qedsymbol}}
\providecommand{\qedsymbol}{\openbox} 

\newcommand{\bproofin}{\begin{proof}}
\newcommand{\eproofin}{\end{proof}}
\newcommand{\bproofend}{\begin{proofnoname}}
\newcommand{\eproofend}{\end{proofnoname}}
\newcommand{\bproofth}{\begin{proofth}}
\newcommand{\eproofth}{\end{proofth}}

\providecommand{\proofof}{Proof of }
\renewcommand{\proofof}{Proof of }

\newcommand{\thsection}{\thesection}

\newcommand{\thsectioneq}{\thesection}

\newcommand{\thsec}{\thsection}
\newcommand{\thseceq}{\thsectioneq}
\providecommand{\thetheorem}{{\bf \thsec.\arabic{theorem}}}
\renewcommand{\thetheorem}{{\bf \thsec.\arabic{theorem}}}
\renewcommand{\theequation}{{\rm \thseceq.\arabic{equation}}}

\makeatletter
   \long
\def\@makecaption#1#2{\vskip 0\p@
   \setbox\@tempboxa\hbox{#1 #2}}
\makeatother

\makeatletter

\providecommand{\l@theorem}{\@dottedtocline{1}{0em}{5em}}
\renewcommand{\l@theorem}{\@dottedtocline{1}{0em}{5em}}

\providecommand{\listttheoremnameb}{\sectitlesize List of Definitions, Assumptions, Propositions and Theorems}

\newcommand\listoftheorems{
 \section*{\listttheoremnameb
           \@mkboth{\MakeUppercase\listttheoremnameb}
           {\MakeUppercase\listttheoremnameb}}
           \@starttoc{lth}
           }
\newcommand{\listoftheoremscont}[1]{
 \providecommand{\listttheoremnameb}{#1}
 \renewcommand{\listttheoremnameb}{#1}
 \section*{\listttheoremnameb
           \@mkboth{\MakeUppercase\listttheoremnameb}
           {\MakeUppercase\listttheoremnameb}}
           \@starttoc{lth}
                      \addcontentsline{toc}{section}{\listttheoremnameb}
           }

\makeatother
\providecommand{\contentshift}{\hspace{-3.5em}}
\renewcommand{\contentshift}{\hspace{-3.5em}}
\providecommand{\contentshiftS}{\hspace{0em}}
\renewcommand{\contentshiftS}{\hspace{0em}}

\newcommand{\sppra}{\hspace{-0.2em}}
\newcommand{\indentprc}{}

\newcommand{\captionproofflex}[2]{}

\newcommand{\captionproofflexc}[2]{
 \sppra\indentprc\textsc{Proof of #1 \protect\ref{#2} }\quad
 \addcontentsline{lth}{theorem}{\protect\numberline{}
  {\contentshift \proofof #1 \protect\ref{#2} }}}

\newcommand{\bAssumptionA}{\begin{assumption}}
\newcommand{\eAssumptionA}{\end{assumption}}

%% file: DufourC1_SeparateThNumberingSecArt.tex
\providecommand{\paraNumbering}{}
\renewcommand{\paraNumbering}{}
\paraNumbering

\newtheorem{theorem}{\sptha \theoremname}[section]
\newtheorem{theoremseq}{\sptha \theoremname}
{\theorembodyfont{\normalfont}%
 \newtheorem{theoremSeqRom}{\sptha \theoremname}}
\newtheorem{acknowledgement}{\sptha \acknowledgementname}[section]
\newtheorem{algorithm}{\sptha \algorithmname}[section]
\newtheorem{assumption}{\sptha \assumptionname}[section]
\newtheorem{assumptionSec}{\sptha \assumptionname}[section]
\newtheorem{assumptionV}{\sptha \assumptionname}[section]
\newtheorem{assumptionLetter}{\sptha \assumptionname}

\newtheorem{axiom}{\sptha \axiomname}[section]
\newtheorem{case}{\sptha \casename}[section]
\newtheorem{claim}{\sptha \claimname}[section]
\newtheorem{conclusion}{\sptha \conclusionname}[section]
\newtheorem{condition}{\sptha \conditionname}[section]
\newtheorem{conjecture}{\sptha \conjecturename}[section]
\newtheorem{corollary}[theorem]{\sptha \corollaryname}
\newtheorem{criterion}{\sptha \criterionname}[section]
\newtheorem{definition}{\sptha \definitionname}[section]
\newtheorem{definitionSec}{\sptha \definitionname}[section]
{\theorembodyfont{\normalfont}%
    \newtheorem{example}{\sptha \examplename}[section]}
{\theorembodyfont{\normalfont}
    \newtheorem{exampleSec}{\sptha \examplename}[section]}
\newtheorem{exercise}{\sptha \exercisename}[section]
\newtheorem{lemma}[theorem]{\sptha \lemmaname}
\newtheorem{lemmaSec}{\sptha \lemmaname}[section]
\newtheorem{notation}{\sptha \notationname}[section]
\newtheorem{problem}{\sptha \problemname}[section]
{\theorembodyfont{\normalfont} 
    \newtheorem{proofth}{\sptha \proofname}[section]}
\newtheorem{property}[theorem]{\sptha \propertyname}
\newtheorem{proposition}[theorem]{\sptha \propositionname}
\newtheorem{propositionSec}{\sptha \propositionname}[section]
{\theorembodyfont{\normalfont} 
    \newtheorem{remark}{\sptha \remarkname}[section]}
    {\theorembodyfont{\normalfont}%
\newtheorem{result}{\sptha \resultname}[section]}
\newtheorem{solution}{\sptha \solutionname}[section]
\newenvironment{statement}{}{}
\newtheorem{summary}{\sptha \summaryname}[section]

\providecommand{\resetcountersSection}{\renewcommand{\thsec}{\thsection} 
  \renewcommand{\thseceq}{\thsectioneq} 
  \setcounter{theorem}{0} \setcounter{definition}{0} \setcounter{equation}{0}}
  
\newcommand{\captiontheorem}[2]{\textsc{ #2}.
 \addcontentsline{lth}{theorem}{{\bf #1 \contentshiftS  \protect\numberline{\thetheorem}}
 {\contentshift   : \contentshiftS #2}}}
\newcommand{\captionassumption}[2]{\textsc{ #2}.
  \addcontentsline{lth}{theorem}{{\bf #1 \contentshiftS  \protect\numberline{\theassumption}}
  {\contentshift  : \contentshiftS #2}}}

\newcommand{\captiondefinition}[2]{\textsc{ #2}.
 \addcontentsline{lth}{theorem}{{\bf #1 \contentshiftS \protect\numberline{\thedefinition}}
 {\contentshift  : \contentshiftS #2}}}

\newcommand{\captionlemma}[2]{\textsc{ #2}.
 \addcontentsline{lth}{theorem}{{\bf #1 \contentshiftS \protect\numberline{\thelemma}}
 {\contentshift  : \contentshiftS #2}}}

\newcommand{\captionproposition}[2]{\textsc{ #2}.
\addcontentsline{lth}{theorem}{{\bf #1 \contentshiftS \protect\numberline{\theproposition}}
 {\contentshift : \contentshiftS #2}}}

\newcommand{\captionresult}[2]{\textsc{ #2}.
 \addcontentsline{lth}{theorem}{{\bf #1 \contentshiftS \protect\numberline{\theresult}}
 {\contentshift  : \contentshiftS #2}}}

\newcommand{\captionproofemptynocontent}[2]{}

\newcommand{\captionproofin}[2]{}


%% file: DufourC1Slides.tex
\providecommand{\sectitlesize}{\huge}
\renewcommand{\sectitlesize}{\huge}
\providecommand{\newpageSlides}{\newpage}
\renewcommand{\newpageSlides}{\newpage}

\providecommand{\vartitleadjust}{\Huge}
\renewcommand{\vartitleadjust}{\Huge}


%% file: HTYRefer.bib
@Article{Raffinetti2015,
  author   = {Raffinetti, Emanuela and Siletti, Elena and Vernizzi, Achille},
  journal  = {Statistical Methods \& Applications},
  title    = {On the {Gini} coefficient normalization when attributes with negative values are considered},
  year     = {2015},
  issn     = {1618-2510, 1613-981X},
  month    = sep,
  number   = {3},
  pages    = {507--521},
  volume   = {24},
  doi      = {10.1007/s10260-014-0293-4},
  language = {en},
  url      = {http://link.springer.com/10.1007/s10260-014-0293-4},
  urldate  = {2019-07-30},
}

@Article{Davidson2009,
  author   = {Davidson, Russell},
  journal  = {Journal of Econometrics},
  title    = {Reliable inference for the {Gini} index},
  year     = {2009},
  issn     = {03044076},
  month    = may,
  number   = {1},
  pages    = {30--40},
  volume   = {150},
  doi      = {10.1016/j.jeconom.2008.11.004},
  language = {en},
  url      = {https://linkinghub.elsevier.com/retrieve/pii/S0304407609000323},
  urldate  = {2019-07-31},
}

@Article{Raffinetti2017,
  author     = {Raffinetti, Emanuela and Siletti, Elena and Vernizzi, Achille},
  journal    = {Social Indicators Research},
  title      = {Analyzing the {Effects} of {Negative} and {Non}-negative {Values} on {Income} {Inequality}: {Evidence} from the {Survey} of {Household} {Income} and {Wealth} of the {Bank} of {Italy} (2012)},
  year       = {2017},
  issn       = {0303-8300, 1573-0921},
  month      = aug,
  number     = {1},
  pages      = {185--207},
  volume     = {133},
  doi        = {10.1007/s11205-016-1354-x},
  language   = {en},
  shorttitle = {Analyzing the {Effects} of {Negative} and {Non}-negative {Values} on {Income} {Inequality}},
  url        = {http://link.springer.com/10.1007/s11205-016-1354-x},
  urldate    = {2019-08-01},
}

@Article{Barrett2009,
  author    = {Garry F. Barrett and Stephen G. Donald},
  journal   = {Journal of Business \& Economic Statistics},
  title     = {Statistical inference with generalized {Gini} indices of inequality, poverty, and welfare},
  year      = {2009},
  number    = {1},
  pages     = {1-17},
  volume    = {27},
  doi       = {10.1198/jbes.2009.0001},
  eprint    = {https://doi.org/10.1198/jbes.2009.0001},
  publisher = {Taylor & Francis},
  url       = {https://doi.org/10.1198/jbes.2009.0001},
}

@Article{Davidson2010a,
  author     = {Davidson, Russell},
  journal    = {Canadian Journal of Economics/Revue canadienne d'économique},
  title      = {Innis {Lecture}: {Inference} on income distributions},
  year       = {2010},
  issn       = {1540-5982},
  number     = {4},
  pages      = {1122--1148},
  volume     = {43},
  copyright  = {© Canadian Economics Association},
  doi        = {10.1111/j.1540-5982.2010.01608.x},
  language   = {en},
  shorttitle = {Innis {Lecture}},
  url        = {http://onlinelibrary.wiley.com/doi/abs/10.1111/j.1540-5982.2010.01608.x},
  urldate    = {2019-08-03},
}

@Unpublished{Bouezmarni2024,
  author = {Bouezmarni, Taoufik and Dufour, Jean-Marie},
  note   = {Working paper},
  title  = {Winners and losers: extended {Lorenz} curves and {Gini} coefficients for possibly negative variables},
  month  = aug,
  year   = {2024},
}

@Article{Cowell1996,
  author    = {Frank A. Cowell and Maria-Pia Victoria-Feser},
  journal   = {Econometrica},
  title     = {{Robustness Properties of Inequality Measures}},
  year      = {1996},
  issn      = {00129682, 14680262},
  number    = {1},
  pages     = {77--101},
  volume    = {64},
  publisher = {[Wiley, Econometric Society]},
  url       = {http://www.jstor.org/stable/2171925},
}

@Book{Yitzhaki2013,
  author    = {Yitzhaki, Shlomo and Schechtman, Edna},
  publisher = {Springer New York},
  title     = {{The Gini Methodology}},
  year      = {2013},
  isbn      = {9781461447207},
  volume    = {272},
  doi       = {10.1007/978-1-4614-4720-7},
  issn      = {0172-7397},
  journal   = {Springer Series in Statistics},
}

@Book{Shao1995,
  author    = {Shao, Jun and Tu, Dongsheng},
  publisher = {Springer Science \& Business Media},
  title     = {The jackknife and bootstrap},
  year      = {1995},
  isbn      = {1-4612-0795-9},
}

@Article{Bhattacharya2007,
  author   = {Bhattacharya, Debopam},
  journal  = {Journal of Econometrics},
  title    = {{Inference on inequality from household survey data}},
  year     = {2007},
  month    = {April},
  number   = {2},
  pages    = {674-707},
  volume   = {137},
  url      = {https://ideas.repec.org/a/eee/econom/v137y2007i2p674-707.html},
}

@Article{Bhattacharya2005,
  author   = {Bhattacharya, Debopam},
  journal  = {Journal of Econometrics},
  title    = {{Asymptotic inference from multi-stage samples}},
  year     = {2005},
  month    = {May},
  number   = {1},
  pages    = {145-171},
  volume   = {126},
  url      = {https://ideas.repec.org/a/eee/econom/v126y2005i1p145-171.html},
}

@TechReport{Dufour2023b,
  author      = {Jean-Marie Dufour and Emmanuel Flachaire and Lynda Khalaf and Abdallah Zalghout},
  institution = {CIRANO},
  title       = {Identification-robust methods for comparing inequality with an application to regional disparities},
  year        = {2023},
  month       = may,
  type        = {CIRANO Working Papers},
  url         = {https://ideas.repec.org/p/cir/cirwor/2020s-23.html},
}

@Article{Gastwirth1971,
  author   = {Gastwirth, Joseph L.},
  journal  = {Econometrica},
  title    = {A {General} {Definition} of the {Lorenz} {Curve}},
  year     = {1971},
  issn     = {00129682},
  month    = nov,
  number   = {6},
  pages    = {1037},
  volume   = {39},
  comment  = {1. def in terms of quantile 2. LC is continuous whether RV is continuous or not !! (LC is an integral not integrand!)},
  doi      = {10.2307/1909675},
  language = {en},
  url      = {https://www.jstor.org/stable/1909675?origin=crossref},
  urldate  = {2019-08-23},
}

@Article{Zheng2002,
  author    = {Zheng, Buhong},
  journal   = {Econometrica},
  title     = {Testing {Lorenz} {Curves} with {Non}-{Simple} {Random} {Samples}},
  year      = {2002},
  issn      = {1468-0262},
  number    = {3},
  pages     = {1235--1243},
  volume    = {70},
  copyright = {The Econometric Society 2002},
  doi       = {10.1111/1468-0262.00325},
  language  = {en},
  url       = {https://onlinelibrary.wiley.com/doi/abs/10.1111/1468-0262.00325},
  urldate   = {2019-08-23},
}

@Article{Beach1983,
  author   = {Charles M. Beach and Russell Davidson},
  journal  = {Review of Economic Studies},
  title    = {Distribution-free statistical inference with {Lorenz} curves and income shares},
  year     = {1983},
  number   = {4},
  pages    = {723-735},
  volume   = {50},
  url      = {https://ideas.repec.org/a/oup/restud/v50y1983i4p723-735..html},
}

@InCollection{Cowell2015,
  author    = {Frank A. Cowell and Emmanuel Flachaire},
  booktitle = {Handbook of Income Distribution},
  publisher = {Elsevier},
  title     = {Chapter 6 - Statistical Methods for Distributional Analysis},
  year      = {2015},
  editor    = {Anthony B. Atkinson and François Bourguignon},
  pages     = {359-465},
  series    = {Handbook of Income Distribution},
  volume    = {2},
  doi       = {https://doi.org/10.1016/B978-0-444-59428-0.00007-2},
  issn      = {1574-0056},
  url       = {https://www.sciencedirect.com/science/article/pii/B9780444594280000072},
}

@Book{Kleiber2003,
  author    = {Kleiber, Christian and Kotz, Samuel},
  publisher = {John Wiley \& Sons},
  title     = {Statistical size distributions in economics and actuarial sciences},
  year      = {2003},
  isbn      = {0-471-45716-7},
  volume    = {470},
}

@Article{McDonald1984,
  author  = {McDonald, James B},
  journal = {Econometrica},
  title   = {Some {Generalized} {Functions} for the {Size} {Distribution} of {Income}},
  year    = {1984},
  number  = {3},
  pages   = {647--665},
  volume  = {52},
}

@Article{Kakwani1998,
  author     = {Kakwani, Nanak and Lambert, Peter J.},
  journal    = {European Journal of Political Economy},
  title      = {On measuring inequity in taxation: a new approach},
  year       = {1998},
  issn       = {01762680},
  month      = may,
  number     = {2},
  pages      = {369--380},
  volume     = {14},
  doi        = {10.1016/S0176-2680(98)00012-3},
  language   = {en},
  shorttitle = {On measuring inequity in taxation},
  url        = {https://linkinghub.elsevier.com/retrieve/pii/S0176268098000123},
  urldate    = {2020-03-25},
}

@Article{Davidson2007a,
  author   = {Russell Davidson and Emmanuel Flachaire},
  journal  = {Journal of Econometrics},
  title    = {Asymptotic and bootstrap inference for inequality and poverty measures},
  year     = {2007},
  issn     = {0304-4076},
  note     = {Semiparametric methods in econometrics},
  number   = {1},
  pages    = {141-166},
  volume   = {141},
  doi      = {https://doi.org/10.1016/j.jeconom.2007.01.009},
  url      = {https://www.sciencedirect.com/science/article/pii/S0304407607000115},
}

@Article{Zheng2001,
  author   = {Zheng, Buhong and J. Cushing, Brian},
  journal  = {Journal of Econometrics},
  title    = {Statistical inference for testing inequality indices with dependent samples},
  year     = {2001},
  issn     = {03044076},
  month    = apr,
  number   = {2},
  pages    = {315--335},
  volume   = {101},
  doi      = {10.1016/S0304-4076(00)00087-7},
  language = {en},
  url      = {https://linkinghub.elsevier.com/retrieve/pii/S0304407600000877},
  urldate  = {2020-06-30},
}

@Article{Hansen2019,
  author   = {Hansen, Bruce E. and Lee, Seojeong},
  journal  = {Journal of Econometrics},
  title    = {Asymptotic theory for clustered samples},
  year     = {2019},
  issn     = {03044076},
  month    = jun,
  number   = {2},
  pages    = {268--290},
  volume   = {210},
  doi      = {10.1016/j.jeconom.2019.02.001},
  language = {en},
  url      = {https://linkinghub.elsevier.com/retrieve/pii/S0304407619300247},
  urldate  = {2020-07-12},
}

@Article{Beran1988,
  author    = {Beran, Rudolf},
  journal   = {Journal of the American Statistical Association},
  title     = {Prepivoting test statistics: a bootstrap view of asymptotic refinements},
  year      = {1988},
  number    = {403},
  pages     = {687--697},
  volume    = {83},
  publisher = {Taylor \& Francis},
}

@InCollection{Dufour1998,
  author    = {Dufour, Jean-Marie and Torr{\`e}s, Olivier},
  booktitle = {Handbook of Applied Economic Statistics},
  publisher = {CRC Press},
  title     = {{Union-Intersection and Sample-Split Methods in Econometrics with Applications to MA and SURE Models}},
  year      = {1998},
  chapter   = {14},
  editor    = {Aman Ullah and David E. A. Giles},
  pages     = {485--491},
}

@Article{Zheng2004,
  author   = {Zheng, Buhong},
  journal  = {Journal of Applied Econometrics},
  title    = {Poverty comparisons with dependent samples},
  year     = {2004},
  number   = {3},
  pages    = {419-428},
  volume   = {19},
  doi      = {https://doi.org/10.1002/jae.779},
  eprint   = {https://onlinelibrary.wiley.com/doi/pdf/10.1002/jae.779},
  url      = {https://onlinelibrary.wiley.com/doi/abs/10.1002/jae.779},
}

@Article{Davidson2012,
  author   = {Davidson, Russell},
  journal  = {The Econometrics Journal},
  title    = {{Statistical inference in the presence of heavy tails}},
  year     = {2012},
  issn     = {1368-4221},
  month    = {02},
  number   = {1},
  pages    = {C31-C53},
  volume   = {15},
  doi      = {10.1111/j.1368-423X.2010.00340.x},
  eprint   = {https://academic.oup.com/ectj/article-pdf/15/1/C31/27677789/ectj0c31.pdf},
  url      = {https://doi.org/10.1111/j.1368-423X.2010.00340.x},
}

@Article{Biewen2002,
  author   = {Martin Biewen},
  journal  = {Journal of Econometrics},
  title    = {Bootstrap inference for inequality, mobility and poverty measurement},
  year     = {2002},
  issn     = {0304-4076},
  number   = {2},
  pages    = {317-342},
  volume   = {108},
  doi      = {https://doi.org/10.1016/S0304-4076(01)00138-5},
  url      = {https://www.sciencedirect.com/science/article/pii/S0304407601001385},
}

@Article{Ibragimov2025,
  author    = {Ibragimov, Rustam and Kattuman, Paul and Skrobotov, Anton},
  journal   = {Econometric Reviews},
  title     = {Robust inference on income inequality: t- statistic based approach},
  year      = {2025},
  issn      = {1532-4168},
  month     = jan,
  number    = {4},
  pages     = {384--415},
  volume    = {44},
  doi       = {10.1080/07474938.2024.2432362},
  publisher = {Informa UK Limited},
}

@Article{Ibragimov2010,
  author    = {Ibragimov, Rustam and M{\"u}ller, Ulrich K.},
  journal   = {Journal of Business \& Economic Statistics},
  title     = {{t-Statistic} based correlation and heterogeneity robust inference},
  year      = {2010},
  number    = {4},
  pages     = {453--468},
  volume    = {28},
  comment   = {Questions:
1) seems addtionally impose assumption that mean of each group is identical ?  If so, isn't it restrictive ? E.g. suppose two groups, say A and B with conditional Gaussian distributions with means mu1 and mu2. Then the unconditional mean is then mu_un = p*mu1 + (1-p)*mu2. Suppose a (simple) random sample, then mean is mu_un. But grouping obs by group A and B will give mu1 and mu2.
2) if 1) is true, the null hypo becomes H0: mu1 = mu2 = mu_cn = c ?? (sufficient condition of orginal H0)},
  publisher = {Taylor \& Francis},
}

@Article{Ibragimov2016,
  author   = {Ibragimov, Rustam and Müller, Ulrich K.},
  journal  = {The Review of Economics and Statistics},
  title    = {{Inference with Few Heterogeneous Clusters}},
  year     = {2016},
  issn     = {0034-6535},
  month    = {03},
  number   = {1},
  pages    = {83-96},
  volume   = {98},
  doi      = {10.1162/REST_a_00545},
  eprint   = {https://direct.mit.edu/rest/article-pdf/98/1/83/1616480/rest\_a\_00545.pdf},
  url      = {https://doi.org/10.1162/REST\_a\_00545},
}

@TechReport{Dufour2024,
  author = {Dufour, Jean-Marie and He, Tianyu},
  title  = {{Measuring inequality for winners and losers: extended Lorenz curves and Gini indices for possibly negative variables}},
  year   = {2024},
  note   = {Working paper},
}

@Book{Wasserman2006,
  author    = {Larry Wasserman},
  publisher = {Springer New York},
  title     = {{All of Nonparametric Statistics}},
  year      = {2006},
  doi       = {https://doi.org/10.1007/0-387-30623-4},
}

@Article{Shi1986,
  author  = {Xiquan Shi},
  journal = {Chinese Journal of Applied Probability and Statistics},
  title   = {{A NOTE ON BOOTSTRAPPING THE U-STATISTICS}},
  year    = {1986},
  number  = {2},
  pages   = {144-148},
  volume  = {2},
}

@TechReport{Dufour2024a,
  author      = {Jean-Marie Dufour and Tianyu He},
  institution = {McGill University},
  title       = {Inference on the difference in the extended inequality measures for possibly negative variables},
  year        = {2024},
  type        = {techreport},
}

@Article{Horowitz2019,
  author    = {Joel L. Horowitz},
  journal   = {Annual Review of Economics},
  title     = {{Bootstrap Methods in Econometrics}},
  year      = {2019},
  month     = {aug},
  number    = {1},
  pages     = {193--224},
  volume    = {11},
  doi       = {10.1146/annurev-economics-080218-025651},
  publisher = {Annual Reviews},
}

@Book{Mammen1992,
  author    = {Enno Mammen},
  publisher = {Springer-Verlag},
  title     = {{When Does Bootstrap Work? Asymptotic Results and Simulations}},
  year      = {1992},
}

@Book{Cochran1977,
  author    = {Cochran, William Gemmell},
  publisher = {John Wiley \& Sons},
  title     = {Sampling techniques},
  year      = {1977},
}

@Article{Zhang2021a,
  author    = {Zhang, Shengyu and Martin, R. Douglas and Christidis, Anthony A.},
  journal   = {Journal of Mathematical Finance},
  title     = {{Influence Functions for Risk and Performance Estimators}},
  year      = {2021},
  issn      = {2162-2442},
  number    = {01},
  pages     = {15--47},
  volume    = {11},
  doi       = {10.4236/jmf.2021.111002},
  publisher = {Scientific Research Publishing, Inc.},
}

@TechReport{Dufour2024d,
  author      = {Jean-Marie Dufour and Tianyu He},
  institution = {McGill University},
  title       = {Finite-sample distribution-free confidence intervals for comparing means of bounded variables with application to poverty measures},
  year        = {2024},
}

@Article{Donaldson1980,
  author    = {Donaldson, David and Weymark, John A.},
  journal   = {Journal of Economic Theory},
  title     = {{A single-parameter generalization of the Gini indices of inequality}},
  year      = {1980},
  issn      = {0022-0531},
  month     = feb,
  number    = {1},
  pages     = {67--86},
  volume    = {22},
  doi       = {10.1016/0022-0531(80)90065-4},
  publisher = {Elsevier BV},
}

@Article{Chakravarty1988,
  author    = {Chakravarty, Satya R.},
  journal   = {International Economic Review},
  title     = {{Extended Gini Indices of Inequality}},
  year      = {1988},
  issn      = {0020-6598},
  month     = feb,
  number    = {1},
  pages     = {147},
  volume    = {29},
  doi       = {10.2307/2526814},
  publisher = {JSTOR},
}

@Article{Hampel1974,
  author    = {Hampel, Frank R.},
  journal   = {Journal of the American Statistical Association},
  title     = {The Influence Curve and its Role in Robust Estimation},
  year      = {1974},
  issn      = {1537-274X},
  month     = jun,
  number    = {346},
  pages     = {383--393},
  volume    = {69},
  doi       = {10.1080/01621459.1974.10482962},
  publisher = {Informa UK Limited},
}

@Book{Lehmann2022,
  author    = {E. L. Lehmann and Joseph P. Romano},
  publisher = {Springer International Publishing},
  title     = {{Testing Statistical Hypotheses}},
  year      = {2022},
  address   = {New York},
  edition   = {3. ed., correct. at 4. print.},
  isbn      = {9783030705787},
  series    = {Springer texts in statistics},
  doi       = {10.1007/978-3-030-70578-7},
  issn      = {2197-4136},
  journal   = {Springer Texts in Statistics},
  pagetotal = {784},
  ppn_gvk   = {1617850829},
}

@Article{MacKinnon2023c,
  author    = {MacKinnon, James G.},
  journal   = {Econometrics and Statistics},
  title     = {Fast cluster bootstrap methods for linear regression models},
  year      = {2023},
  issn      = {2452-3062},
  month     = apr,
  pages     = {52--71},
  volume    = {26},
  doi       = {10.1016/j.ecosta.2021.11.009},
  publisher = {Elsevier BV},
}

@Article{MacKinnon2023d,
  author    = {James G. MacKinnon and Morten Ørregaard Nielsen and Matthew D. Webb},
  journal   = {Journal of Econometrics},
  title     = {Cluster-robust inference: A guide to empirical practice},
  year      = {2023},
  issn      = {0304-4076},
  month     = feb,
  number    = {2},
  pages     = {272--299},
  volume    = {232},
  doi       = {10.1016/j.jeconom.2022.04.001},
  publisher = {Elsevier BV},
}

@Article{Djogbenou2019,
  author    = {Antoine A. Djogbenou and James G. MacKinnon and Morten Ørregaard Nielsen},
  journal   = {Journal of Econometrics},
  title     = {Asymptotic theory and wild bootstrap inference with clustered errors},
  year      = {2019},
  issn      = {0304-4076},
  month     = oct,
  number    = {2},
  pages     = {393--412},
  volume    = {212},
  doi       = {10.1016/j.jeconom.2019.04.035},
  publisher = {Elsevier BV},
}

@Article{MacKinnon2016,
  author    = {MacKinnon, James G. and Webb, Matthew D.},
  journal   = {Journal of Applied Econometrics},
  title     = {{Wild Bootstrap Inference for Wildly Different Cluster Sizes}},
  year      = {2016},
  issn      = {1099-1255},
  month     = feb,
  number    = {2},
  pages     = {233--254},
  volume    = {32},
  doi       = {10.1002/jae.2508},
  publisher = {Wiley},
}

@Article{Davidson2008a,
  author    = {Davidson, Russell and Flachaire, Emmanuel},
  journal   = {Journal of Econometrics},
  title     = {The wild bootstrap, tamed at last},
  year      = {2008},
  issn      = {0304-4076},
  month     = sep,
  number    = {1},
  pages     = {162--169},
  volume    = {146},
  doi       = {10.1016/j.jeconom.2008.08.003},
  publisher = {Elsevier BV},
}

@Article{Davidson2000a,
  author    = {Davidson, Russell and MacKinnon, James G.},
  journal   = {Econometric Reviews},
  title     = {Bootstrap tests: how many bootstraps?},
  year      = {2000},
  issn      = {1532-4168},
  month     = jan,
  number    = {1},
  pages     = {55--68},
  volume    = {19},
  doi       = {10.1080/07474930008800459},
  publisher = {Informa UK Limited},
}

@Article{Mills1997,
  author    = {Jeffrey A. Mills and Sourushe Zandvakili},
  journal   = {Journal of Applied Econometrics},
  title     = {{Statistical Inference Via Bootstrapping for Measures of Inequality}},
  year      = {1997},
  issn      = {1099-1255},
  month     = mar,
  number    = {2},
  pages     = {133--150},
  volume    = {12},
  doi       = {10.1002/(sici)1099-1255(199703)12:2<133::aid-jae433>3.0.co;2-h},
  publisher = {Wiley},
}

@Book{Davidson2004,
  author    = {Davidson, Russell and MacKinnon, James G.},
  publisher = {Oxford Univ. Press},
  title     = {Econometric theory and methods},
  year      = {2004},
  address   = {New York, NY [u.a.]},
  isbn      = {0195123727},
  note      = {Literaturverz. S. 702-721},
  issn      = {0195123727},
  pagetotal = {750},
  ppn_gvk   = {393847152},
}

@TechReport{MacKinnon2016a,
  author      = {MacKinnon, James G.},
  institution = {Queen’s University},
  title       = {{Inference with Large Clustered Datasets}},
  year        = {2016},
  address     = {2016-09},
  number      = {2110-2018-4501},
  doi         = {https://doi.org/10.22004/ag.econ.274691},
  pages       = {18},
  recid       = {274691},
  series      = {Working Paper No. 1365},
  url         = {http://ageconsearch.umn.edu/record/274691},
}

@Book{Hall1992,
  author    = {Hall, Peter},
  publisher = {Springer New York},
  title     = {{The Bootstrap and Edgeworth Expansion}},
  year      = {1992},
  isbn      = {9781461243847},
  doi       = {10.1007/978-1-4612-4384-7},
  issn      = {0172-7397},
  journal   = {Springer Series in Statistics},
}

@Article{Webb2023,
  author    = {Webb, Matthew D.},
  journal   = {Canadian Journal of Economics/Revue canadienne d’économique},
  title     = {Reworking wild bootstrap‐based inference for clustered errors},
  year      = {2023},
  issn      = {1540-5982},
  month     = may,
  number    = {3},
  pages     = {839--858},
  volume    = {56},
  doi       = {10.1111/caje.12661},
  publisher = {Wiley},
}
